\documentclass[conference]{IEEEtran}

\usepackage[dvipsnames]{xcolor}

\usepackage[colorlinks=true,linkcolor=ForestGreen,breaklinks=True,citecolor=brown,urlcolor=purple]{hyperref}

\usepackage{tikz}
\usetikzlibrary{positioning}
\usepackage{graphicx}
\usepackage{xcolor}
\usepackage{caption}
\usepackage{subcaption}

\usepackage{dingbat}

\usepackage{adjustbox}
\usepackage[linesnumbered,algo2e,ruled]{algorithm2e}
\usepackage{multirow}
\usepackage{booktabs,subcaption,dcolumn}
\usepackage{tikz}
\usepackage{fontawesome5}
\usepackage{enumitem}
\setlist[itemize]{noitemsep, topsep=0pt}
\usepackage{tabularx}
\usepackage[normalem]{ulem}
\usepackage{setspace}

\usepackage{pifont}
\usepackage{svg}
\usepackage{soul}
\usepackage[font=footnotesize]{caption}
\usepackage[noadjust]{cite}

\usepackage{amsmath}
\usepackage{amsfonts, amssymb}

\usepackage{colortbl}
\usepackage{tablefootnote}
\usepackage[most]{tcolorbox}
\usepackage[sort&compress, numbers]{natbib}

\AtBeginDocument{%
  \providecommand\BibTeX{{%
    \normalfont B\kern-0.5em{\scshape i\kern-0.25em b}\kern-0.8em\TeX}}}

\newcommand{\cmark}{{\color{ForestGreen}{\ding{51}}\xspace}}

\newcolumntype{?}{!{\vrule width 1.5pt}}

\newcommand{\textbox}[1]{
    \noindent\fbox{%
        \parbox{0.97\columnwidth}{%
            {#1}
        }%
    }
}

\newtcolorbox{cooltextbox}[1][]{%
    colback=black!5,
    colframe=black!5,
    notitle,
    sharp corners,
    borderline west={0pt}{0pt}{red!80!black},
    enhanced,
    breakable,
    left=0pt,
    right=0pt,
    top=0pt,
    bottom=0pt
    }

\newcommand\smamath[1]{{\small $#1$}}
\newcommand\smacal[1]{{\small $\mathcal{#1}$}}
\newcommand\smabb[1]{{\small $\mathbb{#1}$}}

\newcommand\scmath[1]{{\scriptsize $#1$}}

\newcommand\revision[1]{%
  \bgroup
  \hskip0pt\color{blue!75!black}%
  #1%
  \egroup
}

\newcommand{\shield}{\resizebox{!}{0.65em}{\faShield*}} 

\newcommand{\woman}{\raisebox{-0.1ex}{\footnotesize\faUser}}
\newcommand{\system}{\raisebox{-0.1ex}{\scriptsize\faServer}}
\newcommand{\both}{\woman\hspace{0.2em}+\hspace{0.2em}\system}
\newcommand{\toy}{\raisebox{0.1ex}{\scriptsize\faRobot}}
\newcommand{\real}{\raisebox{-0.1ex}{\textcolor{green!70!black}{\scriptsize\faCog}}}
\newcommand{\nomention}{\raisebox{-0.1ex}{\textcolor{red!80!black}{\scriptsize\faCommentSlash}}}
\newcommand{\yesqual}{\raisebox{-0.1ex}{\scriptsize\faComment}}
\newcommand{\yesquant}{\raisebox{-0.1ex}{\scriptsize\faCalculator}}
\newcommand{\yescost}{\raisebox{-0.1ex}{\scriptsize\faCoins}}

\newcommand{\collectedbyauthors}{\resizebox{!}{0.5em}{\faUserCog}}
\newcommand{\publiclyavailable}{\resizebox{!}{0.5em}{\faDesktop}}

\newcommand{\prompt}{\resizebox{!}{0.5em}{\faEdit}}

\definecolor{lightred}{rgb}{1, 0.8, 0.8}
\definecolor{lightblue}{rgb}{0.8, 0.8, 1}
\definecolor{lightgreen}{rgb}{0.88, 1, 0.88}

\newcommand{\mitre}{\cellcolor{lightgreen}}
\newcommand{\otheroai}{\cellcolor{lightblue}}

\newcommand\off{\textdagger}
\newcommand\assist{$\star$}

\begin{document}

\title{SoK: On the Offensive Potential of AI
{\scriptsize \\ \vspace{-4mm} Paper accepted at the 3rd IEEE Conference on Secure and Trustworthy Machine Learning (SaTML'25)}
}\vspace{-3mm}

\author{
\hspace{-0cm}
\IEEEauthorblockN{Saskia Laura Schröer\IEEEauthorrefmark{1}, Giovanni Apruzzese\IEEEauthorrefmark{1}, Soheil Human\IEEEauthorrefmark{2}, Pavel Laskov\IEEEauthorrefmark{1}, \\
Hyrum S. Anderson\IEEEauthorrefmark{3},
\hspace{-0cm}Edward W. N. Bernroider\IEEEauthorrefmark{2},
Aurore Fass\IEEEauthorrefmark{4},
Ben Nassi\IEEEauthorrefmark{5},
Vera Rimmer\IEEEauthorrefmark{6},
Fabio Roli\IEEEauthorrefmark{7},\\
Samer Salam\IEEEauthorrefmark{8},
Ashley Shen\IEEEauthorrefmark{8},
Ali Sunyaev\IEEEauthorrefmark{9},
Tim Wadhwa-Brown\IEEEauthorrefmark{8},
Isabel Wagner\IEEEauthorrefmark{10},
Gang Wang\IEEEauthorrefmark{11}
\\}
\IEEEauthorblockA{{ 
\IEEEauthorrefmark{1}\textit{University of Liechtenstein},
\IEEEauthorrefmark{2}\textit{Vienna University of Economics and Business},
\IEEEauthorrefmark{3}\textit{Robust Intelligence},
\IEEEauthorrefmark{4}\textit{CISPA},
\IEEEauthorrefmark{5}\textit{Technion},
}}

\IEEEauthorblockA{{\IEEEauthorrefmark{6}\textit{KU Leuven},
\IEEEauthorrefmark{7}\textit{University of Genoa},
\IEEEauthorrefmark{8}\textit{CISCO Systems},
\IEEEauthorrefmark{9}\textit{Karlsruhe Institute of Technology},
\IEEEauthorrefmark{10}\textit{University of Basel},
\IEEEauthorrefmark{11}\textit{UIUC}
}
\\
{\small Corresponding authors: saskia.schroeer@uni.li, giovanni.apruzzese@uni.li, soheil.human@wu.ac.at, pavel.laskov@uni.li
}
}}

\pagestyle{plain}
\maketitle

\begin{abstract}
Our society increasingly benefits from Artificial Intelligence (AI). Unfortunately, more and more evidence shows that AI is also used for offensive purposes. Prior works have revealed various examples of use cases in which the deployment of AI can lead to violation of security and privacy objectives. No extant work, however, has been able to draw a holistic picture of the offensive potential of AI. In this SoK paper we seek to lay the ground for a systematic analysis of the heterogeneous capabilities of offensive AI. In particular we (i) account for AI risks to both humans and systems while (ii) consolidating and distilling knowledge from academic literature, expert opinions, industrial venues, as well as laypeople---all of which being valuable sources of information on offensive AI.

To enable alignment of such diverse sources of knowledge, we devise a common set of criteria reflecting essential technological factors related to offensive AI. With the help of such criteria, we systematically analyze: 95 research papers; 38 InfoSec briefings (from, e.g., BlackHat); the responses of a user study (N=549) entailing individuals with diverse backgrounds and expertise; and the opinion of 12 experts. Our contributions not only reveal concerning ways (some of which overlooked by prior work) in which AI can be offensively used \textit{today}, but also represent a foothold to address this threat in the \textit{years to come}.
\end{abstract}


\section{Introduction}
\label{sec:introduction}
\noindent
Artificial Intelligence (AI) is an exemplary use-case of a disruptive technology~\cite{girasa2020,puavualoaia2023artificial}. 
AI has revolutionized the IT ecosystem worldwide, providing cost-effective solutions for new and existing tasks---potentially exceeding the proficiency of humans~\cite{soori2023artificial,bharadiya2023machine,kelly2023factors}.
Unfortunately, the disruptive nature of AI also has gradually materialized in a more literal sense---as a means to \emph{realize, facilitate and enhance cyberattacks}. Such an observation underscores that the potential of AI must be proactively scrutinized from a cybersecurity perspective.

The domains of AI and cybersecurity are, in fact, strongly intertwined. Abundant works highlight the potential of ``AI for cybersecurity''~\cite{apruzzese2022role}, e.g., showing that AI can improve cybersecurity routines~\cite{sayler2020artificial,taddeo2019trusting}; or that AI can perform tasks otherwise unfeasible for security operators~\cite{van2022deepcase}. At the same time, a large body of literature focuses on ``security of AI''~\cite{biggio2018wild,papernot2018sok}, e.g., elucidating that AI methods can be broken with tiny perturbations~\cite{carlini2017towards}; or that some confidential information pertaining to AI solutions (i.e., training data, or the AI model itself) can be leaked~\cite{hitaj2017deep} or stolen~\cite{tramer2016stealing}. There is another use case, however, that links AI and cybersecurity, but has not received the same degree of attention so far: ``offensive~AI.''

\begin{figure}[!htbp]
    \vspace{-2mm}
    \centering
    \centerline{
    \includegraphics[width=0.82\columnwidth]{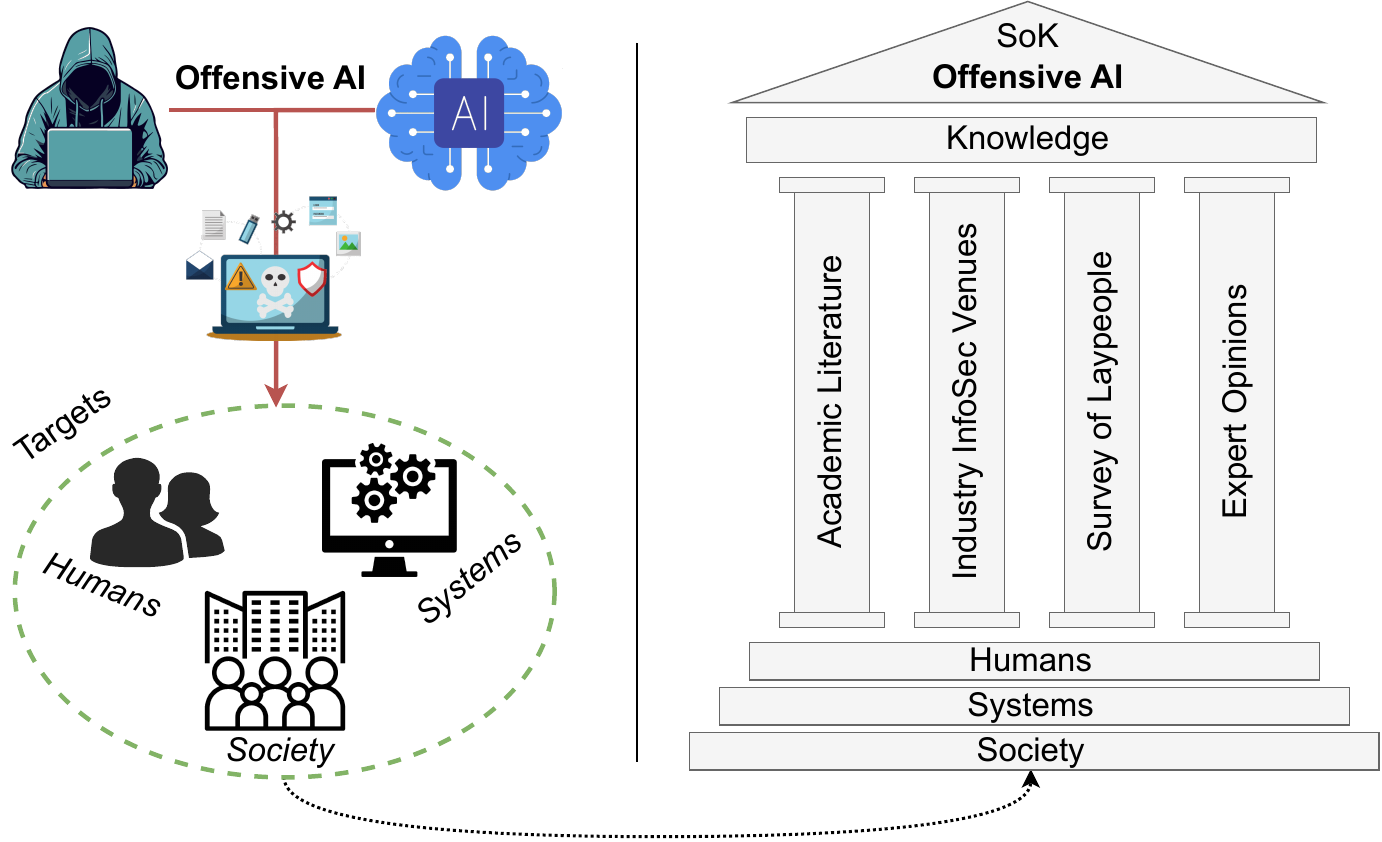}
    }
    \vspace{-1mm}
    \caption{\textbf{Targets and knowledge sources related to \textit{offensive AI}.}
    } 
    \label{fig:sok}
    \vspace{-2mm}
\end{figure}

Some prior works have considered scenarios wherein AI is used as an offensive tool. For instance, using Large Language Models (LLM) to write phishing emails~\cite{langford2023phishing} is cheap and effective~\cite{heiding2024devising}, and evidence shows that this is already happening~\cite{violino2023chatgpt,slashnext2023phishing}. However, no prior work has systematically analyzed the topic of offensive AI, examining a broad range of attack targets and accounting for diverse sources of knowledge. Indeed, prior systematizations (e.g.,~\cite{aiyanyo2020systematic,mirsky2021threat}) mostly accounted for the viewpoint of academic literature, which is a profound but not the only source of information. Briefings of \textit{industrial conferences}, opinions of \textit{experts}, and even \textit{laypeople} provide complementary perspectives on the ins-and-outs of offensive AI. Furthermore, the offensive potential of AI poses a threat not only to IT \textit{systems} in a narrow sense, as primarily considered in prior work, but also to any stakeholder relying on them, e.g., \textit{humans}, or even the \textit{society} as a whole.  Hence, to tackle today's unforeseen risks of offensive AI, a broader scope must be considered for systematization of knowledge, as suggested schematically in Fig.~\ref{fig:sok}. New knowledge sources and versatile use-cases  should be taken into account for a comprehensive analysis of this inescapable threat.

\noindent
In this SoK, we seek to establish a foundation 
for understanding and mitigating the (current and future) offensive potential of AI. To this end, we make three high-level \textbf{contributions}: 
\setlist{nolistsep}
\begin{itemize}[noitemsep]
    \item[\smabb{C}1:] We present a \textbf{snapshot of the current landscape} of offensive AI, by accounting for its three crucial stakeholders: systems, humans, and society. This contribution serves to \textit{review its various use cases}, including those potentially overlooked by prior surveys on offensive AI.
    
    \item[\smabb{C}2:] We devise an original \textbf{long-term classification of key technological factors} related to offensive AI. This contribution serves to examine and compare selected works on offensive AI according to a \textit{common set of criteria}. We develop an online tool~\cite{tool} to make this contribution applicable to any relevant works (potentially omitted in our literature analysis), including the future ones. 

    \item[\smabb{C}3:] We outline an actionable research agenda for future work, pinpointing \textbf{open problems and concerns to be addressed by the research community} in various areas. This contribution serves as a \textit{guide for any stakeholder} interested in mitigating the threat of offensive AI.
\end{itemize}

\noindent
\textbf{\textsc{Organization.}} 
We define the scope of our SoK and describe the methods used to carry out our systematization in Section §\ref{sec:2-preliminaries}. There, we also present an original \textit{checklist} providing the groundwork for our analyses (and also for \smabb{C}2). 

Section §\ref{sec:3-academic} is devoted to the analysis of \textit{academic literature}. We scrutinize 95 peer-reviewed papers on offensive AI (identified through a systematic search across more than 3000 works) under the lens of our checklist. Our findings reveal several gaps in prior work; for instance, that certain offensive AI use-cases targeting humans, e.g., attribute inference attacks~\cite{gong2016you, golbeck2011predicting}, were left out in prior literature reviews.

In Section §\ref{sec:4-infosec}, we focus on the \textit{industrial perspective}. We survey the landscape of major InfoSec outlets (BlackHat, DefCon) and identify 38 briefings related to offensive AI. We systematize these works through our checklist, underscoring the offensive potential of AI revealed in practical venues.

In Section §\ref{sec:5-userSurvey}, we consider the \textit{viewpoint of laypeople}. We present the results of a user study (n=549) exposing the perception of offensive AI by ``non-experts.'' We observe that a large share of participants (\smamath{84}\%) are concerned about the potential offensive use of AI and qualitatively analyze the reasons for such concerns---revealing some potential misconceptions.  

Finally, in Section §\ref{sec:experts}, we study the \textit{opinion of experts} on the offensive potential of AI. We reached out to 12 experts~in cybersecurity, privacy, and information systems. First, we asked them to complete the questionnaire we used for the general public study. Then, we provided these experts with an early draft of this paper, and requested to write statements defining ``three open problems in the field of offensive AI.'' We systematically review and coalesce these statements (reported verbatim in Appendix~\ref{app:expert-statements}) into ten open problems and fundamental concerns of offensive AI---the basis of \smabb{C}3.

We wrap up our systematization in Section §\ref{sec:discussion} by reflecting on the takeaways on \textit{all hitherto analyzed sources of knowledge} (which collectively form the snapshot of \smabb{C}1, set up the stage for \smabb{C}2, and are used for \smabb{C}3). We also identify and discuss limitations, and compare our contributions with previous related work. We conclude our SoK in Section §\ref{sec:conclusions}.

\vspace{-2mm}

\begin{cooltextbox}
    \textbf{\textsc{Scope}.} In our SoK, we consider offensive AI (OAI) as \textit{the means of using AI to accomplish a task that violates security and privacy objectives}. Such a broad notion covers a wide array of risks, stemming from an attacker who is deliberately trying to cause harm---and does not cover cases in which, e.g., an AI leads to harm due to negligence or misconfiguration. Specifically, our notion encompasses cases when AI is used to amplify existing threats (e.g., disinformation is a well-known problem which can be made much worse via AI~\cite{lohn2021disinformation}) or develop previously unseen threats (e.g., attribute inference attacks are essentially enabled by AI~\cite{gong2016you}).\footnote{\textbf{Prior work.} We summarize the evolution of the term ``offensive AI'' in the literature in Appendix~\ref{sapp:background-OAI}. Some works associate techniques for generation of ``adversarial examples'' to OAI~\cite{mirsky2021threat}. According to our definition, \textit{some instances} of such techniques can be considered as OAI (e.g., if an attack involves Generative Adversarial Networks~\cite{hu2022generating, lin2022idsgan}, which clearly rely on AI), while others are orthogonal to OAI (e.g., some ``evasion'' attacks~\cite{vsrndic2014practical} not necessarily rely on AI to be staged---i.e., computing FGSM to generate an adversarial perturbation can be done algorithmically without leveraging any AI technique~\cite{biggio2018wild}). We stress that we use ``AI'' to denote techniques within the machine-learning (ML) domain~\cite{jordan2015machine}.}
    
    \vspace{-2mm}
    
\end{cooltextbox}

\vspace{-2mm}

\section{Research Methods and Checklist}
\label{sec:2-preliminaries}

\noindent
We introduce the research methods applied in our SoK: the systematization of scientific literature~(\S\ref{ssec:2.1-literature}) and of InfoSec briefings~(\S\ref{ssec:2.2-industrial}), the user study with non-experts~(\S\ref{ssec:2.3-userstudy}), and the elicitation and systematization of expert knowledge~(\S\ref{ssec:2.4-expert}). We also present our \textit{OAI Assessment Checklist}, which provides the means for alignment and systematization of diverse classes of prior work considered in our SoK~(§\ref{ssec:checklist}).
Some details of our methods are in the Appendix, including a timeline (in Fig.~\ref{fig:timeline}) encompassing all our research activities. 

\subsection{Systematic Literature Review (Methodology)}
\label{ssec:2.1-literature}

\begin{figure}[!htbp]
    \vspace{-3mm}
    \centering
    \centerline{
    \includegraphics[width=0.42\textwidth]{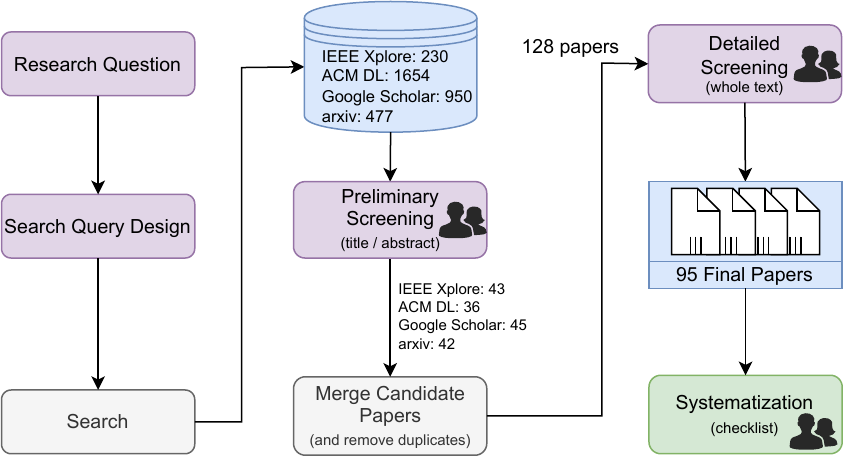}
    }
    \vspace{-2mm}
    \caption{\textbf{Systematic Literature Review.}
    \textmd{\footnotesize We collect over 3000 papers from various repositories. After filtering, screening and inter-researcher discussions, we coalesce 95 papers on OAI which we consider in this SoK.}} 
    \label{fig:literature-reivew}
    \vspace{-3mm}
\end{figure}

\noindent
Prior surveys on OAI~\cite{mirsky2021threat, kaloudi2020ai, guembe2022emerging} are grounded in academic literature. Hence, to ensure continuity, we consider research papers as our first source of knowledge. 
We perform a \textit{systematic literature review}, following established guidelines~\cite{keele2007guidelines}, illustrated in Fig.~\ref{fig:literature-reivew}. We describe the pivotal points below.

\textbf{Search Queries.} 
We began our literature review by asking ourselves ``how has prior work envisioned \textit{offensive AI}?'' 
To systematically encompass a broad spectrum of prior art, we search for related papers indexed by four popular databases (until Nov. 2023): IEEE Xplore, Google Scholar, ACM DL, and arXiv (similarly to Ladisa et al.~\cite{ladisa2023sok}). 
We carried out our search by formulating queries corresponding to two macro-search queries. Specifically, the first, straightforward, macro search-query entails ``offensive AI.'' However, we are confident that there are other ways in which prior work has conceived AI-based applications that can fall within our definition of OAI. Hence, as an exemplary use-case to extend our search, we consider another macro-query revolving around ``AI in offensive security'': indeed, the idea of using AI in penetration testing (e.g.,~\cite{erdHodi2021simulating, zennaro2023modelling}) \textit{can} also be leveraged by real attackers to bypass a given system. We have even reached out to the authors of some well-known publications (e.g.,~\cite{valea2020towards, ceccato2016sofia, bhattacharya2020automated}) who confirmed that their AI-based tool could be used maliciously. Overall, based on the our two macro-search queries, we devise 36 search queries, condensed in the box below:

\vspace{1mm}
{\setstretch{0.8}
\textbox{{\small 
\textit{Macro query \#1}:} {\footnotesize "offensive" $\land$ ("AI" $\lor$ "artificial intelligence" $\lor$ "ML" $\lor$ "machine learning") $\land$ ("security" $\lor$ "cyber security" $\lor$ "cybersecurity" $\lor$ "cyber attack" $\lor$ "attack" $\lor$ "privacy" $\lor$ "threat");} 

{\small \textit{Macro query \#2}:} {\footnotesize ("AI" $\lor$ "artificial intelligence" $\lor$ "ML" $\lor$ "machine learning") $\land$ ("penetration testing" $\lor$ "red teaming")}.}}

\vspace{-1mm}

\noindent
Importantly, we are aware that our search strategies have limitations and cannot capture ``all'' works that have considered OAI applications. This is, however, not our goal. We elaborate on how to overcome this limitation in §\ref{ssec:checklist} and §\ref{sec:discussion}.

\textbf{Dual Reviewing (with adjudication).}
Our search returned 3311 papers. To mitigate bias, these papers have been reviewed by two authors who worked independently and later compared their findings to find a consensus; in cases of disagreement, a senior reviewer acted as adjudicator~\cite{hersh2015outcomes}. Such a system was used for two steps of our literature analysis:
\begin{itemize}[leftmargin=*]
    \item \textit{Screening.} First, we identified unique works that fall in our definition of OAI. This was done mostly by inspecting the title and abstract of the papers, which was typically sufficient to remove papers outside our scope; if we were uncertain, we also looked at the entire content of the papers.\footnote{We omitted: {\scriptsize \textit{(i)}}~papers that do not present any OAI capability---e.g., AI for security (e.g.~\cite{jana2019appmine}), or security of AI (e.g.~\cite{apruzzese2022spacephish}), or unrelated to security (e.g.,~\cite{van2016wavenet}); {\scriptsize \textit{(ii)}}~literature reviews on a specific sub-field, e.g., AI in penetration testing~\cite{mckinnel2019systematic}; and {\scriptsize \textit{(iii)}}~grey literature/white papers~\cite{kubovivc2018can}.} After removing duplicates, we derived a set of 95 papers. 
    
    \item \textit{Systematization.}  We systematically assess these 95 works. First, we differentiate ``technical'' from ``non-technical'' papers: technical papers \textit{must} demonstrate a practical implementation/usage of an AI model; in contrast, non-technical papers encompass case studies, user/expert surveys, conceptual papers, opinion papers, or similar. This classification yielded 16 non-technical and 79 technical papers. Then, we systematize these works according to our checklist (described in §\ref{ssec:checklist}), and we further scrutinize technical papers to underscore technical aspects of their implementation.
\end{itemize}

\subsection{Systematic Analysis of InfoSec Briefings (Methodology)}
\label{ssec:2.2-industrial}
\noindent
Since the emergence of ChatGPT, OAI has become a focal point of discussion, often featured in the news such as by Forbes, Economist, or CNN~\cite{forbes2023oai,economist2023oai,cnn2023oai}. 
While non-academic literature can include different types of works, such as news articles, or security reports, our objective is to \textit{also} scrutinize prior work that has not undergone an academic publishing process, but that {\small \textit{(i)}}~still allows us to identify OAI use cases, {\small \textit{(ii)}}~is highly relevant to practical and real-world threats, and {\small \textit{(iii)}}~has been subject to some kind of review process. Hence, we consider the content of two renown security events: BlackHat and DefCon. These venues are highly competitive: for instance, the acceptance rate for the AI track of BlackHat Asia'24 was 7\%~\cite{blackhat2024acceptancerate}. To the best of our knowledge, this is the first SoK to consider the perspective of InfoSec ``briefings,'' i.e., presentations of 30--40m with slides and abstract. 

We examined the entire history of these venues, from 1993 to 2023 for DefCon, and from 1997 to 2023 for BlackHat.
For BlackHat, we assessed the content of the events held in the USA (27), Europe (23), and Asia (19), for a total of 69 events. For DefCon, we assessed all 31 events. We then followed a similar procedure for our literature analysis, rooted in the dual reviewing with adjudication system. First, we looked at all the briefings trying to identify which were related to OAI. To this end, we first inspected the title and abstract; then we performed a deeper analysis by going through the slides, the video, and even the captions of the recording (if available; we could not find any of these resources for~\cite{blackhat2018socialmapper} which we exclude).
Ultimately, we identified 38 briefings related to OAI, which we scrutinize according to our checklist (§\ref{ssec:checklist}).

\vspace{1mm}

{\setstretch{0.8}
\textbox{{\small 
We show in Fig.~\ref{fig:works} (Appendix~\ref{app:extra_figures}) the yearly distribution of the works (95 papers and 38 briefings) considered in our SoK. Intriguingly, the earliest work for each category appeared in 2008~\cite{golle2008machine, clarke2008hacking}}
}}

\subsection{Study of ``non-expert'' Opinion (Methodology)}
\label{ssec:2.3-userstudy}

\noindent
Literature and briefings provide extensive knowledge on OAI; yet, they may not capture what OAI-related concerns are predominant in the real world. To provide a complementary perspective that allows one to ascertain more transient forms of knowledge (as also done in other SoKs, e.g.,~\cite{thomas2021sok,stephenson2022sok}), we also investigate the perception of OAI among individuals who are not necessarily subject-matter experts.

\begin{figure}[!htbp]
    \vspace{-2mm}
    \centering
    \centerline{
    \includegraphics[width=0.45\textwidth]{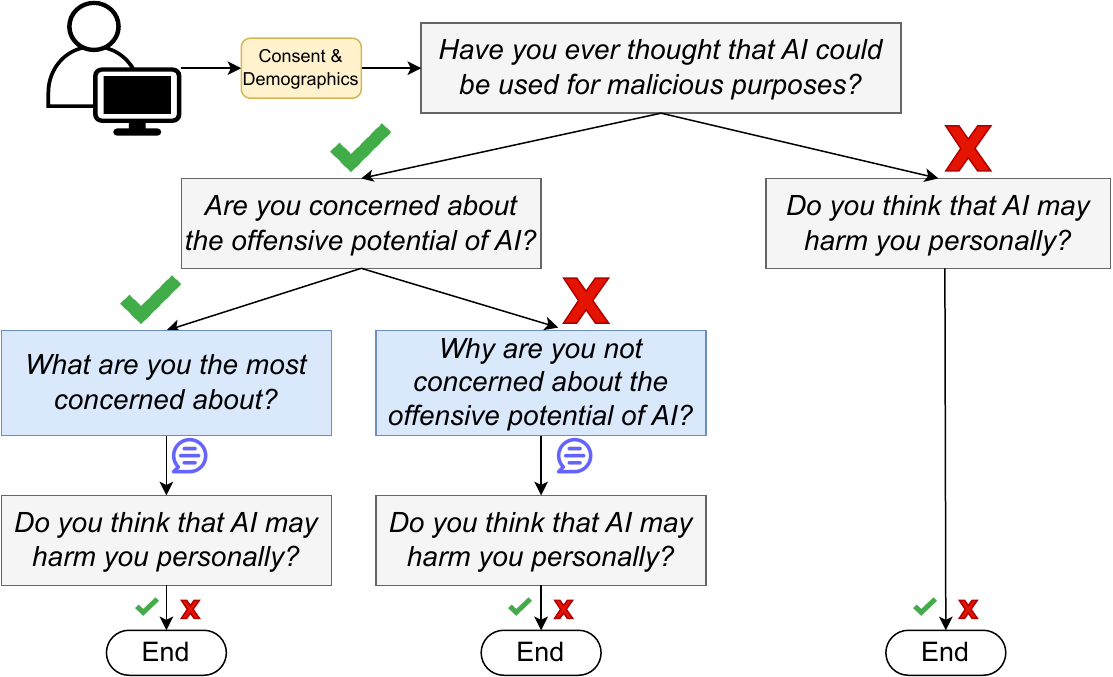}
    }
    \vspace{-2mm}
    \caption{\textbf{Questionnaire.}
    \textmd{\footnotesize Depending on the answers, participants have to respond to up to four broad questions (e.g., no specific time frame is given)}. Some questions expect open answers, ensuring freedom to share any concern.}
    \label{fig:survey-design}
    \vspace{-5mm}
\end{figure}

\textbf{Questionnaire Design.} 
We devise an anonymous questionnaire covering various aspects related to OAI.  Our questionnaire is \textit{short} (\smamath{\sim}5 minutes according to five pilot tests) to maximize the response rate and improve the quality of responses. After informing our participants of their rights and collecting some (optional) demographic details (we do not, e.g., ask for specific employment information), we ask up to four questions, visualized in Fig.~\ref{fig:survey-design}. Potentially, the questionnaire may end after just the first question---which serves as a ``screening,'' so that only participants who have thought about OAI are requested to elaborate their concerns/ideas. Our questionnaire\footnote{\textbf{Ethical Statement:} we treated our participants ethically, following the Menlo report~\cite{bailey2012menlo}. We did not ask for personally identifiable information, and our participants can ask us to delete their data if they so desire. Our institutions are aware of our research. Participation in our questionnaire was voluntary and we did not offer any form of compensation.} is provided (verbatim) in our repository~\cite{repository}.

\textbf{Dissemination.} To ensure a diverse respondent pool, we distributed the survey over various channels, spanning across online social networks (e.g., one author made posts on LinkedIn---ensuring not to mention ``OAI'' anywhere in the text) as well as educational events (e.g., lectures and workshops---not related to OAI), and we also relied on convenience sampling~\cite{emerson2015convenience,antoun2016comparisons}. Importantly, we \textit{never primed} our participants: the events were not about OAI, the posts used to share our questionnaire did not link to external resources related to OAI, and the questionnaire did not provide any specific information about OAI. Indeed, our goal was to collect the genuine opinion of each participant about their own vision of OAI. Moreover, our dissemination channels ensured that our survey would reach subjects who (despite being highly educated) hardly possess expertise in OAI-related themes.
We collected a total of 549 valid responses during Sep.--Dec. 2023.

\textbf{Analysis.} After collecting our responses, we split the analysis into a quantitative and a qualitative part---the latter reliant on the dual reviewing with adjudication system. The \textit{quantitative analysis} includes demographics, expertise, and an initial overview of participants’ ideas about OAI. We additionally perform a correlation analysis to understand the relationship between technical expertise and concerns about OAI. 
In the \textit{qualitative analysis}, we adhere to the constructivist grounded theory methodology \cite{charmaz2014constructing}, incorporating four iterative rounds of coding. First, an ``initial coding'' is carried out to identify emergent themes and concepts. Then, a more ``focused coding'' allows us to derive broader categories of findings. Next, an ``axial coding'' step delineates the relationships between categories and subcategories. Finally, the ``theoretical coding'' integrates these categories into a comprehensive framework.

\subsection{Systematization of Expert Opinion (Methodology)}
\label{ssec:2.4-expert}
\noindent
One of the main objectives of our SoK is to identify open research problems on OAI (\smabb{C}3). Inspired by~\cite{dwivedi2023so, clark2013whatever, kairouz2021advances}, we do so by leveraging the collaboration with experts from the field.

\textbf{Selection of Experts.} To ensure the coverage of interdisciplinary perspectives of OAI, we seek to collect the opinion of experts having diverse backgrounds in terms of: area of expertise (i.e., AI, security, privacy, information systems), application domain (research or practice), institution (academia or industry), as well as gender and work experience (junior or senior).
In our ``recruitment'' process involving private communication (with no priming), we have converged on a set of 12 experts.
Eight experts stem from academia, and four from industry; eight are males, and four are females; five have a background in both security and AI, three in security, two in information systems, one in AI, and one in privacy.
We did not know any expert's thoughts about OAI beforehand.

\textbf{Opinion Collection and Analysis.}
We could have increased participation by, e.g., using interviews---but this would have introduced bias in the data collection phase~\cite{chenail2011interviewing}. Hence, to collect the opinion of experts in a bias-free way, we proceed in three steps. First, we ask experts to fill in the questionnaire for the non-expert survey (\S\ref{ssec:2.3-userstudy}), which we have slightly modified by adding one open question requesting input about ``open problems on OAI.'' After all 12 experts completed the questionnaire, we send them a draft of the paper and ask them to {\small \textit{(i)}}~read the paper, and {\small \textit{(ii)}}~write a paragraph of 300--500 words to describe three ``open problems of OAI'' based on their impressions from the draft paper as well as perception of the field. Each expert had 3 weeks to perform this task. Finally, having collected all expert responses (reported verbatim in Appendix~\ref{app:expert-statements}), we systematically analyze them quantitatively and qualitatively. Our goal is twofold: to infer open problems in OAI from the experts' input, and to compare the experts' initial input (provided \textit{before} reading the draft paper)\footnote{We analyzed the experts' responses to the survey in the same way as we did for the non-expert survey (§\ref{ssec:2.3-userstudy}) and we also use our checklist (§\ref{ssec:checklist}).} with their final opinion (formulated \textit{after} reading the draft paper). 

\textbf{Rationale.} The reason why we ask our experts to provide their opinion both before and after having read our paper is twofold. 
On the one hand, we want to see if our systematization (and corresponding findings) ``poisoned their mind'': if so, this means that one of our objectives (i.e., raising awareness on some concealed aspects of OAI) was met. 
On the other hand, we want to analyze their opinion after providing them with {\small \textit{(i)}}~a comprehensive overview of OAI which {\small \textit{(ii)}}~aligns with our notion of OAI: indeed, the notion of OAI has been used in various way by prior work (see Fig.\ref{fig:oai-term}), and some experts may see it differently. Hence, asking the experts to read our SoK ensures that their opinion has maximum usefulness for the sake of defining open problems. Nevertheless, we showed our manuscript to the experts after integrating and systematizing their statements in our paper, asking for further feedback (which has been used to substantially enhance this SoK).

\subsection{OAI Assessment Checklist (Original Contribution)}
\label{ssec:checklist}

\noindent
We now present the method used to derive the checklist underpinning our contribution \smabb{C}2. 

\vspace{-2mm}

\begin{cooltextbox}
    \textbf{\textsc{Motivation}:} Given the vast space of AI use cases it is \textit{impossible} to identify all works that may---explicitly or implicitly---describe a \emph{potentially offensive} AI application.
    Hence, our aim is to \emph{provide the means} for systematization of knowledge in the field of OAI based on the ``snapshot'' of the current OAI landscape obtained as a result of our literature search~(§\ref{ssec:2.1-literature}) and analysis of InfoSec events~(§\ref{ssec:2.2-industrial}). To this end, we develop a consistent set of criteria, implemented as a checklist, that can be matched against any relevant work (past or future). To ensure the sustainability of our checklist, we have developed a tool~\cite{tool} for downstream research.     
\end{cooltextbox}

\vspace{-2mm}

\noindent
Our checklist revolves around three fundamental questions. Two are inspired by prior work~\cite{mirsky2021threat, apruzzese2023real}, and serve to align our SoK with existing summaries; whereas one is driven by the overarching goal of our paper. Let us describe them.

\subsubsection{\textbf{``What is the OAI use-case?''}}
\label{sssec:usecase}
This question, inspired by~\cite{mirsky2021threat}, enables the \emph{systematization of the OAI use cases}. To address this question, we follow three steps.
\textbf{(I)}~We identify the (primary) OAI use-case, i.e., the ``context'' in which AI is used offensively. We begin by scrutinizing whether the work can be mapped to any of the scenarios covered in MITRE~\cite{mitre-tactics}, and specifically in ATT\&CK for Enterprise~\cite{mitre-ent}. If we cannot find any match, we consider other MITRE matrices: Mobile~\cite{mitre-mobile} and ICS~\cite{mitre-ics}. If no valid match can still be found, we assign a special category. This is common for works envisioning, e.g., attacks against privacy or society, cyberwar, or considering autonomous agents---all of which being scenarios not covered by MITRE. 
\textbf{(II)}~We infer the overarching \textit{purpose} of the OAI-related work. We consider the following three categories: offensive security~(\shield), denoting cases wherein AI is used in, e.g., penetration testing; hacking assistant~(\assist), denoting cases wherein AI is used to facilitate some procedures that could be maliciously abused; and novel attack~(\off), denoting works that propose full-fledged attacks reliant on AI.
\textbf{(III)}~We examine if the work addresses mitigation of the potential harm caused by OAI. Specifically, for works showcasing novel attacks (\off), we assess whether there is an evaluation of potential defensive mechanisms; whereas for works using AI for offensive security~(\shield), we inspect whether there is any statement warning downstream users that the proposed method could be maliciously exploited; we do both of these checks for works wherein AI is used as hacking assistant (\assist).

\subsubsection{\textbf{``What is the target of OAI?''}}
\label{sssec:targetimpact}
This is an original question of our SoK, and serves to explore the impact of OAI on the three stakeholders indicated in Fig.~\ref{fig:sok}. We address this question in three steps. \textbf{(I)}~We discriminate whether the OAI use-case \textit{targets} a human (e.g., a privacy violation~\cite{tricomi2023attribute}), a system (e.g., CAPTCHA cracking~\cite{yu2019low}), or both (e.g., a phishing scenario wherein both the detector and end-users must be deceived~\cite{draganovic2023dousers}). 
\textbf{(II)}~For works considering attacks against \textit{systems}, we determine whether the evaluation entailed a ``real'' system (e.g., an operational product) or a ``toy'' system (e.g., a simplified version of a real system~\cite{xu2015my}).
\textbf{(III)}~To understand the potential relevance of the social perspective within each paper, we count the occurrences ({which is a well-known practice to objectively study the focus of a document~\cite{sampagnaro2023keyword}}) of the terms ``society,'' ``social,'' ``societal,'' and ``socio.''

\subsubsection{\textbf{``What is the \textit{cost/benefit} of OAI?''}}
\label{sssec:costbenefit}
This question stems from the consideration that real attackers are primarily driven by economical motives. We address this question through three steps, mostly inspired by~\cite{apruzzese2023real}.
\textbf{(I)}~First, we analyze if the respective work assesses potential \textit{benefits} for attackers from using the specific AI technique. Four outcomes are considered: {\small \textit{(a)}}~explicit mentioning of ``financial'' benefits (e.g., monetary gains, resources saved); {\small \textit{(b)}}~quantitative analysis via ML metrics (e.g., accuracy); {\small \textit{(c)}}~a qualitative discussion; {\small \textit{(d)}}~no mentioning at all.
\textbf{(II)}~Next, we assess whether the \textit{cost} of employing OAI is taken into account, considerng the same four possible outcomes as for ``benefits.''
\textbf{(III)}~Finally, we consider if the work performed a comparison with a non-AI baseline---to determine the ``contribution'' of using AI for offensive purposes. Three cases are possible: {\small \textit{(a)}}~quantitative comparison; {\small \textit{(b)}}~qualitative considerations; {\small \textit{(c)}}~none.

\vspace{1mm}
{\setstretch{0.7}
\textbox{{\small \textbf{Remark:} 
In Appendix~\ref{app:checklist}, we provide a low-level description of the many elements we considered when analysing each work. Such a description is to ensure scientific transparency and reproducibility, thereby contributing to the ``long-term'' aspect of our checklist.}}}
\vspace{-1mm}
\begin{table*}[!htbp]
    \centering
    \caption{\textbf{Literature Review -- Technical OAI Papers.}
    \textmd{\footnotesize 
    We report 79 technical OAI papers, including discipline (see Appendix~\ref{app:sok}; CS=Computer Science; Eco=Economics; Eng=Engineering) and the number of citations as of Jan. 2024 (taken from Google Scholar), scrutinized based on our OAI checklist~(§\ref{ssec:checklist}). Our search (done until Nov. 2023) identified some arXiv preprints: we report the reference to their published version (even if it appeared after Nov. 2023---e.g.~\cite{antonelli2024dirclustering} originally uploaded on arxiv in 2021 and published in 2024). For the specific OAI use case, we map most papers to the \sethlcolor{lightgreen}\hl{MITRE ATT\&CK}~Enterprise matrix and five papers to the ICS and Mobile matrices (indicated with *); the use cases not covered by MITRE are highlighted in \sethlcolor{lightblue}\hl{blue}. We assign each use case (``Purpose'') either to the category ``defense'' (\shield), ``assisted-hacking'' (\assist), or ``attack'' (\off). ``Def.?'' denotes whether the paper considered countermeasures, ``Pot. Abuse'' stands for ``considerations of potential malicious abuse of a defensive tool''. For ``Targ.'' (target), we use \woman{} to denote ``humans'' and \system{} for ``system''; if the target is a system, we use \toy{} to denote a ``toy'' system, and \real{} for a ``real'' system. For the cost/benefit column, \yescost{} denotes a monetary assessments of the costs, \yesquant{} a quantitative assessment, \yesqual{} a qualitative discussion, \nomention{} is no mention. For the ``Code'' column, an ``x'' denotes if the code is available, and \prompt{} only denotes prompts for AI models; the icon also embeds an hyperlink to the repository (if available).}}
    \label{tab:technical}
    \vspace{-2mm}
    \resizebox{1.78\columnwidth}{!}{
        \begin{tabular}{c|c|c|c|c|c|c|c|c|c|c|c|c|c|c}
            \toprule
            \multicolumn{4}{c|}{} & \multicolumn{4}{|c|}{\textbf{OAI Use Case}} & \multicolumn{3}{|c|}{\textbf{Target/Impact}} & \multicolumn{3}{|c|}{\textbf{Cost/Benefit}} \\

            \cmidrule(lr){5-7} \cmidrule(lr){8-10} \cmidrule(lr){11-13} 

            \textbf{Paper} & \textbf{Year} & \textbf{Discipline} & \textbf{Cit.} & \textbf{Specific OAI Use Case} & \textbf{Purpose} & \textbf{Def.?} & \textbf{Pot. Abuse?} & \textbf{Targ.} & \textbf{Real/Toy} & \textbf{Social Persp.} & \textbf{Benef.} & \textbf{Cost} & \textbf{Base.} & \textbf{Code} \\
            \midrule
            
            Antonelli~\cite{antonelli2024dirclustering} & 2024 & CS & 5 & \mitre Init.Acc.  & \shield & ~ & ~ & \system & \toy & 0 & \yesquant & \yesqual & \yesquant & ~ \\

            \hline
            
            AlMajali~\cite{almajali2023vulnerability} & 2023 & CS & 0 &  \otheroai Autonomous Agents & \shield & ~ & \cmark & \system & \toy & 0 & \nomention & \nomention & \nomention & ~ \\

            Chen~\cite{chen2023gail} & 2023 & CS & 3 &  \otheroai Autonomous Agents & \shield & ~ & ~ & \system & \toy & 0 & \yesquant & \yesquant & \yesqual & \href{https://github.com/Shulong98/GAIL-PT}{x} \\
            
            Chowdhary~\cite{chowdhary2023generative} & 2023 & CS & 0 & \mitre Init.Acc.  & \shield & ~ & ~ & \system & \real & 0 & \nomention & \nomention & \nomention & ~ \\
            
            Gallus~\cite{gallus2023generative} & 2023 & CS & 0 & \mitre Init.Acc.  & \assist & ~ & ~ & \system & \toy & 1 & \nomention & \nomention & \nomention & \prompt \\
            
            Ghanem~\cite{ghanem2023hierarchical} & 2023 & CS & 13 &  \otheroai Autonomous Agents & \shield & ~ & ~ & \system & \toy & 0 & \yesqual & \yesqual & \nomention & ~ \\
            
            Happe~\cite{happe2023getting} & 2023 & CS & 8 & \otheroai  Autonomous Agents & \assist & ~ & \cmark & \system & \toy & 0 & \nomention & \nomention & \nomention & \href{https://github.com/ipa-lab/hackingBuddyGPT}{x} \\

            Iqbal~\cite{iqbal2023chatgpt} & 2023 & CS & 3 &  \otheroai  Autonomous Agents & \assist & ~ & ~ & \system & \toy & 4 & \yesqual & \nomention & \nomention & \prompt \\
            
            Karinshak~\cite{karinshak2023working} & 2023 & CS & 17 & \otheroai Atk on soc. & \off & ~ & ~ & \woman & ~ & 20 & \yesqual & \nomention & \yesqual & \prompt \\
            
            Ozturk~\cite{ozturk2023new} & 2023 & CS & 2 & \mitre Disc.  & \assist & ~ & ~ & \system & \toy & 1 & \yesquant & \nomention & \yesquant & \prompt \\
            
            Pa Pa~\cite{pa2023attacker} & 2023 & CS & 9 & \mitre Res.Dev.  & \assist & \cmark & ~ & \system & \toy & 3 & \yesqual & \yesquant & \nomention & \prompt \\
            
            Zennaro~\cite{zennaro2023modelling} & 2023 & CS & 25 &  \otheroai Autonomous Agents & \shield & ~ & \cmark & \system & \toy & 3 & \yesqual & \nomention & \nomention & \href{https://github.com/FMZennaro/CTF-RL}x \\
        
            \hline
            
            Auricchio~\cite{auricchio2022automated} & 2022 & CS & 6 & \mitre Init.Acc.  & \shield & ~ & ~ & \system & \toy & 0 & \yesquant & \yesqual & \yesquant & ~ \\
            
            Biesner~\cite{biesner2022combining} & 2022 & CS & 3 & \mitre Cred.Acc.  & \off & ~ & ~ & \system & \toy & 1 & \yesquant & \nomention & \yesquant & \href{https://github.com/fraunhofer-iais/password_generation}{x} \\

            Cody~\cite{cody2022discovering} & 2022 & CS & 15 &  \mitre Exfil.  & \shield & ~ & ~ & \system & \toy & 0 & \nomention & \nomention & \nomention &  ~\\
            
            Confido~\cite{confido2022reinforcing} & 2022 & Eng. & 2 &  \otheroai Autonomous Agents & \shield & ~ & ~ & \system & \toy & 4 & \yesquant & \yesquant & \yesqual & ~ \\

            Gangupantulu~\cite{gangupantulu2022using} & 2022 & CS & 27 & \otheroai Autonomous Agents & \shield & ~ & ~ & \system & \toy & 1 & \yesqual & \nomention & \nomention &  ~ \\

            Hu~\cite{hu2022generating} & 2022 & CS & 635 & \mitre Def.Ev. & \off & \cmark & ~ & \system & \toy & 0 & \nomention & \nomention & \nomention &  ~ \\
            
            Jagamogan~\cite{jagamogan2022penetration} & 2022 & CS & 2 & \mitre Init.Acc. & \shield & ~ & ~ & \system & \toy & 0 & \yesquant & \nomention & \yesquant & \href{https://github.com/gyoisamurai/GyoiThon}{x} \\
            
            Karanatsiou~\cite{karanatsiou2022my} & 2022 & CS & 18 & \otheroai Priv.atk. & \off & ~ & ~ & \woman & ~ & 59 & \nomention & \nomention & \nomention & ~ \\
            
            Lee~\cite{lee2022link} & 2022 & CS & 16 & \mitre Init.Acc. & \off & ~ & ~ & \system & \real & 0 & \yescost & \nomention & \yesquant & \href{https://github.com/WSP-LAB/Link}{x} \\
            
            Li~\cite{li2022deep} & 2022 & CS & 6 & \otheroai Autonomous Agents(ICS) & \shield & ~ & ~ & \system & \toy & 0 & \nomention & \nomention & \nomention &  ~ \\
            
            Lin~\cite{lin2022idsgan} & 2022 & CS & 332 & \mitre Def.Ev. & \off & \cmark & ~ & \system & \toy & 0 & \yesquant & \nomention & \yesquant &  ~ \\

            Nhu~\cite{nhu2022leveraging} & 2022 & CS & 3 &  \otheroai Autonomous Agents & \shield & ~ & ~ & \system & \toy & 0 & \nomention & \nomention & \nomention &  ~\\
            
            Pagnotta~\cite{pagnotta2022passflow} & 2022 & CS & 8 & \mitre Cred.Acc.  & \off & ~ & ~ & \system & \toy & 0 & \yesquant & \nomention & \nomention &  ~ \\
            
            Tran~\cite{tran2022cascaded} & 2022 & CS & 8 &  \otheroai Autonomous Agents & \shield & ~ & ~ & \system & \toy & 0 & \yesquant & \nomention & \nomention &  ~ \\
            
            Yao~\cite{yao2022intelligent} & 2022 & CS & 0 &\otheroai  Autonomous Agents & \shield & ~ & ~ & \system & \toy & 0 & \nomention & \nomention & \nomention &  ~\\
            
            \hline
            
            Caturano~\cite{caturano2021discovering} & 2021 & CS & 29 & \mitre Init.Acc.  & \assist & ~ & ~ & \system & \toy & 0 & \nomention & \nomention & \yesquant &  ~ \\
            
            Erdődi~\cite{erdHodi2021simulating} & 2021 & CS & 38 & \mitre Init.Acc.  & \shield & ~ & \cmark & \system & \toy & 0 & \nomention & \nomention & \nomention & \href{https://github.com/FMZennaro/CTF-SQL}{x} \\
            
            Gangupantulu~\cite{gangupantulu2021crown} & 2021 & CS & 13 & \mitre Disc.  & \shield & ~ & ~ & \system & \toy & 1 & \nomention & \nomention & \nomention &  ~ \\
            
            Khan~\cite{khan2021offensive} & 2021 & Eco & 5 & \mitre Init.Acc. & \off & ~ & ~ & \both & \real & 4 & \yesqual & \yesqual & \nomention & ~ \\
            
            Kujanpää~\cite{kujanpaa2021automating} & 2021 & CS & 8 & \mitre Priv.Esc.  & \off & \cmark & ~ & \system & \toy & 0 & \yesqual & \nomention & \nomention &  ~ \\
            
            Lee~\cite{lee2021offensive} & 2021 & CS & 4 & \mitre Cred.Acc.  & \off & ~ & ~ & \system & \toy & 0 & \yesquant & \nomention & \yesquant &  ~ \\
            
            Maeda~\cite{maeda2021automating} & 2021 & CS & 49 & \mitre Priv.Esc.  & \shield & ~ & \cmark & \system & \toy & 0 & \yesqual & \nomention & \yesqual & ~ \\ 
            
            Neal~\cite{neal2021reinforcement} & 2021 & Eng. & 12 & \mitre Procc.Contr.*  & \shield & ~ & ~ & \system & \toy & 0 & \nomention & \nomention & \nomention &  ~ \\
            
            Sharevski~\cite{sharevski2021regulation} & 2021 & CS & 1 &  \otheroai Atk on soc. & \off & ~ & ~ & \woman & ~ & 15 & \yesqual & \yesqual & \nomention &  ~ \\
            
            Standen~\cite{standen2021cyborg} & 2021 & CS & 52 & \mitre Priv.Esc.  & \shield & ~ & ~ & \system & \toy & 0 & \nomention & \nomention & \nomention &  ~ \\
            
            Toemmel~\cite{toemmel2021catch} & 2021 & CS & 1 & \mitre Pers.  & \off & ~ & ~ & \both & \toy & 0 & \nomention & \nomention & \nomention &  ~  \\
            
            Tran~\cite{tran2021deep} & 2021 & CS & 31 &  \otheroai Autonomous Agents & \shield & ~ & ~ & \system & \toy & 0 & \yesquant & \nomention & \nomention & ~\\
            
            \hline

            Al-Hababi~\cite{al2020man} & 2020 & CS & 9 & \mitre Recon. & \off & ~ & ~ & \system & \toy & 11 & \nomention & \nomention & \nomention &  ~ \\
            
            Bhattacharya~\cite{bhattacharya2020automated} & 2020 & CS & 14 &  \otheroai Autonomous Agents(ICS) & \shield & ~ & ~ & \system & \toy & 0 & \yesqual & \yesqual & \nomention &  ~ \\
            
            Chowdhary~\cite{chowdhary2020autonomous} & 2020 & CS & 49 &  \otheroai Autonomous Agents & \shield & ~ & ~ & \system & \toy & 0 & \yesqual & \yesqual & \yesqual & \href{https://github.com/ankur8931/asap}{x} \\
            
            Halimi~\cite{halimi2020efficient} & 2020 & CS & 6 &  \otheroai  Priv.atk. & \off & ~ & ~ & \woman & ~ & 36 & \yesquant & \nomention & \nomention & ~ \\
            
            Hu~\cite{hu2020automated} & 2020 & CS & 74 &  \otheroai Autonomous Agents & \shield & ~ & ~ & \system & \toy & 0 & \nomention & \nomention & \nomention & ~ \\
            
            Lee~\cite{lee2020cybersecurity} & 2020 & CS & 18 & \mitre Cred.Acc.  & \off & ~ & ~ & \system & \toy & 2 & \yesqual & \nomention & \nomention &  ~ \\
            
            Lee~\cite{lee2020improved} & 2020 & CS & 4 & \mitre Cred.Acc.  & \off & ~ & ~ & \system & \toy & 2 & \yesquant & \nomention & \yesquant & ~ \\
            
            Liu~\cite{liu2020deepsqli} & 2020 & CS & 42 & \mitre Init.Acc. & \shield & ~ & ~ & \system & \real & 0 & \yescost & \yesquant & \yesquant & \href{https://github.com/COLA-Laboratory/issta2020}{x} \\
            
            Pearce~\cite{pearce2020machine} & 2020 & CS & 5 & \mitre Def.Ev. & \off & \cmark & ~ & \system & \toy & 0 & \yesqual & \nomention & \nomention &  \href{https://github.com/moohax/Deep-Drop}{x} \\
            
            Sharevski~\cite{sharevski2020wikipediabot} & 2020 & CS & 3 & \otheroai Atk on soc. & \off & \cmark & ~ & \both & \toy & 13 & \nomention & \nomention & \nomention &  ~ \\
            
            Shu~\cite{shu2020generative} & 2020 & CS & 46 & \mitre Def.Ev. & \off & \cmark & ~ & \system & \toy & 1 & \yesqual & \nomention & \nomention &  ~ \\
            
            Song~\cite{song2020mab} & 2020 & CS & 33 & \mitre Def.Ev. & \off & \cmark & ~ & \system & \real & 0 & \yesquant & \nomention & \nomention & \href{https://github.com/weisong-ucr/MAB-malware}{x} \\
            
            Valea~\cite{valea2020towards} & 2020 & CS & 27 & \otheroai Autonomous Agents & \shield & ~ & ~ & \system & \toy & 0 & \yesqual & \yesqual & \nomention  & ~ \\
            
            Yu~\cite{yu2020ai} & 2020 & CS & 8 & \mitre Disc. & \off & \cmark & ~ & \system & \real & 0 & \yesqual & \yesqual & \yesquant & ~ \\
            
            \hline
        
            Basu~\cite{basu2019generating} & 2019 & CS  & 0 & \mitre Init.Acc.  & \off & \cmark & ~ & \both & \real & 8 & \nomention & \nomention & \yesquant &  ~ \\
            
            Cecconello~\cite{cecconello2019skype} & 2019 & CS & 12 & \mitre Recon. & \off & \cmark & ~ & \both & \real & 5 & \yesqual & \yesqual & \nomention &  ~\\
            
            Chung~\cite{chung2019availability} & 2019 & CS  & 31 & \mitre Evas.*  & \off & \cmark & ~ & \system & \toy & 0 & \yesqual & \yesqual & \nomention &  ~\\
            
            Das~\cite{das2019x} & 2019 & CS & 115 & \mitre Recon. & \off & ~ & ~ & \system & \real & 0 & \yesquant & \nomention & \yesquant &  \href{https://github.com/SparcLab/X-DeepSCA}{x} \\
            
            Ghanem~\cite{ghanem2019reinforcement} & 2019 & CS & 104 & \otheroai Autonomous Agents & \shield & ~ & ~ & \system & \toy & 0 & \yesqual & \yesqual & \nomention &  ~\\
            
            Tshimula~\cite{tshimula2019har} & 2019 & CS & 10 & \otheroai Priv.atk. & \off & ~ & ~ & \woman & ~ & 24 & \nomention & \nomention & \nomention & ~ \\
            
            Yu~\cite{yu2019low} & 2019 & CS & 26 & \mitre Init.Acc. & \off & ~ & ~ & \system & \real & 0 & \nomention & \yesqual & \nomention & ~ \\
            
            Zhang~\cite{zhang2019using} & 2019 & CS  & 18 & \mitre Cred.Acc.* & \off & \cmark & ~ & \system & \real & 0 & \nomention & \nomention & \nomention & ~ \\
            
            \hline

            Anand~\cite{anand2018keyboard} & 2018 & CS & 19 & \mitre Cred.Acc. & \off & \cmark & ~ & \system & \toy & 1 & \yesqual & \nomention & \yesquant  & ~ \\
            
            Bahnsen~\cite{bahnsen2018deepphish} & 2018 & CS & 78 & \mitre Def.Ev. & \off & \cmark & ~ & \system & \toy & 0 & \yesquant & \nomention & \yesquant & \href{https://github.com/albahnsen/DeepPhish_BlackHat_Demo}{x} \\
            
            Kronjee~\cite{kronjee2018discovering} & 2018 & CS & 51 & \mitre Init.Acc. & \off & \cmark & ~ & \system & \real & 0 & \yesquant & \yesqual & \yesqual & \href{https://github.com/jorkro/wirecaml}{x} \\
            
            Rigaki~\cite{rigaki2018bringing} & 2018 & CS & 133 & \mitre Def.Ev. & \off & \cmark & ~ & \system & \real & 0 & \yesqual & \nomention & \nomention &  ~ \\
            
            Zhuo~\cite{zhou2018deeplink} & 2018 & CS & 173 & \otheroai  Priv.atk. & \off & ~ & ~ & \woman & ~ & 29 & \yesquant & \nomention & \yesquant & \href{https://github.com/KDD-HIEPT/DeepLink}{x} \\
            
            \hline

            Yao~\cite{yao2017automated} & 2017 & CS & 210 & \otheroai Atk on soc. & \off & \cmark & ~ & \both & \toy & 24 & \yesquant & \yesqual & \yesquant &  ~ \\
            \hline
            
            Anderson~\cite{anderson2016deepdga} & 2016 & CS & 230 & \mitre C2 & \off & \cmark & ~ & \system & \toy & 0 & \nomention & \nomention & \nomention & \href{https://github.com/roreagan/DeepDGA}{x} \\
            
            Ceccato~\cite{ceccato2016sofia} & 2016 & CS & 45 & \mitre Init.Acc. & \shield & ~ & ~ & \system & \real & 0 & \yesqual & \yesqual & \yesquant &  ~ \\
            
            Grieco~\cite{grieco2016toward} & 2016 & CS & 276 & \mitre Disc. & \shield & ~ & ~ & \system & \real & 0 & \nomention & \nomention & \yesquant & \href{https://github.com/CIFASIS/VDiscover}{x} \\
            
            \hline
            
            Freitas~\cite{freitas2015reverse} & 2015 & CS & 181 & \otheroai Atk on soc. & \off & ~ & ~ & \both & \real & 230 & \yesquant & \nomention & \nomention  & ~ \\
            
            \hline

            Bursztein~\cite{bursztein2014end} & 2014 & CS & 178 & \mitre Init.Acc. & \off & \cmark & ~ & \system & \real & 0 & \nomention & \yesqual & \yesquant & ~ \\
            
            \hline
            
            Adali~\cite{adali2012predicting} & 2012 & CS & 151 & \otheroai Priv.atk. & \off & ~ & ~ & \woman & ~ & 39 & \nomention & \nomention & \nomention & ~ \\
            
            Malhotra~\cite{malhotra2012studying} & 2012 & CS & 307 & \otheroai Priv.atk. & \off & ~ & ~ & \woman & ~ & 53 & \nomention & \nomention & \nomention &  ~ \\
            
            Sumner~\cite{sumner2012predicting} & 2012 & CS & 423 & \otheroai Priv.atk. & \off & ~ & ~ & \woman & ~ & 55 & \nomention & \nomention & \nomention & ~ \\
            
            \hline

            Goldbeck~\cite{golbeck2011predicting} & 2011 & CS & 852 & \otheroai Priv.atk. & \off & ~ & ~ & \woman & ~ & 35 & \nomention & \nomention & \nomention & ~ \\
            
            Yamaguchi~\cite{yamaguchi2011vulnerability} & 2011 & CS & 212 & \mitre Disc. & \shield & ~ & \cmark & \system & \real & 0 & \nomention & \nomention & \nomention & ~ \\
            
            \hline
            
            Bursztein~\cite{bursztein2009decaptcha} & 2009 & CS & 103 & \mitre Init.Acc. & \off & \cmark & ~ & \system & \real & 0 & \yesquant & \nomention & \nomention & ~ \\
            
            \hline
            
            Golle~\cite{golle2008machine} & 2008 & CS & 394 & \mitre Init.Acc. & \off & \cmark & ~ & \system & \real & 0 & \nomention & \nomention & \nomention & ~\\
            
            \hline
            
            \bottomrule
        \end{tabular}
    }
    \vspace{-5mm}
\end{table*}

\section{Overview of Academic Literature on OAI}
\label{sec:3-academic}

\noindent
We present the first part of our primary contribution (\smabb{C}1) by focusing the attention on academic publications. Our literature search yielded 95 papers on OAI (see §\ref{ssec:2.1-literature}).\footnote{Among the works reviewed in this SoK, only two \cite{yao2017automated, golle2008machine} are from the big four conferences (three are from collocated workshops~\cite{bursztein2014end, yamaguchi2011vulnerability, bursztein2009decaptcha}).} In what follows, we first discuss the 79 technical papers (§\ref{ssec:3.1-technical}) and then the 16 non-technical papers (§\ref{ssec:3.2-nontechnical}), which we analyze under the lens of our checklist (refer to §\ref{ssec:checklist}).

\subsection{Technical Papers}
\label{ssec:3.1-technical}

\noindent 
We present in Table~\ref{tab:technical} (in the Appendix) the systematization of the 79 technical papers, in accordance with our checklist (§\ref{sssec:usecase_technical} to §\ref{sssec:costbenefit_technical}). Furthemore, in §\ref{sssec:technical_technical} we analyze the technical requirements pertaining to this class of works.

\subsubsection{\textbf{OAI Use Case}}
\label{sssec:usecase_technical}
We begin by considering the OAI use case envisioned by each work, aligning it to MITRE ATT\&CK. 

\begin{figure}[!htbp]
    \centering
    \centerline{
    \includegraphics[width=0.44\textwidth]{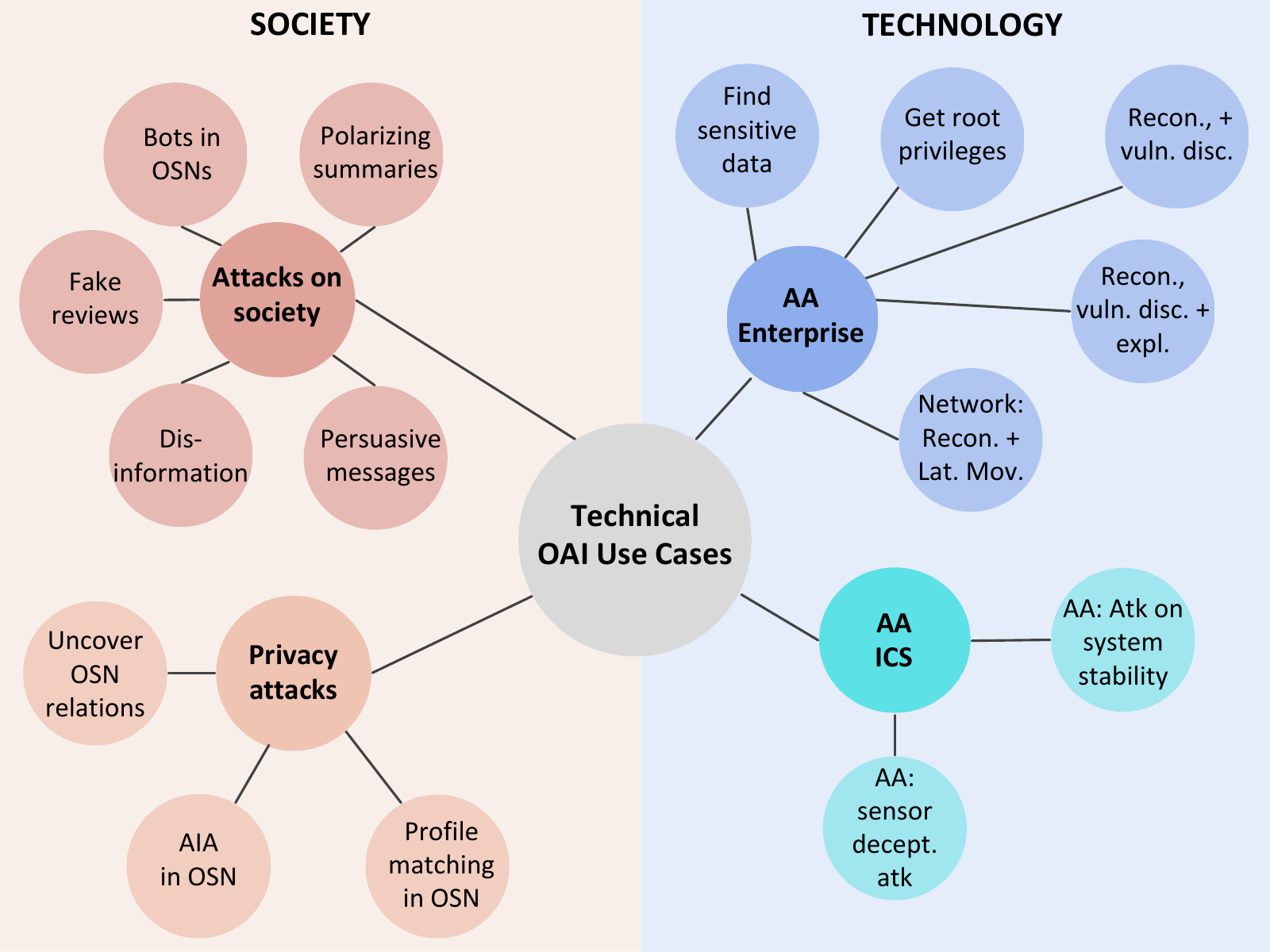}
    }
    \vspace{-1mm}
    \caption{\textbf{OAI use cases not covered by MITRE ATT\&CK (technical papers).}
    \textmd{\footnotesize Some focus on society and privacy (OSN=online social networks).}} 
    \label{fig:tech-other}
    \vspace{-5mm}
\end{figure}

\begin{itemize}
    \item \textit{We mapped 48 papers (\smamath{61\%}) to the use-cases covered by MITRE}~\cite{mitre-tactics}. 
    Among these, 2 papers were mapped to the ICS matrix (one paper focusing on Evasion and another on Process Control), whereas 1 paper was mapped to the Mobile matrix (addressing Credential Access). The remaining 45 papers aligned with the  Enterprise matrix. In general, among these 48 papers, most works focus on Initial Access (\smamath{22\%}). Other common goals of OAI are Defense Evasion (\smamath{9\%}), Credential Access (\smamath{9\%}), and Discovery (\smamath{6\%}). Only \smamath{4\%} of the papers focus on exploiting OAI for Reconnaissance: this is likely due to the fact that this step can be carried out via various well-known means (e.g., port scanning, or OSINT) which do not require OAI and which are not easily detected~\cite{kelly2019adversarially}. We could not find any paper that specifically proposed OAI techniques for Impact or Lateral Movement (some, however, do use autonomous attack agents to carry out also these operations; e.g.,~\cite{bhattacharya2020automated}).
       
    \item \textit{The remaining (\smamath{39\%}) papers envisioned use-cases not covered by MITRE ATT\&CK.}  These papers mainly address attacks on society, privacy, or focus on autonomous attack agents (which involves automating various MITRE tactics, as done in~\cite{valea2020towards}). 
    Attacks on society cover, e.g., polarizing summaries~\cite{sharevski2020wikipediabot} or crowdturfing attacks in online review systems~\cite{yao2017automated}, while privacy attacks include attribute inference attacks~\cite{karanatsiou2022my}, or profile matching across multiple social networks~\cite{halimi2020efficient, zhou2018deeplink}. 
    We report in Fig.~\ref{fig:tech-other} the groups (and corresponding relationships) of the OAI use cases not covered by MITRE.
\end{itemize}

\noindent
With respect to the purpose, out of 79 papers, 43 (\smamath{54\%}) propose novel attacks~(\off); 30 (\smamath{38\%}) focus on offensive security~(\shield), and 6 (\smamath{8\%}) use AI as a hacking assistant~(\assist).
Some works focusing on novel attacks (\off) consider potential defenses against the proposed attacks: e.g.,~\cite{chowdhary2023generative} considers Web Application Firewalls to protect the targeted application. Only half (\smamath{51\%}) of the novel attack (\off)\ papers, however, propose/evaluate a potential countermeasure, which is alarming. Moreover, for those papers that focus on offensive security (\shield), we found that only \smamath{14\%} of these considered the possibility that their tool could be used for malicious purposes. Notably, only one work (\cite{yamaguchi2011vulnerability}) did so before 2020. 
We believe that the publication of the ``Autonomous Weapon Open Letter'' from Future of Life~\cite{futureLifeLetter} in 2016 may have encouraged more works to acknowledge potential abuse of AI-based tools (for instance, this letter was explicitly mentioned in both~\cite{erdHodi2021simulating,zennaro2023modelling}). 
Nonetheless, to shed further light on this aspect, \textit{we reached out to the authors of 25 offensive security} (\shield) papers in which we found no explicit mention of potential abuse of the proposed AI-based tool, inquiring if such a scenario would open up potential security concerns. We received a response from the authors of 11 of these papers (44\% response rate): while all responses did not deny such a possibility, the overall sentiment is that a considerable amount of effort is necessary to exploit the proposed AI-based tool for real-world attacks. 

\vspace{-2mm}

\begin{cooltextbox}
    \textbf{\textsc{Lessons Learned}:}  \textbf{(1)}~Many reported OAI techniques transcend the use-cases covered by MITRE ATT\&CK, which was primarily used in prior summaries, such as~\cite{mirsky2021threat}. \textbf{(2)}~Many works reporting OAI techniques do not adequately consider countermeasures. 
    \textbf{(3)}~Even ethical statements that merely acknowledge potential abuse risks for considered AI techniques are quite uncommon in prior OAI research. 
\end{cooltextbox}
\vspace{-1mm}

\subsubsection{\textbf{Target/Impact}}
\label{sssec:targetimpact_technical}
For the second set of criteria in our checklist, we first assess whether the reported OAI techniques target a human (\woman), a system (\system), or both; whether attacks against systems target ``toy'' (\toy) or ``real'' (\real) systems.
Among the analyzed 79 papers, 62 (\smamath{78\%}) consider attacks against systems, while 10 (\smamath{13\%}) propose attacks against humans, and 7 (\smamath{9\%}) use AI to attack both systems and humans. For the 69 papers that attack systems, \smamath{70\%} consider ``toy'' (\toy) systems, whereas only 21 papers target ``real'' (\real) systems.\footnote{For example, the authors of~\cite{cecconello2019skype} attack a Voice-over-IP software (Skype) which we consider as a ``real'' system; whereas the authors of~\cite{anderson2016deepdga} attack a DGA detector envisioned by prior work (\cite{yadav2012detecting,antonakakis2012throw}), i.e. a ``toy'' system.} Among these 21, the majority (\smamath{52}\%) use OAI for ``Initial Access'' (more precisely, on phishing, web applications, and CAPTCHAs). 
Then, we assess the emphasis given by the paper on the societal aspect of its contribution, and we counted the number of occurrences of the words ``society,'' ``social,'' ``societal,'' or ``socio'' in the main text of the paper. We observed that 49 papers (\smamath{62\%}) \textit{never mention any of these terms}. Among those that do use these terms, most works belong to the ``privacy'' domain (e.g., \cite{freitas2015reverse} mentions these terms 230 times). Finally, for the 17 papers considering attacks against humans, we investigated whether they carried out any validation (e.g., a user study) on real people. None of these 17 papers did. A likely explanation for this is the difficulty to carry out such experiments due to, e.g., ethical concerns.

\vspace{-2mm}

\begin{cooltextbox}
    \textbf{\textsc{Lessons Learned}:}  
    \textbf{(1)}~Most prior technical papers have targeted systems rather than humans; however, attacks against \textit{real systems} are scarce. This finding (which echoes~\cite{apruzzese2023real}) suggests that the impact of OAI techniques on real systems may still be unknown.
    \textbf{(2)}~Most works do not emphasize the impact of their findings on our society. This may suggest that the societal impact of OAI on our society is difficult to estimate in this domain.
\end{cooltextbox}

\vspace{-1mm}

\subsubsection{\textbf{Cost/Benefit}} 
\label{sssec:costbenefit_technical}
To shed light on the practicality of the identified techniques, we examined whether the authors considered attackers' economic incentives for applying the proposed method (benefits) or highlighted potential costs that may deter attackers from doing so. Among our 79 papers, 32 (\smamath{41\%}) did not provide any analysis (\nomention) of the attacker's benefits when leveraging AI: a typical conclusion in these papers is that the ``proposed method works.'' Among the remaining 47 papers, 23 (\smamath{29\%}) provided a qualitative evaluation (\yesqual) of the benefits, and 22 (\smamath{28\%}) even a quantitative evaluation (\yesquant). Only \smamath{2\%} explicitly mentioned (\yescost) monetary benefits or time saved according to metrics that go beyond sheer accuracy and precision, e.g.~\cite{liu2020deepsqli}. Concerning the \textit{costs}, 58 (\smamath{73\%}) papers do not provide any analysis, while 17 (\smamath{22\%}) only provide a brief qualitative evaluation. The remaining 4 papers (\smamath{5\%}) provide a quantitative analysis.  In general, no paper (among our 79 technical papers) has precisely quantified the required investment to launch the attack, or the return on such investment if the attack were successful. Finally, another measure of the attack practicality is whether the same objective could be achieved without AI. The majority of papers (\smamath{65\%}) did not consider any non-AI-baseline for comparison.

\vspace{-1mm}

\begin{cooltextbox}
    \textbf{\textsc{Lessons Learned}:} The economical aspect tends to be neglected by most technical papers. Such a finding raises a question, to what extent OAI represents a tangible threat in practice and if so, what threat actors (from individuals to state-sponsored groups) are likely to deploy such techniques. Future work should take such cost factors into account.
\end{cooltextbox}

\vspace{-1mm}

\begin{table*}[!htbp]
    \centering
    \caption{\textbf{Literature Review -- Non-Technical OAI Papers.}
    \textmd{\footnotesize We report 16 non-technical OAI papers (scrutinized under the same criteria as those in Table~\ref{tab:technical}).}} 
    \label{tab:related-non-tech}
    \vspace{-2mm}
    \resizebox{1.65\columnwidth}{!}{
        \begin{tabular}{c|c|c|c|c|c|c|c|c|c|c|c|c|c|c}    
            \toprule
            \multicolumn{4}{c|}{} & \multicolumn{4}{|c|}{\textbf{OAI Use Case}} & \multicolumn{3}{|c|}{\textbf{Target/Impact}} & \multicolumn{3}{|c|}{\textbf{Cost/Benefit}}  \\

            \cmidrule(lr){5-7} \cmidrule(lr){8-10} \cmidrule(lr){11-13} 

            \textbf{Paper} & \textbf{Year} & \textbf{Discipline} & \textbf{Cit.} & \textbf{Specific OAI Use Case} & \textbf{Purpose} & \textbf{Def.} & \textbf{Pot. Abuse?} & \textbf{Targ.} & \textbf{Real/Toy} & \textbf{Social Persp.} & \textbf{Benef.} & \textbf{Cost} & \textbf{Base.} & \textbf{Code} \\
            \midrule

            Dall'Agnol~\cite{Dall2023Artificial} & 2023 & Soc. Sciences & 12 &  \otheroai  Atk.in (cyber) war & \off & \cmark & ~ & \both & ~ & 0 & \nomention & \nomention & \nomention \\
            
            De Angelis~\cite{de2023chatgpt}    & 2023 & Medicine & 166 & \otheroai Atk.on society & \off & \cmark & ~ & \woman  & ~   & 7   & \yesqual  & \yesqual  & \nomention          \\

            Illiashenko~\cite{illiashenko2023security}   & 2023 & CS  & 4 & \otheroai Atk.on society & \off & \cmark & ~ & \woman  & ~   & 2   & \yesqual  & \nomention          & \nomention          \\
            
            Pashentsev~\cite{pashentsev2023destabilization}    & 2023 & Soc. Sciences & 0 & \otheroai Atk.on society & \off & \cmark & ~ & \both  & ~   & 161 & \yesqual  & \nomention          & \nomention          \\

            Rickli~\cite{rickli2023artificial} & 2023 & Soc. Sciences & 3 & \otheroai Atk.in (cyber) war & \off & ~ & ~ & \system & ~ & 0 & \nomention & \nomention & \nomention \\
            
            \hline
            
            Hao~\cite{hao2022intelligent}           & 2022 & CS & 0 &  \otheroai Autonomous Agents             & \shield & ~ & \cmark & \both  & ~ & 1   & \nomention          & \nomention          & \nomention          \\
            
            Kasim~\cite{kasim2022cybersecurity}  & 2022 & CS & 0 &  \otheroai Autonomous Agents             & \off & \cmark & ~ & \both  & ~ & 0   & \yesquant & \yesquant & \yesquant \\
            
            McIlroy-Young~\cite{mcilroy2022mimetic} & 2022 & CS & 7 &  \otheroai Atk.on society & \off & \cmark & ~ & \woman  & ~   & 17  & \yesqual  & \yesqual  & \nomention          \\

            \hline

            Nica~\cite{nica2020using} & 2020 & Soc. Sciences & 1 &  \otheroai Atk.in (cyber) war & \off & ~ & ~ & \both & ~ & 18 & \yesqual & \yesqual & \nomention \\
            
            Skeba~\cite{skeba2020informational}  & 2020 & CS & 11 & \otheroai Priv.atk       & \off & \cmark & ~ & \woman  & ~   & 28  & \nomention          & \nomention          & \nomention          \\
            
            \hline

            Easttom~\cite{easttom2019methodological} & 2019 & CS & 5 &\otheroai Atk.in (cyber) war & \off & ~ & ~ & \both & ~ & 0 & \nomention & \nomention & \nomention \\

            Burton~\cite{burton2019autonomous} & 2019 & CS & 5 &  \otheroai Atk.in (cyber) war  & \off & ~ & ~ & \system & ~ & 0 & \yesqual & \nomention & \nomention \\

            Burton~\cite{burton2019understanding} & 2019 & CS  & 47 &  \otheroai Atk.in (cyber) war  & \off & ~ & ~ & \both & ~ & 15 & \yesqual & \yesqual & \nomention \\
            
            Giaretta~\cite{giaretta2019community}   & 2019 & CS & 11 & \mitre Init.Acc.   & \off & ~ & ~ & \woman  & ~   & 4   & \nomention          & \nomention          & \nomention          \\

            \hline
            
            Maus~\cite{maus2015decoding}    & 2015 & CS & 7 &  \otheroai Atk.on society & \off & \cmark & ~ & \system. & ~ & 64  & \yesqual  & \nomention          & \nomention          \\
            
            \hline

            Guarino~\cite{guarino2013autonomous} & 2013 & CS & 22 &  \otheroai Atk.in (cyber) war  & \off & ~ & ~ & \both & ~ & 0 & \yesqual & \yesqual & \nomention \\

            \bottomrule
        \end{tabular}
   }
    \vspace{-1mm}
\end{table*}

\subsubsection{\textbf{Technical Requirements}}
\label{sssec:technical_technical}
We conclude this section by analyzing the technical requirements of the 79 ``technical'' papers. Specifically, we address the question ``What is the degree of technical effort required to set up the proposed OAI tool?'' At a high level, we proceed as follows.
\begin{itemize}
    \item First, we examine whether the tool is based on a pre-existing ML model (e.g., ChatGPT) or if the ML model must be developed from scratch. Only 9 (\smamath{11\%}) papers rely on a pre-existing model (e.g.,~\cite{khan2021offensive} used a pre-trained Large Language Model to generate spear-phishing emails with the intent of deceiving the system and the user). 
    
    \item If the ML model must be developed from scratch (which is the case for 70 papers out of 79) we scrutinize the availability of the data required to train such an ML model. 25 papers (\smamath{36\%}) rely on publicly available data (e.g., benchmarks, such as~\cite{biesner2022combining}); for 6 papers (\smamath{9\%}), the authors needed special access rights to obtain their training data (e.g.,~\cite{chung2019availability}); the remaining 39 papers (\smamath{57\%}), entailed creation of a custom training dataset (e.g.,~\cite{chung2019availability}).

    \item Finally, we review the reproducibility of the implementation. Only 17 (\smamath{22\%}) papers released their source code, and 5 papers (\smamath{6\%}) release the exact prompts used to realize the attack. In contrast, 57 papers (\smamath{72\%}) do not provide such low-level details (a result which echoes~\cite{apruzzese2023real,olszewski2023get}).
\end{itemize}
A detailed explanation of this analysis is provided in Appendix~\ref{sapp:technical-algorihm} (and these results are also shown in Table~\ref{tab:technical-algorithm}).

\vspace{-2mm}

\begin{cooltextbox}
    \textbf{\textsc{Lessons Learned}:} 
    Most technical papers implement their OAI tool from scratch (and few release their code/data), suggesting that implementing/launching the attack by third parties is not trivial in practice. However, we can expect this trend to change given the increasing availability of LLM---which could benefit both researchers and attackers.
\end{cooltextbox}

\subsection{Non-technical Papers}
\label{ssec:3.2-nontechnical}
\noindent
We now analyze the 16 ``non-technical'' papers (in Table~\ref{tab:related-non-tech} in the Appendix) through our checklist and discuss our findings.

\subsubsection{\textbf{OAI Use Case}}
\label{sssec:usecase_nontechnical}
We could only map one paper~\cite{giaretta2019community} to MITRE. 
The remaining 15 (\smamath{94\%}) papers consider use cases not covered by MITRE. For instance, \cite{kasim2022cybersecurity} uses game theory to analyze whether ``human defenders'' can withstand ``AI attackers,'' and conclude that well-trained AI agents are almost impossible to beat. Intriguingly, 7 papers (\smamath{46\%}) focus on \textit{cyber warfare} (not covered by MITRE). For instance, \cite{easttom2019methodological} provides theoretical arguments on how AI could be used to develop malware bypassing the detection mechanisms of the attacked entity in a cyber war. In terms of purpose, 15 papers envision using AI for novel attacks (\off), and 1 for offensive security~(\shield).

\subsubsection{\textbf{Target/Impact}}
\label{sssec:targetimpact_nontechnical}
Most papers (13, \smamath{81\%}) consider OAI targeting humans (\woman), and the overall occurrence of society-related terms is higher than for technical papers (\smamath{38\%} of papers in Table~\ref{tab:related-non-tech} have ten or more occurrences, compared to \smamath{18\%} for those in Table~\ref{tab:technical}). Finally, none of these 16 papers carry out user studies with real humans to validate any given hypothesis.

\subsubsection{\textbf{Cost/Benefit}} 
\label{sssec:costbenefit_nontechnical}
Most papers provide a shallow analysis of benefits (38\% \nomention, 56\% \yesqual) and costs (63\% \nomention, 31\% \yesqual).

\vspace{-2mm}

\begin{cooltextbox}
    \textbf{\textsc{Lessons Learned}:} Non-technical papers on OAI put a greater emphasis on the human perspective and envision scenarios not covered by MITRE, such as cyber warfare. However, these works do not carry out user studies to validate their hypotheses or assess the opinion of real people. Lack of such a validation may either over- or under-estimate the relevance of the envisioned OAI scenario to our society. 
\end{cooltextbox}

\begin{table*}[!htbp]
    \centering
    \caption{\textbf{Analysis of InfoSec Briefings.}
    \textmd{\footnotesize 
    We report 38 industrial ``InfoSec'' briefings from BlackHat and DefCon (scrutinized under the same criteria as the works in Table~\ref{tab:technical}). Column ``Acad.?'' denotes whether a briefing also has a corresponding publication (we denote * with preprints).}}
    \label{tab:related-industry}
    \vspace{-2mm}
    \resizebox{1.6\columnwidth}{!}{
        \begin{tabular}{c|c|c|c|c|c|c|c|c|c|c|c|c|c}
            \toprule
            \multicolumn{2}{c|}{} & \multicolumn{4}{|c|}{\textbf{OAI Use Case}} & \multicolumn{3}{|c|}{\textbf{Target/Impact}} & \multicolumn{3}{|c|}{\textbf{Cost/Benefit}} & \multicolumn{1}{|c}{~} & \multicolumn{1}{|c}{~}\\
            
             \cmidrule(lr){3-6} \cmidrule(lr){7-9} \cmidrule(lr){10-12} 

            \textbf{Author} & \textbf{Year} & \textbf{Specific OAI Use Case} & \textbf{Purpose} & \textbf{Def.?} & \textbf{Pot. Abuse?} & \textbf{Targ.} & \textbf{Real/Toy} & \textbf{Social Persp.} & \textbf{Benef.} & \textbf{Cost} & \textbf{Base.} & \textbf{Code} & \textbf{Acad.?}\\
            \midrule

            Scheiner~\cite{Scheiner2023} & 2023 & \otheroai Atk. on society & \off & \cmark & ~ & \woman & ~ & 7 & \yesqual & \yesqual & \yesqual & ~ &  ~ \\
            Canham~\cite{Canham2023}  & 2023 & \otheroai Atk. on society & \off & \cmark & ~ & \woman & ~ & 11 & \yesqual & \yesqual & \yesqual & ~ &  ~ \\
            Heiding~\cite{Heiding2023} & 2023 & \mitre Init.Acc. & \off & \cmark & ~ & \both & \real & 0 & \yesquant & \yesqual  &  \yesquant & ~ & \cite{heiding2023devising}* \\ 
            
            Herbert-Voss~\cite{Herbert-Voss2023} & 2023 & \mitre Init.Acc. & \assist & ~ & ~ & \system & \real & 2 & \yesqual & \yesqual & \yesqual & ~ &  ~ \\
            
            Waligóra~\cite{Waligora2023} & 2023 & \mitre Recon. & \off & ~ & ~ & \system & \real & 0 & \yesqual  & \yescost & \nomention & ~ & ~ \\
            Gibson~\cite{Gibson2023} & 2023 & \otheroai  Atk. on society & \off & \cmark & ~ & \woman & ~ & 4 & \yesqual & \yesqual & \nomention & ~ & ~ \\
            Zror~\cite{Zror2023} & 2023 & \mitre Recon. & \off & ~ & ~ & \woman & ~ & 7  & \yesqual & \yesqual & \yesqual &      & ~  \\
            
            \hline
            
            Xing~\cite{Xin2022} & 2022 & \mitre Init.Acc. & \off & \cmark & ~ & \both & \toy & 2 & \yesquant & \nomention & \yesquant & \href{https://github.com/dstsmallbird/SiF-DeepVC_Dataset}{x} & ~ \\
            
            Chi~\cite{Chi2022} & 2022 & \mitre Init.Acc. & \shield & ~ & \cmark & \system & \real & 0 & \yesqual & \nomention & \yesqual & ~ & ~ \\

            \hline
            
            Lim~\cite{Lim2021} & 2021 & \mitre Init.Acc. & \off & \cmark & ~ & \both & \real & 8 & \yesquant & \yesqual & \yesquant & ~ & ~ \\
            Lohn~\cite{Lohn2021} & 2021 & \otheroai  Atk. on society & \off & \cmark & ~ & \woman & ~ & 3 & \yesquant & \yesquant & \nomention & ~ & ~ \\

            \hline
            
            Tully~\cite{Tully2020} & 2020 & \otheroai  Atk. on society & \off & \cmark & ~ & \both & \real & 20 & \yesqual & \yesqual & \nomention & ~ & ~ \\
            Basu~\cite{Basu2020} & 2020 & \mitre Init.Acc. & \off & ~ & ~ & \woman & ~ & 2 & \nomention & \nomention & \nomention & \href{https://github.com/titanlambda/identity-cloning-toolkit-ICT}{x} & ~ \\
            Sharma~\cite{Sharma2020} & 2020 & \mitre Disc. & \shield & ~ & ~  & \system  & \real & 2 & \yesqual & \nomention & \nomention & ~ & \cite{chen2020automated} \\ 

            \hline
            
            Takaesu~\cite{Takaesu2019} & 2019 & \otheroai Autonomous Agents & \shield & ~ & \cmark & \system  & \real & 0 & \nomention & \nomention & \nomention & \href{https://github.com/gyoisamurai/GyoiThon}{x} & ~ \\
            Botwicz~\cite{Botwicz2019} & 2019 & \mitre Recon. & \shield & ~ & \cmark & \system & \real & 0 & \nomention & \nomention & \nomention & \href{https://github.com/Samsung/cotopaxi}{x} & ~\\
            Bursztein~\cite{Bursztein2019}  & 2019 & \mitre Recon. & \off & ~ & ~ & \system & \real & 0 & \yesqual & \yesqual & \nomention & \href{https://github.com/google/scaaml/tree/main/scaaml}{x} & ~ \\
            Ding~\cite{Ding2019} & 2019 & \mitre Init.Acc. & \shield & ~ & ~ & \system & \real & 0 & \yesqual & \nomention & \nomention & ~ & ~ \\
            Price~\cite{Price2019} & 2019 & \otheroai Atk. on society & \off & \cmark & ~ & \both & \real & 1 & \nomention & \nomention & \nomention & \href{https://github.com/zerofox-oss/deepstar}{x} & ~ \\

            \hline
            Bahnsen~\cite{Bahnsen2018-BlackHat} & 2018 & \mitre Dev.Ev. & \off & \cmark & ~& \system & \toy & 1 & \yesquant & \nomention & \yesquant & \href{https://github.com/albahnsen/DeepPhish_BlackHat_Demo}{x} & \cite{bahnsen2018deepphish} \\
            Greenstadt~\cite{Greenstadt2018} & 2018 & \otheroai  Priv.atk & \off & ~ & ~ & \woman & ~ & 0 & \yesquant & \nomention & \nomention          & ~ & \cite{caliskan2015coding} \\ 
            Kirat~\cite{Kirat2018}  & 2018 & \mitre Def.Ev. & \off & \cmark & ~ & \both & \real & 2 & \yesquant & \nomention & \nomention & ~ & ~ \\
            Perin~\cite{Perin2018} & 2018 & \mitre Recon. & \off &  ~ & ~ & \system & \real & 0 & \yesquant & \nomention & \yesqual & ~ & ~\\
            Gomez~\cite{Gomez2018} & 2018 & \otheroai  Priv.atk & \off & ~ & ~ & \woman & ~ & 4 & \nomention & \nomention & \nomention & \href{https://github.com/ffr4nz/UnknownUnknowns}{x} & ~          \\

            \hline

            Anderson~\cite{Anderson2017} & 2017 & \mitre Def.Ev. & \off  & \cmark & ~ & \system  & \toy & 0  & \yesquant  & \yesquant & \nomention          & \href{https://github.com/endgameinc/gym-malware}{x} & \cite{anderson2018learning} \\ 
            Lain~\cite{Lain2017-BlackHat} &  2017 & \mitre Recon. & \off & \cmark & ~ & \both & \real & 0 & \yesqual & \yesqual & \nomention & \href{https://github.com/SPRITZ-Research-Group/Skype-Type}{x} & \cite{cecconello2019skype} \\
            Morris~\cite{Morris2017} & 2017 & \mitre Init.Acc. & \off & ~ & ~ & \system &  \real & 22  & \nomention  & \nomention  & \nomention  & ~ & ~ \\
            Tully~\cite{Tully2017}      & 2017 &       \otheroai Atk. on society & \off   & \cmark        &          & \both  & \real          & 69     & \yesqual          & \yesqual           & \nomention          &      & ~          \\
            Singh~\cite{Singh2017}      & 2017 & \mitre Recon.        & \off   &   \cmark       &          & \both   &      \real      & 13     & \nomention          & \nomention          & \nomention          &      & ~          \\

            \hline
            
            Polakis~\cite{Polakis2016}    & 2016 & \mitre Init.Acc.     & \off   & \cmark        &          & \system  &    \real        & 0      & \yescost  & \yescost  & \yesquant          &      & \cite{sivakorn2016robot} \\ 
            Argyros~\cite{Argyros2016}    & 2016 & \mitre Init.Acc.    & \shield   &          &          & \system  &    \real        & 0      & \nomention          & \nomention          & \nomention          & \href{https://github.com/lightbulb-framework/lightbulb-framework}{x}    & \cite{argyros2016sfadiff} \\
            Seymour~\cite{Seymour2016}    & 2016 & \mitre Init.Acc.       & \off  & \cmark        &          & \both   & \real          & 22     & \yesquant & \yesquant & \yesquant &      & ~          \\
            Wolff~\cite{Wolff2016}      & 2016 & \mitre Exfil.       & \off   &          &          & \system  &    \real        & 0      & \nomention          & \nomention          & \nomention          &      & ~          \\

            \hline
            
            Bursztein~\cite{Bursztein2014}  & 2014 &   \otheroai Atk. on society & \off  &          &          & \both   & \real          & 0      & \yesqual  & \nomention          & \nomention          & \href{https://github.com/dstsmallbird/SiF-DeepVC_Dataset}{x}    & ~          \\
            Fu~\cite{Fu2014}        & 2014 &  \otheroai  Priv.atk & \off   & \cmark        &          & \both   & \real          & 0      & \yesqual  & \nomention          & \nomention          &      & \cite{yue2014blind} \\ 

            \hline
            
            Vanned~\cite{Vanned2013}    & 2013 & \mitre Res.Dev.  & \off  &          &          & \system  &   \real         & 0      & \yescost  & \yesquant & \yesquant & \href{https://github.com/soen-vanned/forced-evolution}{x}    & ~          \\
            Espinhara~\cite{Espinhara2013}  & 2013 & \mitre Recon.     & \off  &          &          & \both   & \real          & 41     & \nomention          & \nomention          & \nomention          & \href{https://github.com/urma/microphisher}{x}    & ~          \\

            \hline
            
            Clarke~\cite{Clarke2008}     & 2008 &  \otheroai  Atk. on society & \off   &          &          & \woman   & ~          & 0      & \yesquant & \nomention          & \nomention          &      & ~          \\
            
            \bottomrule
        \end{tabular}
    }
    \vspace{-2mm}
\end{table*}

\section{OAI in InfoSec Briefings}
\label{sec:4-infosec}
\noindent
We identified 38 non-academic works (also known as ``briefings'') related to OAI at BlackHat and DefCon. We first analyze these 38 briefings (§\ref{ssec:4.1-infosec_analysis}), and then compare them with the 95 papers from academic literature (§\ref{ssec:4.2-infosec_comparison}).

\subsection{Analysis}
\label{ssec:4.1-infosec_analysis}

\noindent
To finalize our primary contribution (\smabb{C}1), we assess our 38 briefings through our checklist (§\ref{ssec:checklist}), and show the results in Table~\ref{tab:related-industry} (in the Appendix), which also reports briefings related to a scientific paper (which occurs for 9 out of 38 briefings---two of which \cite{bahnsen2018deepphish, cecconello2019skype} are also included in Table~\ref{tab:related-non-tech}). 
The first briefing on OAI we found (discussing how to ``hack human desire'') dates back to 2008~\cite{clarke2008hacking}; many briefings on OAI appeared in 2023---likely due to the rollout of ChatGPT. 

\subsubsection{\textbf{OAI Use Case}}
\label{sssec:usecase_infosec}
Most briefings (25, \smamath{69\%}) can be mapped to MITRE, for which ``Initial Access'' (11, such as using LLM to write phishing emails~\cite{Heiding2023}) and ``Reconnaissance'' (8, e.g., via side-channel~\cite{Waligora2023, Bursztein2019, Perin2018}) are the most prominent use cases. The remaining 13 briefings are not covered by MITRE. These briefings focus mostly on attacks against society (9, such as using AI for virtual kidnapping~\cite{Gibson2023}) and privacy (3, such as deanonymizing developers based on their code~\cite{Greenstadt2018}). 
Overall, only one briefing~\cite{Herbert-Voss2023} used AI as a hacking assistant~(\assist). In contrast, 6 briefings envisioned an offensive security application (\shield), half of which explicitly mention that the proposed tool can be maliciously exploited also by attackers. The remaining 31 briefings proposed a new attack reliant on AI (\off): among these, 18 consider a countermeasure.

\subsubsection{\textbf{Target/Impact}}
\label{sssec:targetimpact_infosec}

9 briefings consider OAI applications targeting humans (\woman), 16 a system (\system), and 13 both. Among those that target systems, 26 attack real systems, and 3 toy systems (e.g., \cite{Xin2022}  generates AI-synthesized speech samples and tests them against three fake-voice detectors proposed by prior academic literature---i.e., a ``toy'' system). 
The term ``society'' is never mentioned in 18 (\smamath{47\%}) briefings; the most occurrences are found in \cite{Tully2017} which considers using AI to hide data in images posted on social networks. 
Only one briefing entails a user study: \cite{Heiding2023} compares the performance of humans and LLM to write phishing emails, and tests which ``author'' was more effective at fooling end-users.

\subsubsection{\textbf{Cost/Benefit}}
\label{sssec:costbenefit_infosec}
Only 6 (\smamath{16\%}) briefings make a clear analysis (\yesquant\ or \yescost) of the \textit{costs} required to realize the attack: for instance,~\cite{Waligora2023} tests the attack on real hardware, quantifying the costs of the proposed attack as: ``a laptop + \$100 PicoScope (software).'' In contrast, 21 briefings (\smamath{55\%}) do not mention the cost/benefit aspect at all, and 11 (\smamath{29\%}) make a shallow qualitative assessment. The opposite holds for the \textit{benefits}: only 10 briefings (\smamath{26\%}) do not mention this aspect, whereas 13 (\smamath{34\%}) perform actual measurements (\yesquant\ or \yescost) and 15 (\smamath{40\%}) make some qualitative analyses.

\subsection{Comparison: Scientific Literature vs InfoSec Briefings}
\label{ssec:4.2-infosec_comparison}

\noindent
Intriguingly, our search revealed that the initial efforts on OAI in both literature and InfoSec date back to 2008. While academic literature has consistently tackled OAI since then, interest in InfoSec venues only started to increase in 2013. 

Let us elucidate the main similarities and differences among these two sources. In regard to ``\emph{1) OAI Use Case},'' technical papers from academia are more similar to Infosec briefings than to non-technical works. Indeed, \smamath{69\%} of InfoSec briefings and \smamath{61\%} of technical papers can be mapped to MITRE (albeit the specific use-cases differ), while this holds for only \smamath{6\%} of non-technical papers. At the same time, InfoSec briefings and technical papers have not focused on cyber warfare, while 46\% of non-technical papers examined this topic. One notable difference, however, is that only 6\% of technical papers focus on attacks on society---whereas the share is 24\% and 31\% for InfoSec briefings and non-technical papers, respectively. 

Related to ``\emph{2) Target/Impact},'' \smamath{78\%} of technical papers consider attacks against systems, compared to \smamath{42\%} for InfoSec briefings, which consider humans in the remaining attacks (either isolated or combined with systems). However, InfoSec briefings consider more attacks on real systems (\smamath{90\%}) compared to technical papers (\smamath{30\%}). For \emph{3) Cost/Benefit} the economic aspect is mostly neglected by academic works, proposing attacks that need to be built from scratch and with custom datasets (\smamath{57\%}). Among InfoSec briefings \smamath{16\%} make a clear analysis of the costs required to carry out the attack, and \smamath{34\%} make actual measurements of the benefits. While academic technical works encompass compelling use cases, InfoSec briefings tend to provide a more comprehensive examination of risks (\emph{in the wild}). Although this observation appears intuitive, integrating both perspectives drives a comprehensive assessment of the risks posed by OAI to society.

\vspace{-2mm}

\begin{cooltextbox}
    \textbf{\textsc{Lessons Learned}:} Most OAI use cases of InfoSec briefings are not discussed in research papers. Only 47\% briefings consider countermeasures or emphasize that attackers can use their tool. These results highlight blind spots exploitable by attackers, to be addressed by future research.
\end{cooltextbox}
\section{User Survey: what do laypeople think of OAI?}
\label{sec:5-userSurvey}
\noindent
We now present the results of our user survey with non-experts, revealing what laypeople think about the offensive potential of AI. We first describe the demographics of our participants (§\ref{ssec:demographics}), and then present the results (§\ref{ssec:userstudy_results})

\subsection{Demographics}
\label{ssec:demographics}

\noindent
We received 570 responses but removed 21 because they were from underage participants or clearly not informative. Hence, our results reflect the opinion of 549 individuals.

Overall, \smamath{63\%} of our respondents identify themselves as ``male,'' \smamath{36\%} as ``female'' and less than \smamath{1\%} as ``other.'' Our respondents are from 42 countries, mostly from Europe (\smamath{70\%}), North America (\smamath{21\%}), and Asia (\smamath{7\%}). Over \smamath{90\%} are between 18--54 years old, and \smamath{2\%} are over 65. With respect to educational qualifications, \smamath{80\%} of participants hold at least a Bachelor's degree. Employment status among the respondents varies, with \smamath{83\%} employed (full or part-time), \smamath{15\%} students, and \smamath{2\%} retired or unemployed. \smamath{75\%} of the participants are engaged in IT-related work. In terms of knowledge of cybersecurity (or AI), \smamath{23\%} (\smamath{17\%}) consider themselves as beginners, \smamath{43\%} (\smamath{53\%}) as intermediate, and \smamath{34\%} (\smamath{30\%}) as advanced or experts. Additional demographic details are provided in Appendix~\ref{app:user-survey}.

Compared to the OECD population, our sample has fewer women and individuals over 55. 
However, we appreciate that our sample consists of highly educated individuals, of which only a minority consider themselves as experts in AI or security. Therefore, even though we cannot claim representativeness of the world's population, our sample is likely to provide valuable insights for the goal of our SoK. To our knowledge, this is the first survey of this kind on OAI, hence its findings are useful (also) for future studies.

\subsection{Results}
\label{ssec:userstudy_results}

\noindent
We systematically analyze our results quantitatively (§\ref{sssec:user_qualitative}) and qualitatively (§\ref{sssec:user_quantitative}) before drawing our conclusions.

\subsubsection{\textbf{Quantitative Analysis}}
\label{sssec:user_quantitative}
\begin{figure}[!htbp]
\vspace{-4mm}
    \centering
    \centerline{
    \includegraphics[width=0.45\textwidth]{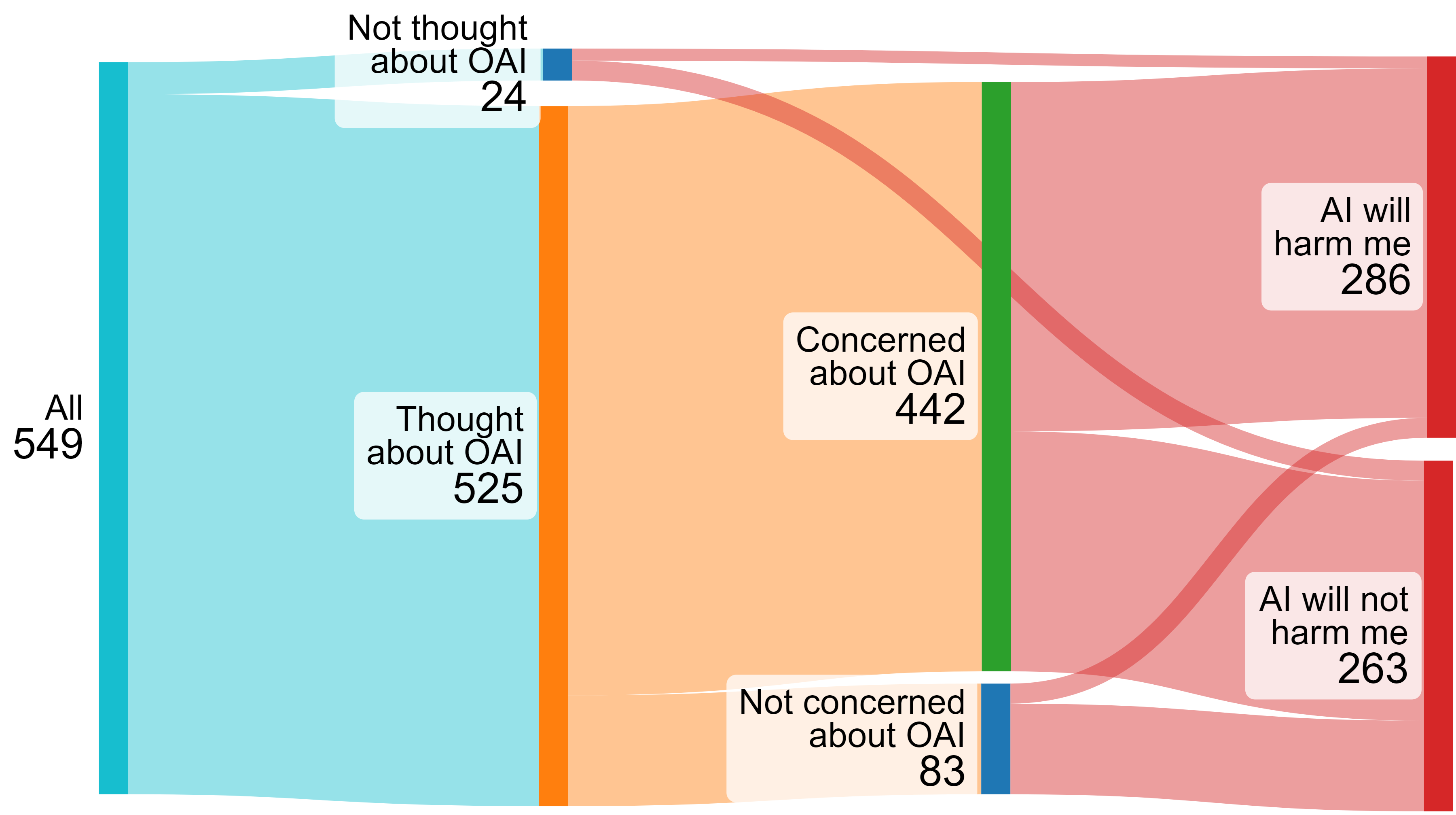}
}
\vspace{-2mm}
    \caption{\textbf{Quantitative results (laypeople).}
    \textmd{\footnotesize Sankey chart of the closed questions.}} 
    \label{fig:survey-responses-1}
    \vspace{-2mm}
\end{figure}

We report the results of our binary questions in Figure~\ref{fig:survey-responses-1}. Let us analyze it at a high-level.

\begin{itemize}[leftmargin=0.5cm]
    \item \textit{``Have you thought about OAI?''}
    The majority (525, \smamath{96\%}) have already considered that AI could be used for malicious purposes. However, the remaining 24 (\smamath{4\%}) have never considered AI's offensive potential. Among these, 14 people work in an IT-related field, and 13 of these have at least intermediate knowledge in cybersecurity or AI. 
    \item \textit{``Are you concerned about OAI?''}
    Among the 525 participants that have considered the offensive potential of AI, 442 are concerned about it (and 83 are not concerned).
    \item \textit{``Do you think that AI will harm you?''}
    A slight majority (\smamath{52\%}) believe that AI may personally harm them. Intriguingly, nearly \smamath{40\%} of the 442 participants who are concerned about OAI do not believe that AI will harm them; whereas \smamath{18\%} of the 83 participants that are not concerned about OAI believe that AI will harm them.
\end{itemize}

\noindent
We have carried out correlation analyses (in Appendix~\ref{app:user-survey}) to determine whether an individual's background affects their views on OAI. We found no correlation between having a job in IT and having concerns about OAI or believing that AI may inflict personal harm. We found a weak positive correlation between the level of expertise and concerns about OAI: in particular, knowledge of AI has a stronger impact than knowledge in cybersecurity (Pearson's $\rho$=\smamath{0.27} vs. \smamath{0.11}). Yet, we found no correlation between the expertise in either cybersecurity or AI and the belief that AI will inflict harm.

\subsubsection{\textbf{Qualitative Analysis}}
\label{sssec:user_qualitative}
We review the answers to the open questions wherein we ask participants to provide reasons why they are (or are not) concerned about OAI. Here, we present the results of our coding-based analyses (discussed in Appendix~\ref{app:user-survey-qualitative}). Every respondent could mention more than one source of concern (or lack thereof).

\begin{itemize}[leftmargin=0.5cm]
    \item \textit{``What are you most concerned about OAI?'' (442 respondents)} Most respondents (141) are concerned about the broad category of ``cyberattacks'' or provide unspecific generic concerns (24). Among those that do provide clear scenarios, 98 mention ``spread of misinformation,'' 61 ``deepfakes,''  21 about ``military applications of AI'' and 19 about ``privacy attacks.'' Some even mentioned concerns that are orthogonal to our vision of OAI: for instance, many are concerned about improper usage of AI by its developers (i.e., 20 mentioned ``losing control of AI,'' 12 ``data misuse,'' 10 ``unintended errors,'' 9 ``regulatory deficiencies,'' 8 ``negligent AI development'') including ``ethical concerns'' (19); whereas others are concerned about ``job displacement,'' (16) ``reduction in human learning'' (15), or ``AI surpassing human performance''  (14).
    \item \textit{``Why are you not concerned about OAI?'' (83 respondents)} Most respondents (26) simply do not express any clear reason for their lack of concerns. Many (21) believe that current protection mechanisms are enough to handle the threat of OAI. Some (9) are more concerned about human-centered issues (e.g., ethics) of AI. A minority believes that AI is not yet mature (5) or that the benefits of AI outweigh potential issues (8). Some even suggested proactivity, stating that ``Just being concerned is not helpful'' or ``we should be prepared and not just concerned.'' These results echo those of a 2023 survey among 199 participants, wherein 35\% of participants stated that generative AI will not provide an advantage to attackers or defenders~\cite{wef2023survey}.
\end{itemize}
We further scrutinize the responses of those (141) participants that mentioned concerns about ``cyberattacks.'' While 58 do not provide additional details, 78 mention use-cases that we were able to map to MITRE ATT\&CK: specifically, 26 consider ``reconnaissance,'' 20 ``resource development,'' 18 ``initial access,'' 9 ``defense evasion,'' 2 ``credential access'' and 1 ``lateral movement.'' Finally, 5 are concerned about autonomous attacks or adversarial ML attacks (which are unrelated to OAI).

\vspace{-2mm}

\begin{cooltextbox}
    \textbf{\textsc{Lessons Learned}:} Over 80\% of our respondents are concerned about OAI. However, our sample may have misconceptions about ``offensive AI,'' since some concerns relate to problems that are orthogonal to how evildoers may use AI to cause  harm. This suggests that non-experts may be oblivious of the offensive potential of AI, thereby underscoring the necessity of proper awareness campaigns.
\end{cooltextbox}

\section{Expert Opinion: What is the future of OAI?}
\label{sec:experts}
\noindent
In this section, we report the systematization of the expert opinions (refer to §\ref{ssec:2.4-expert}). We begin by presenting their responses to the survey (§\ref{ssec:expert-survey}); then, we summarize their input after they reviewed this paper (§\ref{ssec:expert-statements}); finally, we compare their opinions before and after having read our paper (§\ref{ssec:expert-comparison}).

\subsection{Expert Survey (opinions before reviewing this paper)}
\label{ssec:expert-survey}

\noindent
Recall that we first inquired the 12 experts to participate in a (slightly modified) version of our survey. Hence, their responses reveal their \textit{unbiased opinion} on OAI. In terms of demographics, 5 (42\%) identify as an expert in AI, 4 (33\%) as advanced, and 3 (25\%) as intermediate; whereas 9 (75\%) deem to be an expert in cybersecurity, and 3 (25\%) identify as having advanced knowledge in cybersecurity. This is in line with our expected target of ``experts'' in these fields.

\subsubsection{\textbf{Quantitative Analysis}}
\label{sssec:experts_quantitative}
We first report the results to the binary questions (as we did in §\ref{sssec:user_quantitative}). All 12 experts have already considered that AI could be used for malicious purposes, and also all experts are concerned about the offensive potential of AI. For the last question, 9 (\smamath{75\%}) experts think that AI could harm them personally, while 3 (\smamath{25\%}) think otherwise.

\subsubsection{\textbf{Qualitative Analysis}}
\label{sssec:experts_qualitative}
Next, we qualitatively analyze the answers to the open questions---starting from the one that was also included in the survey with the general population.

\begin{itemize}[leftmargin=0.5cm]
    \item \textit{``What are you most concerned about OAI?''} The most prevalent concerns are ``cyberattacks'' (4), and the ``speed, automation and ease of use'' (4). Some experts are concerned about ``privacy attacks'' (2), ``deepfakes'' (1), and the ``spread of misinformation'' (1). 
    
    \item \textit{``Can you think of (or do you know of) some ways in which AI can be used offensively?'' [expert-only]} We mapped the answers to the OAI use cases of our checklist (§\ref{ssec:checklist}). Most experts (6) highlighted multiple use cases. For MITRE ATT\&CK: 7 experts mentioned ``initial access,'' 3 ``resource development,'' and 1 ``reconnaissance.'' 4 experts highlighted attacks against society, and 2 experts autonomous attacks. None of the experts mentioned privacy attacks.
    
    \item \textit{``What are some means that can be used to counter AI-powered cyberattacks?'' [expert-only]} Most experts (8) stated multiple countermeasures. Intriguingly, 4 experts recommend to use AI-powered countermeasures, whereas 6 think that AI is not necessarily required in such defenses. Additional solutions include: user education and awareness (4); mechanisms to recognize ``AI behavior'' (4), such as CAPTCHAs; data anonymization (1); as well as banning generative AI for public figures (1). Finally, 2 experts think that OAI threats can be tackled through the same mechanisms as traditional cyber attacks. 
    
    \item \textit{``Which stakeholders should be responsible for implementing/realizing/advertising such countermeasures?'' [expert-only]} To analyze these answers, we grouped the mentioned stakeholders in: system providers (industry, technical service providers, vendors); sovereign entities (government, institutions, regulations); and generic humans (individuals, users). Among our experts, 6 consider system providers as the primary responsible parties; whereas 5 indicate system providers \textit{and} sovereign entities; one considers everyone as equally responsible (i.e., system providers, sovereign entities, and the general human population).

\end{itemize}
We will draw some comparisons between these results and those of the survey among the general population in §\ref{ssec:findings}.

\subsection{Expert Statements (after reading our paper)}
\label{ssec:expert-statements}

\noindent
We now focus on the statements that the 12 experts contributed after reviewing our draft paper. 
We first objectively analyze these texts via natural language processing (NLP) techniques~(§\ref{sssec:expert-overview}). Then, we coalesce the experts' opinions into 7 open problems and 3 fundamental concerns about OAI~(§\ref{sssec:expert-open}), serving as a basis for future work on OAI~(\smabb{C}3).

\subsubsection{\textbf{Preliminary Analysis}}
\label{sssec:expert-overview}
Overall, the statements written by our 12 experts span across \smamath{\approx}5k words and \smamath{\approx}35k characters. To provide an objective foundation for a systematic assessment, we carry out a preliminary analysis for which we rely on well-known NLP techniques for text mining. Specifically, we first extract the 20 most common bi-/tri-grams across the entire statements; then, we perform a more fine-grained analysis and apply KeyBERT~\cite{grootendorst2020keybert} to extract the 5 most relevant keywords for each expert statement. Finally, we apply topic modeling via BERTopic~\cite{grootendorst2022bertopic} to identify the most relevant topics across the entire statements. We provide in the Appendix~\ref{sapp:expert-overview} a more low-level description of these procedures (including how they work and why they are relevant), as well as the detailed results (the full source code is available in our repository~\cite{repository}). At a high-level, our analyses with BERTopic revealed that certain topics were more prevalent than others. For instance, words such as ``bias,'' ``cognitive,'' and ``exploit,'' had a lot of weight, suggesting the topic of ``cognitive bias manipulation;'' the same can be said for ``picture,'' ``video,'' and ``generated,'' suggesting the theme of ``AI generated content.'' At an individual level, the results of KeyBERT showed that many experts think about ``defenses'' (or ``detection'' or ``countermeasures'' or ``protection''); intriguingly, the most relevant keyword for one expert, ``privacy,'' had the second-most highest weight among the most relevant keywords for all other experts.

\subsubsection{\textbf{Open Problems and Concerns of OAI} (\smabb{C}\smamath{3})}
\label{sssec:expert-open}
We use our preliminary analysis as a scaffold and further inspect all expert statements to derive open problems and concerns on OAI. Such a summarization was carried out by four authors who independently formulated their conclusions after reading the statements and then discussed the resulting viewpoints to reach a consensus. While the experts were asked to write statements ``describing three open problems of OAI,'' our analysis showed that some of the statements revealed \textit{specific research problems}, whereas others pertained to more \textit{generic concerns} connected with the OAI threat. Below, we present these two categories of findings separately.

We have identified the following \textbf{seven open problems}~(\smacal{P}):

\begin{itemize}[leftmargin=0.8cm,noitemsep]
    \item[\smacal{P}1:] \textit{Differentiating AI from Reality.} Content generated by AI has reached quality comparable with reality. This development poses substantial problems for \text{society}. Falsification of content by means of AI may have grave consequences for the democratic order, the rule of law, education and numerous other faces of our society. It is of vital importance for the research community to understand the potential effects of such fake audio-visual content and develop corresponding countermeasures. 

    \item[\smacal{P}2:] \textit{Privacy threats of AI.} AI facilitates the extraction of private information about \text{humans}, e.g., via attribute inference attacks, linking of separate data items, cross-device tracking, fingerprinting of encrypted traffic. Substantial advances in privacy-related research are needed to counter novel privacy threats enabled by offensive AI. 

    \item[\smacal{P}3:] \textit{Management of offensive AI risks.} The operational implications of OAI in the context of \text{systems} cannot be resolved by technical means alone. This puts the problem into a management perspective. To enable decision making, quantification of various risks is required. Attention should be given to the skill level needed for (ab)using certain AI tools, and to the benefits of deployment of AI techniques w.r.t. conventional attacks.

    \item[\smacal{P}4:] \textit{Implications of offensive AI for social engineering.}
    Besides a general impact of offensive AI on humans in the societal context, such impact has specific implications for security systems. The risk of humans being the weakest link in a security chain is widely recognized. Offensive AI brings scalability of social engineering attacks to a new level. To counter this threat, both \text{human}-centric research (e.g., new methods for awareness training), and \text{system}-centric research (e.g., minimizing the likelihood and impact of human error), needs to be pursued.

    \item[\smacal{P}5:] \textit{Expansion of AI governance.} Offensive AI may damage \text{humans} also by impacting specific institutions they interact with. Alongside state regulation (potentially complemented at the international level), new governance mechanisms should be explored to induce/enable institutions to prevent malicious abuse of their AI infrastructures.

    \item[\smacal{P}6:] \textit{Understanding the pros and cons of offensive AI.} There are a lot of ``success stories'' about AI-powered attacks. However, not much is known about cases in which such AI-powered attacks resulted in failure---which could be used to shed light on the limitations of AI as an offensive tool. Future work should discuss such negative results as well, which could also entail attacks that are successful, but which are unreasonably expensive to stage in reality.

    \item[\smacal{P}$7$:] \textit{Cognitive bias and its implications.}
    AI can cause cognitive bias in human decisions. An exogenous positive bias induced by offensive AI can strengthen people's existing beliefs and deter critical thinking. AI can also affect behavioral economics, thus eroding economic theories based on the assumption of rational decision making. Understanding of the cognitive impact of offensive AI, as well as of potential ``collaboration'' between humans and machines to counter such threats, is strongly desirable.

\end{itemize}
We stress that: {\small \textit{(i)}}~the above mentioned problems are listed in no specific order of importance; and {\small \textit{(ii)}}~we do not use ``majority'' to identify any given problem---even if multiple experts share similar views, every opinion has the same value.

Further, we have identified \textbf{three fundamental concerns on OAI} expressed by our experts. Such concerns do not necessarily constitute specific research areas but rather affect the entire realm of research in AI and information security. 
\textbf{(I)}~First, \textit{AI is a double-edged sword}. AI was conceived to make human life easier. Unfortunately, it also makes attackers' life easier. 
Hence the potential dual use of AI must be addressed at various levels, from methodical research to legal regulation, management, and compliance. 
\textbf{(II)}~Second, the implications of OAI being able to target \textit{humans who detain different roles and responsibilities}. For instance, tricking an employee has different consequences from tricking a decision maker. Hence, a proper evaluation of OAI threats requires to model all such scenarios and foresee the corresponding effects---which requires assessments of the risks from various viewpoints.
\textbf{(III)}~Third, \textit{countermeasures are needed but challenging to deliver}. On the one hand, there is still little that is known about AI, making it hard to find effective solutions to OAI (and AI being a ``double edged sword'' further aggravates this challenge); on the other hand, from a research viewpoint, there is a higher incentive in showcasing ``novel attacks'' rather than on studying, evaluating, and implementing appropriate defenses. Hence, future work should put countermeasures in higher regard---potentially by focusing on techniques that, despite not addressing the OAI threat universally, may just increase the cost to sustain an OAI-based attack.

\vspace{1mm}
{\setstretch{0.7}
\textbox{{\small \textbf{Remark:} 
We acknowledge that the problems/concerns mentioned above may appear ``well-known'' in the security community. However, to the best of our knowledge, our SoK is the first scientific work wherein the opinions of 12 experts on OAI have been systematically coalesced into a set of avenues for future research.}}}
\vspace{-1mm}

\subsection{Comparison of Opinions: Pre and Post Reading this SoK}
\label{ssec:expert-comparison}

\noindent
While summarizing our experts' opinions we have observed some changes in their views which can potentially be attributed to the findings of our paper. 
Specifically, two topics were stressed more often in the experts' statements (provided after reading our paper) than in the initial responses to the survey.

The first topic is \textit{privacy.} In the initial responses, two experts essentially described OAI use cases that could be related to privacy, but never explicitly associated them with the term ``privacy''. In written statements, the term ``privacy'' is explicitly mentioned eight times (by four experts). This may be the result of becoming aware of privacy as an important use-case of OAI, which is revealed in our Fig.~\ref{fig:tech-other}.

The second topic is \textit{cost.} During our survey, only one expert mentioned ``cost'' (twice), and the term ``econom-'' was never mentioned. In contrast, in the statements, three experts mentioned ``cost''  (the word ``cost'' occurs 6 times in total), whereas ``econom-'' was mentioned by four experts (and it occurs 8 times). 
This fact clearly bears some correlation with the cost/benefit analysis being an essential systematization criterion in our checklist, suggesting that our findings led to an increased awareness on the economical factor of OAI.

\section{Discussion}
\label{sec:discussion}

\noindent
We now reflect on the findings (§\ref{ssec:findings}) and limitations~(§\ref{ssec:limitations}) of our paper and compare it with related meta-research (§\ref{ssec:related}). Our intention is to demonstrate the importance of analyzing various sources of knowledge.

\subsection{Findings}
\label{ssec:findings}

\noindent
We summarize our findings by making explicit reference to the three-fold contributions of our paper (\smabb{C}\smamath{1}, \smabb{C}\smamath{2}, \smabb{C}\smamath{3}).

Our SoK provides a snapshot of the OAI landscape (\smabb{C}1). Such a snapshot, however, has been made possible only thanks to the collective ``contributions'' of four sources of knowledge---each of which covers the potential \textit{blind spots} of the others. For instance, our literature review (§\ref{sec:3-academic}) showed that previous taxonomization of offensive AI use-cases (e.g., reliant on MITRE) are insufficient to cover the landscape of the OAI threat. At the same time, the ``limited practicality'' exhibited even by technical papers (which mostly attack ``toy'' systems) may suggest that OAI does not represent a tangible threat---but, perhaps unfortunately, the analysis of the InfoSec briefings (§\ref{sec:4-infosec}) revealed that OAI can be practically exploited in the real world. In contrast, no InfoSec briefing considered OAI in warfare, but we found many papers covering such use cases---which also seem to worry non-experts~(§\ref{sec:5-userSurvey}), despite not having been mentioned by any of our experts (§\ref{sec:experts}). Finally, despite our extensive analyses, we acknowledge that our review of prior work may have missed some OAI use cases: one such example are website fingerprinting~\cite{rimmer2018automated} attacks, which we overlooked in Table~\ref{tab:technical}, but which were mentioned by one expert in their statements. Altogether, these findings show that there is a need of a perpetual and collective effort to monitor the threat of OAI, since we expect more OAI use cases (existing or new) to be identified in the future (see, e.g.,~\cite{dragan2024frontier}).

Our SoK provides a foundation for a long-term classification of OAI works (\smabb{C}\smamath{2}). It is obvious that the field of potential offensive use-cases for AI is vast. Our simple checklist (§\ref{ssec:checklist}) encapsulates clear criteria that can be used to systematically analyze works on OAI, thereby aligning the corresponding findings to those provided in this paper. Such a checklist represents a methodological stepping stone for keeping track of future discoveries (technical or theoretical) in the OAI context---which serves to identify potential treatments to the threat of OAI. To facilitate the usage of our checklist by downstream research, we have integrated it in an online website~\cite{tool} in which we {\small \textit{(i)}}~\textbf{maintain a curated and vetted} archive of OAI-related works, and {\small \textit{(ii)}}~allow interested individuals to add more works to the archive by \textbf{submitting ``new entries'' after applying our checklist}. We have recorded a 60s \href{https://github.com/hihey54/sok_oai/blob/main/demo.mp4}{video} (in our repository~\cite{repository}) showcasing how to use our tool to add new OAI-related works (e.g., the previously mentioned~\cite{rimmer2018automated}).

Our SoK provides intriguing avenues for future work (\smabb{C}\smamath{3}).
The simple characterization of essential features of 133 OAI works and our survey with 549 laypeople enabled to distill many ``lessons learned'' for future work (scattered through §\ref{sec:3-academic}--§\ref{sec:5-userSurvey}), which have been complemented by the shortlist of problems and concerns derived by analyzing the statements of 12 experts (§\ref{sec:experts}). The latter include: the need for research on countermeasures, the lack of ethical statements and of general understanding of issues related to the dual use of AI, limited focus on societal impact of OAI, the necessity of cost/benefit analysis as part of risk management, and the importance of the ``human dimension'' in OAI. In some cases, the issues envisioned by the experts (in their statements and survey) align with those that emerged in our user study with non-experts (e.g., deepfakes, cyberattacks, manipulation); however, it is interesting to see that some ``educated'' laypeople (most of our respondents have degrees, see Appendix~\ref{app:user-survey}) may have different thoughts (e.g., the above mentioned ``warfare''); moreover, some non-experts appear to be more worried of how AI may negatively impact our lives in the general sense, rather than due to an explicit abuse by attackers---which is a valid concern, despite falling outside our scope. Hence, we argue that future work should tackle OAI-related themes by accounting for different perspectives---all being equally important.

\subsection{Limitations}
\label{ssec:limitations}
\noindent
We identify three main limitations that may affect the validity of our findings and discuss them below. 

The first limitation pertains to our \textit{search for the literature review and Infosec venues} (§\ref{sec:3-academic} and §\ref{sec:4-infosec}). For the latter, we only included BlackHat and DefCon, but there are more InfoSec venues which do accept briefings on AI (e.g., the RSA conference~\cite{siliconangle2024rsa}). For the former, our search queries mostly revolved around ``offensive AI'' and ``offensive security'' (see §\ref{ssec:2.1-literature}), but there may have been other terms that could have been used to identify works that fall into our definition of OAI; moreover, querying repositories also has limitations on its own~\cite{bramer2016variation}. 
Therefore, the works considered in our SoK (95 research papers and 38 InfoSec briefings) may under-represent the overall number of works on OAI. 
Our goal, however, was not to attain complete coverage of existing works (which is clearly unfeasible---as we showed in this SoK). Instead, we provide our ``snapshot'' by analyzing a subset of works drawn from a systematic search of prior work (academic and industrial) and complemented it with the systematic analysis of knowledge distilled from a broad set of sources. 

The second limitation entails the \textit{bias in the population of our user studies} (§\ref{sec:5-userSurvey} and §\ref{sec:experts}). For the non-expert survey, our sample clearly cannot cover all laypeople in the world. For the expert opinions, we reached out to 12 individuals with different expertise in fields related to OAI. Again, the lack of complete coverage of such opinions should not pose a major threat to the validity of our claims, since we clearly stated that every opinion is equally valid and reported even the concerns shared by few individuals. Nonetheless, carrying out user studies is notoriously difficult (e.g., some top-tier security papers collect the opinion of 10--20 individuals~\cite{alahmadi202299,apruzzese2023sok,mink2023everybody}).

The third limitation is the potential \textit{subjectivity of qualitative analyses}~\cite{madill2000objectivity}. Indeed, reviewing each work (paper or briefing), and reviewing the responses of our surveys as well as the expert statements---all these methodical instruments employed in our SoK rely on analyses carried out by its authors. To mitigate this potential shortcoming, we discussed our findings to clarify doubts and to reach a consensus (§\ref{ssec:2.1-literature} and §\ref{ssec:2.3-userstudy}); and we also relied on well-known practices and technical algorithms~(§\ref{sssec:expert-overview}). Moreover, our extensive appendix provides additional information for reproducing most of our results.

\subsection{Comparison with Related Work}
\label{ssec:related}

\noindent
We found no prior work that systematically analyzed the theme of offensive AI to the same extent as done in our SoK. 

\subsubsection{\textbf{Prior Work (Surveys/Summarizations) on OAI}}
\label{sssec:prior}
First, we found no SoK paper that specifically addressed offensive AI. By turning the attention at ``literature reviews'' (or similar papers), we found that most such papers considered a \textit{single target} (e.g., only ``organizations''~\cite{mirsky2021threat}; or only ``humans''~\cite{arora2021review}, or only ``systems''~\cite{aiyanyo2020systematic}). A notable exception is the recent paper by Malatji and Tolah~\cite{malatji2024artificial}, which accounts for offensive AI from a socio-technical perspective. Yet, the analysis in~\cite{malatji2024artificial} (as well as those in~\cite{aiyanyo2020systematic,arora2021review}) is \textit{rooted only on the findings of academic literature}. Remarkably, the work by Mirsky et al.~\cite{mirsky2021threat} also accounts for the perspective of practitioners, but it does not account for industrial venues, nor investigate the opinion of laypeople---which are among the primary targets of OAI and represent a valuable source of knowledge to pinpoint some of the most perceived concerns. Such a narrower scope may lead to some oversights. 

\subsubsection{\textbf{Novel Findings}}
\label{sssec:novel}
Our SoK underscored OAI use-cases that have not been emphasized in prior summaries. For instance, Mirsky et al. focus on MITRE ATT\&CK, meaning that anything outside such matrix was outside their scope, e.g., there is no mention of ``cyber warfare'' or ``privacy'' in the main body of their paper~\cite{mirsky2021threat}. In contrast, these terms have been mentioned in, e.g.,~\cite{malatji2024artificial, aiyanyo2020systematic}; however, these works overlooked the potential of ``attribute inference attacks'' or ``fingerprinting,'' both of which can be perpetrated via AI. However, we reiterate that the reason why we captured these additional use cases is due to our SoK having a broader scope.
Hence, our SoK extends all such prior work by providing a systematization of \textit{all potential targets} of OAI by accounting for \textit{diverse knowledge sources} (see Fig.~\ref{fig:sok}). Moreover, we also claim that our approach is unprecedented in extant SoK papers. 

\subsubsection{\textbf{Advancing the state of the art in SoK}}
\label{sssec:sok}
We have studied the 270 SoK papers listed in Shujun Li's online bibliography of SoK papers~\cite{li2023sokpapers}, from 2010 to Jan. 2024. We could not find any paper that considers expert opinion as one of the knowledge sources and presents verbatim such expert opinions.
The majority of prior SoK papers draw their conclusions from the scientific literature. Some also carry out user studies (either with experts~\cite{ladisa2023sok}, or among the general public~\cite{stephenson2022sok}). Yet, we found no SoK that considered both of these dimensions---and, specifically, no SoK paper that reached out to experts in an attempt to \emph{draw avenues for future research}. Hence, our contribution can be inspiring also for future SoK papers. 
Importantly, some SoKs carry out technical experiments (e.g., by reproducing prior work, such as~\cite{apruzzese2022sok}): these SoKs are orthogonal to ours. However, future ``technical SoK'' can also benefit from our intuitions, e.g., by systematizing the techniques proposed in InfoSec venues.
\section{Conclusions}
\label{sec:conclusions}

\noindent
We consider this paper as a first step in laying down a scientific groundwork for investigating various facets of offensive AI. 

In short, we found that the offensive capabilities of AI are very heterogeneous and can adversely affect systems, humans and the society as a whole. 
Due to this heterogeneity, offensive AI use cases cannot be classified into a single framework, such as MITRE ATT\&CK, but require a broader systematization which we provide with the help of our OAI assessment checklist (§\ref{ssec:checklist}) and in our online tool~\cite{tool}. 

We hope that the insights obtained in this SoK paper enable security and privacy researchers to better appreciate the societal impact of problems related to offensive AI.
Our findings also underscore the necessity of interdisciplinary collaboration with the areas of cognitive science, psychology, economics, political science, law, ethics, and perhaps many other, to fully comprehend and mitigate the offensive potential of AI.

\textbf{\textsc{Acknowledgement.}}
The authors would like to thank the anonymous reviewers for their feedback. We also extend our gratitude to all participants of our non-expert user survey.
Part of this research has been funded by the Hilti Foundation. Moreover, we acknowledge the financial support from the projects SERICS (PE00000014) and FAIR (PE00000013) under the MUR National Recovery and Resilience Plan funded by the European Union - NextGenerationEU. This research has also been partially funded by the Research Fund KU Leuven, and by the Cybersecurity Research Program Flanders.

\bibliographystyle{IEEEtran}

{\footnotesize

}

\appendices

\section{Checklist for analyzing OAI-related works}
\label{app:checklist}
\noindent
We explain at a low-level the criteria embedded in our proposed offensive AI checklist. %
We use our checklist to analyze each work considered in this SoK, but our checklist can also be used by future research to carry out analyses that align with ours and expand the findings of our systematization. 

\subsection{What is the OAI use-case?}

\noindent
First, we map each paper to one of the MITRE ATT\&CK Tactics for Enterprise, Mobile or ICS~\cite{mitre-tactics}.\footnote{We did not use \href{https://atlas.mitre.org/}{MITRE ATLAS}, since ATLAS focuses on vulnerabilities in AI-enabled systems, which are orthogonal to our focus (§\ref{sec:2-preliminaries}).} We associate each paper with the primary MITRE Tactic covered: For instance, if the goal of a side-channel attack is password stealing~\cite{anand2018keyboard}, we map it to Credential Access. Some OAI use cases cannot be mapped to the MITRE ATT\&CK tactics, as they entail techniques beyond the matrices' scope. These OAI use cases either: encompass multiple tactics---which we classify under the category of autonomous agents; or focus more broadly on attacks against ``society'' or ``privacy,'' as well as applications in (cyber) ``warfare.''

Second, we review the original focus of the paper, which we do by asking ourselves the following three questions: 
\begin{itemize}[leftmargin=*]
    \item Does the paper focus on offensive security? If so, we assign it to the category ``defense'' (\shield).
    \item Does the paper explore whether AI could be an effective hacking assistant? If so, we assign it to the category ``assisted-hacking'' (\assist).
    \item Does the paper showcase a new attack? If so, we assign it to the category ``attack'' (\off).
\end{itemize}  
Third, we further explore two classes of papers.

If a paper proposed a novel attack (\off), we examine whether the authors proposed any countermeasures (``Def.''), resulting in a binary conclusion depending on whether they \emph{explicitly} state such countermeasures. For instance, Anderson et al.~\cite{anderson2016deepdga} designed a deep learning-based DGA to intentionally bypass the DGA detector.  Yet, they also explore approaches to enhance the detector's security.

If a paper is designed for offensive security (\shield), we check whether the authors \emph{explicitly} stated that attackers could abuse the proposed technique. We report the offensive potential of a technique in the column ``Pot. Abuse.'' One example of such an explicit statement is provided by Zennaro et al.~\cite{zennaro2023modelling} within the ethical considerations, acknowledging the potential for malicious use and condemning any such implementation to attack or harm.\footnote{The authors additionally refer to the Autonomous Weapons Open Letter: AI \& Robotics Researchers of \href{https://futureoflife.org/open-letter/open-letter-autonomous-weapons-ai-robotics/}{Future of Life}.}

\subsection{``What is the target of OAI?''}

\noindent
First, we evaluate whether the objective of the OAI use case is to impact a human (\woman), a system (\system), or both (\both). 
\begin{itemize}[leftmargin=*]
    \item We categorize an attack as targeting a human (\woman) if {\small \textit{(a)}}~it exploits a ``vulnerability'' in human behavior, such as susceptibility to misinformation (e.g., polarizing summaries~\cite{sharevski2021regulation}) or {\small \textit{(b)}}~it infringes on privacy (e.g., by inferring sensitive data from non-sensitive data~\cite{karanatsiou2022my}).
    \item If the attack is directed at a system (\system), it implies that its success does not strictly depend on human involvement (e.g., bypassing an Intrusion Detection System~\cite{lin2022idsgan}, or CAPTCHA cracking~\cite{yu2019low}. 
    \item In instances where an attack is directed at both (\both) successful execution requires deceiving both human and system. For example, in a phishing scenario (e.g.,~\cite{khan2021offensive}), success relies on the system failing to detect the phishing email and the human falling for the deception. 
\end{itemize}
Furthermore, for attacks directed against a system (either \system{} or \both{}), we review whether the attack was directed against a real system (\real) or a toy system (\toy). A toy system is a simplified, smaller-scale version of a real-world system used for experimentation. It can mirror specific aspects of a complex system with reduced user involvement and cost, as defined by Xu et al.~\cite{xu2015my}. For instance, Hu et al.~\cite{hu2022generating} developed their own black-box detector, serving as a ``toy system'' to assess the proposed evasion technique.

Finally, to understand the role of the social aspect within each paper, we counted the occurrences of the words ``society,'' ``social,'' ``societal,'' or ``socio.''

\subsection{``What is the \textit{cost/benefit} of OAI?''}

\noindent
First, we analyze each paper to assess \emph{potential benefits} that may motivate attackers to utilize the proposed AI technique (benefit). Second, we evaluate each paper to identify any \emph{associated costs} that could discourage attackers from adopting the proposed technique (cost). We scrutinize each paper for the below criteria (inspired by~\cite{apruzzese2023real}):
\begin{enumerate}[label={(\Roman*)}, leftmargin=0.5cm]
    \item Did the authors analyze the \emph{benefits} of the attacker leveraging the proposed AI technique? More precisely, did the authors provide any supplementary evidence/analysis/discussion of the attacker's benefits beyond simply stating that ``the proposed method works''? Four answers are possible:
        \begin{itemize}
            \item No, no mention (\nomention).
            
            \item Yes, qualitative (\yesqual): Just a discussion. For instance, Iqbal et al.~\cite{iqbal2023chatgpt}, in addition to showcasing technical use cases, discuss the benefits of  ChatGPT for cybercriminals to refine their skills and facilitate more effective attacks.
            
            \item Yes, quantitative (\yesquant) (e.g., based on accuracy/precision). For instance, Ozturk et al.~\cite{ozturk2023new} evaluate static code analyzers and ChatGPT to detect OWASP vulnerabilities, comparing them based on success rate and accuracy.
            
            \item Yes, clear mention of monetary benefit or time/resources saved according to some metrics that go beyond sheer accuracy/precision (\yescost). E.g., Lee et al.~\cite{lee2022link} compare their technique with non-AI techniques to detect cross-site scripting vulnerabilities. They fixed the time budget to 3 hours, and then compare the number of detected vulnerabilities and the number of attack requests required to identify the vulnerabilities (required resources).
            
        \end{itemize}
        
    \item Did the authors analyze the \emph{costs} for designing/building/implementing the proposed technique? To examine this, we reviewed the papers individually and searched for keywords such as ``cost,'' ``money,'' ``time,'' or ``resources.'' Four answers are possible: 
        \begin{itemize}
            \item No, no mention (\nomention). E.g., Goldbeck et al.~\cite{golbeck2011predicting} leverage publicly available information to predict a user's personality, but do not mention the time/cost for data collection, algorithm development, runtime or similar.
            
            \item Yes, qualitative (\yesqual): Just a discussion. As an example, Yu et al.~\cite{yu2019low} argue that their proposed method for CAPTCHA cracking is low-cost since: it uses an open-source library; it can be implemented on a personal computer; unlimited training is available.

            \item Yes, quantitative (\yesquant). This includes a clear numeric estimation of the costs based on some metrics, e.g., required resources or time (beyond sheer accuracy/precision). Liu et al.~\cite{liu2020deepsqli}, for instance, compare wall clock time, and Pa Pa et al.~\cite{pa2023attacker} delineate that the development of functional malware with ChatGPT or AutoGPT, including debugging, takes around 90 minutes. 
            
            \item Yes, clear mention of the required \$\$\ to launch the attack~(\yescost). We did not identify any paper in this category.
            
        \end{itemize}
        
    \item For \emph{Non-AI-baseline comparison}, we ask ourselves whether, e.g., the authors consider if the same objective could be achieved without AI. Three cases are possible:
        \begin{itemize}
            \item No, no mention (\nomention) of any alternative.
            
            \item Yes, qualitative (\yesqual). For instance,~\cite{chowdhary2020autonomous} provides a qualitative comparison to manual penetration tests.
            
            \item Yes, quantitative (\yesquant). For instance, Bahnsen et al.~\cite{bahnsen2018deepphish} quantitatively compare the effectiveness of DeepPhish in creating phishing URLs to non-AI techniques. 
        \end{itemize}
\end{enumerate}
Additionally, we check whether each paper shares the source code or prompts (\prompt) if an LLM was used. We include the link to the repository, if available.

\subsection{Academic Literature (Techn. Papers): Additional Review}
\label{sapp:technical-algorihm}
\noindent
For the technical papers from the academic literature (discussed in §\ref{ssec:3.2-nontechnical}) we additionally analyzed the \emph{technical requirements} (e.g., algorithm, or data) to set up the attack. As identified by~\cite{apruzzese2023real}, most papers on adversarial machine learning do not consider the human effort required to technically implement the attack. Inspired by this observation, we extend the review of the technical papers by performing additional analyses, the results of which are presented in Table~\ref{tab:technical-algorithm}. Let us explain how we derived this table.
\begin{enumerate}[leftmargin=*]
    \item First, we scrutinize whether the applied algorithm/technique is publicly available (e.g., ChatGPT) and/or can be easily re-used (such as in~\cite{khan2021offensive}) If so, we mark the column with a ``yes.'' Otherwise, if we do not mark the column, this means that the attacker has to develop the algorithm from scratch, such as in~\cite{bahnsen2018deepphish}.\footnote{Khan et al.~\cite{khan2021offensive} re-use a pre-trained large language model to generate spear-phishing emails, while Bahnsen et al.~\cite{bahnsen2018deepphish} design DeepPhish, a Long Short-Term Memory Network, that learns the structure of effective URLs and then generates new synthetic URLs to create ``better phishing attacks.``}

    \item If the algorithm needs to be developed from scratch, we review the availability of training data. We distinguish: 
    \begin{itemize}
        \item publicly available data (\publiclyavailable), e.g., Biesner et al~\cite{biesner2022combining} use publicly available data sets of leaked passwords; 
        \item data collected by the authors (\collectedbyauthors), e.g., Lee et al.~\cite{lee2021offensive} constructed six datasets of keyboard inputs; 
        \item and data collected by the authors with special access (\collectedbyauthors*), e.g., Chung et al.~\cite{chung2019availability} collected the operational data of building automation systems protecting a compute infrastructure---but they could only collect the data because they had \emph{access} to such a system.
    \end{itemize}
    Moreover, we examine whether shallow learning or deep learning techniques are used (also done in~\cite{apruzzese2023real}). This allows a basic distinction between the required amount of training data and resources, since (deep) neural networks typically require more computational effort to be set up.
\end{enumerate}
Finally, we review whether the authors publicly released their code and add the link to the repository, if available.

\begin{table}[!htbp]
    \centering
    \caption{\textbf{Literature Review---Technical OAI Papers (Algorithm).}
    \textmd{\footnotesize 
    We report 79 technical OAI papers and the analysis of the \emph{technical requirements} to set up the attack: We assess the availability of the applied algorithm (A-av.), ---if the algorithm needs to be developed from scratch---the availability of training data (D-av.)), and whether deep learning techniques are used (DL). * indicates that special access for the data collection is required.}}
    \label{tab:technical-algorithm}
    \vspace{-2mm}
    \resizebox{0.6\columnwidth}{!}{
        \begin{tabular}{c|c|c|c|c}
            \toprule
            \multicolumn{2}{c|}{} & \multicolumn{3}{c}{\textbf{Algorithm}} \\
            
            \cmidrule(lr){1-2} \cmidrule(lr){3-5} 

            \textbf{Paper} & \textbf{Year} & \textbf{A-av.} & \textbf{D-av.} & \textbf{DL} \\
            \midrule
            
            Antonelli~\cite{antonelli2024dirclustering} & 2024 & ~ & \collectedbyauthors & \cmark \\

            \hline
            
            AlMajali~\cite{almajali2023vulnerability} & 2023 & ~ & \collectedbyauthors & ~ \\

            Chen~\cite{chen2023gail} & 2023 & ~ & \collectedbyauthors & \cmark  \\
            
            Chowdhary~\cite{chowdhary2023generative} & 2023 &  ~ & \collectedbyauthors & \cmark \\
            
            Gallus~\cite{gallus2023generative} & 2023 & \cmark & ~ & \cmark \\
            
            Ghanem~\cite{ghanem2023hierarchical} & 2023 & ~ & \collectedbyauthors & ~  \\
            
            Happe~\cite{happe2023getting} & 2023 & \cmark & ~ & \cmark \\

            Iqbal~\cite{iqbal2023chatgpt} & 2023 & \cmark & ~ & \cmark \\
            
            Karinshak~\cite{karinshak2023working} & 2023 & \cmark & ~ & \cmark \\
            
            Ozturk~\cite{ozturk2023new} & 2023 & \cmark & ~ & \cmark \\
            
            Pa Pa~\cite{pa2023attacker} & 2023 & \cmark & ~ & \cmark \\
            
            Zennaro~\cite{zennaro2023modelling} & 2023 & ~ & \publiclyavailable & ~ \\
        
            \hline
            
            Auricchio~\cite{auricchio2022automated} & 2022 & ~ & \publiclyavailable & \cmark \\
            
            Biesner~\cite{biesner2022combining} & 2022 & ~ & \publiclyavailable & \cmark\\

            Cody~\cite{cody2022discovering} & 2022 & ~ & \collectedbyauthors & \cmark \\
            
            Confido~\cite{confido2022reinforcing} & 2022 & ~ & \publiclyavailable & \cmark \\
            
            Gangupantulu~\cite{gangupantulu2022using} & 2022 & ~ & \publiclyavailable & \cmark \\

            Hu~\cite{hu2022generating} & 2023 &  ~ & \collectedbyauthors & \cmark \\
            
            Jagamogan~\cite{jagamogan2022penetration} & 2022 & \cmark & ~ & ~  \\
            
            Karanatsiou~\cite{karanatsiou2022my} & 2022 & ~ & \collectedbyauthors & ~ \\
            
            Lee~\cite{lee2022link} & 2022 &  ~ &  \collectedbyauthors* & \cmark \\
            
            Li~\cite{li2022deep} & 2022 & ~ & \collectedbyauthors & \cmark \\
            
            Lin~\cite{lin2022idsgan} & 2022 & ~ & \publiclyavailable & \cmark \\

            Nhu~\cite{nhu2022leveraging} & 2022 &  ~ & \collectedbyauthors & \cmark \\
            
            Pagnotta~\cite{pagnotta2022passflow} & 2022 & ~ & \publiclyavailable & \cmark \\
            
            Tran~\cite{tran2022cascaded} & 2022 & ~ & \collectedbyauthors & \cmark \\
            
            Yao~\cite{yao2022intelligent} & 2022 & ~ & \collectedbyauthors & \cmark \\
            
            \hline
            
            Caturano~\cite{caturano2021discovering} & 2021 & ~ & \collectedbyauthors & ~ \\
            
            Erdődi~\cite{erdHodi2021simulating} & 2021 & ~ & \collectedbyauthors & \cmark \\
            
            Gangupantulu~\cite{gangupantulu2021crown} & 2021 & ~ & \publiclyavailable & \cmark \\
            
            Khan~\cite{khan2021offensive} & 2021 & \cmark & ~ & \cmark \\
            
            Kujanpää~\cite{kujanpaa2021automating} & 2021 & ~ & \collectedbyauthors & \cmark \\
            
            Lee~\cite{lee2021offensive} & 2021 & ~ & \collectedbyauthors & ~ \\
            
            Maeda~\cite{maeda2021automating} & 2021 & ~ & \publiclyavailable & \cmark \\ 
            
            Neal~\cite{neal2021reinforcement} & 2021 & ~ &  \collectedbyauthors* & ~ \\
            
            Sharevski~\cite{sharevski2021regulation} & 2021 & ~ & \publiclyavailable & ~ \\
            
            Standen~\cite{standen2021cyborg} & 2021 &  ~ & \publiclyavailable & \cmark \\
            
            Toemmel~\cite{toemmel2021catch} & 2021 & ~ & \publiclyavailable & \cmark \\
            
            Tran~\cite{tran2021deep} & 2021 & ~ & \publiclyavailable & \cmark \\
            
            \hline

            Al-Hababi~\cite{al2020man} & 2020 & ~ & \collectedbyauthors & ~  \\
            
            Bhattacharya~\cite{bhattacharya2020automated} & 2020 & ~ & \collectedbyauthors & ~ \\
            
            Chowdhary~\cite{chowdhary2020autonomous} & 2020 & ~ & \collectedbyauthors & \cmark \\
            
            Halimi~\cite{halimi2020efficient} & 2020 & ~ & \collectedbyauthors & ~  \\
            
            Hu~\cite{hu2020automated} & 2020 & ~ & \collectedbyauthors & \cmark  \\
            
            Lee~\cite{lee2020cybersecurity} & 2020 & ~ & \collectedbyauthors & ~  \\
            
            Lee~\cite{lee2020improved} & 2020 & ~ & \collectedbyauthors & ~  \\
            
            Liu~\cite{liu2020deepsqli} & 2020 & ~ & \publiclyavailable & \cmark  \\
            
            Pearce~\cite{pearce2020machine} & 2020  & ~ & \collectedbyauthors & ~ \\
            
            Sharevski~\cite{sharevski2020wikipediabot} & 2020  & ~ &  \collectedbyauthors* & ~ \\
            
            Shu~\cite{shu2020generative} & 2020 & ~ & \publiclyavailable & \cmark \\
            
            Song~\cite{song2020mab} & 2020 &  ~ & \collectedbyauthors & ~ \\
            
            Valea~\cite{valea2020towards} & 2020 & ~ & \publiclyavailable & ~ \\
            
            Yu~\cite{yu2020ai} & 2020 & ~ & \publiclyavailable & \cmark \\
            
            \hline
        
            Basu~\cite{basu2019generating} & 2019 & ~ &  \collectedbyauthors* & \cmark  \\
            
            Cecconello~\cite{cecconello2019skype} & 2019 & ~ & \collectedbyauthors & ~ \\
            
            Chung~\cite{chung2019availability} & 2019 & ~ &  \collectedbyauthors* & ~ \\
            
            Das~\cite{das2019x} & 2019 & ~ & \collectedbyauthors & \cmark \\
            
            Ghanem~\cite{ghanem2019reinforcement} & 2019 & ~ & \collectedbyauthors & ~ \\
            
            Tshimula~\cite{tshimula2019har} & 2019 & ~ & \collectedbyauthors* & ~ \\
            
            Yu~\cite{yu2019low} & 2019 & ~ & \publiclyavailable & \cmark \\
            
            Zhang~\cite{zhang2019using} & 2019 & ~ & \collectedbyauthors & \cmark \\
            
            \hline

            Anand~\cite{anand2018keyboard} & 2018 & ~ & \collectedbyauthors & ~  \\
            
            Bahnsen~\cite{bahnsen2018deepphish} & 2018 &  ~ & \publiclyavailable & \cmark  \\
            
            Kronjee~\cite{kronjee2018discovering} & 2018 &  ~ & \publiclyavailable & ~  \\
            
            Rigaki~\cite{rigaki2018bringing} & 2018 & ~ & \collectedbyauthors  & \cmark  \\
            
            Zhuo~\cite{zhou2018deeplink} & 2018 & ~ & \publiclyavailable  & \cmark \\
            
            \hline

            Yao~\cite{yao2017automated} & 2017 & ~ & \publiclyavailable & \cmark \\
            \hline
            
            Anderson~\cite{anderson2016deepdga} & 2016 & ~ & \publiclyavailable & \cmark \\
            
            Ceccato~\cite{ceccato2016sofia} & 2016 &  ~ & \collectedbyauthors & ~ \\
            
            Grieco~\cite{grieco2016toward} & 2016 & ~ & \publiclyavailable & ~ \\
            
            \hline
            
            Freitas~\cite{freitas2015reverse} & 2015 & \cmark & ~ & ~ \\
            
            \hline

            Bursztein~\cite{bursztein2014end} & 2014 & ~ & \collectedbyauthors & ~ \\
            
            \hline
            
            Adali~\cite{adali2012predicting} & 2012 & ~ & \collectedbyauthors & ~ \\
            
            Malhotra~\cite{malhotra2012studying} & 2012 & ~ & \collectedbyauthors & ~ \\
            
            Sumner~\cite{sumner2012predicting} & 2012 &  ~ & \collectedbyauthors & ~ \\
            
            \hline

            Goldbeck~\cite{golbeck2011predicting} & 2011 & ~ & \collectedbyauthors & ~ \\
            
            Yamaguchi~\cite{yamaguchi2011vulnerability} & 2011 &  ~ & \publiclyavailable & ~ \\
            
            \hline
            
            Bursztein~\cite{bursztein2009decaptcha} & 2009 & ~ & \collectedbyauthors & ~ \\
            
            \hline
            
            Golle~\cite{golle2008machine} & 2008 & ~ & \collectedbyauthors & \cmark \\

            \bottomrule
        \end{tabular}
    }
    \vspace{-5mm}
\end{table}

\section{User Survey: Supplementary Information}
\label{app:user-survey}

\noindent
We provide more low-level details on our user survey with non-experts (§\ref{sec:5-userSurvey}). We implemented the survey via \href{https://www.qualtrics.com/uk/}{Qualtrics} for both web and mobile. Participation required user consent for anonymous data use; data was stored only after answering all questions and submitting the responses. Participants could exit and resume the survey at any time.

\begin{figure}[!htbp]
\vspace{-3mm}
    \centering
    \centerline{
    \includegraphics[width=0.4\textwidth]{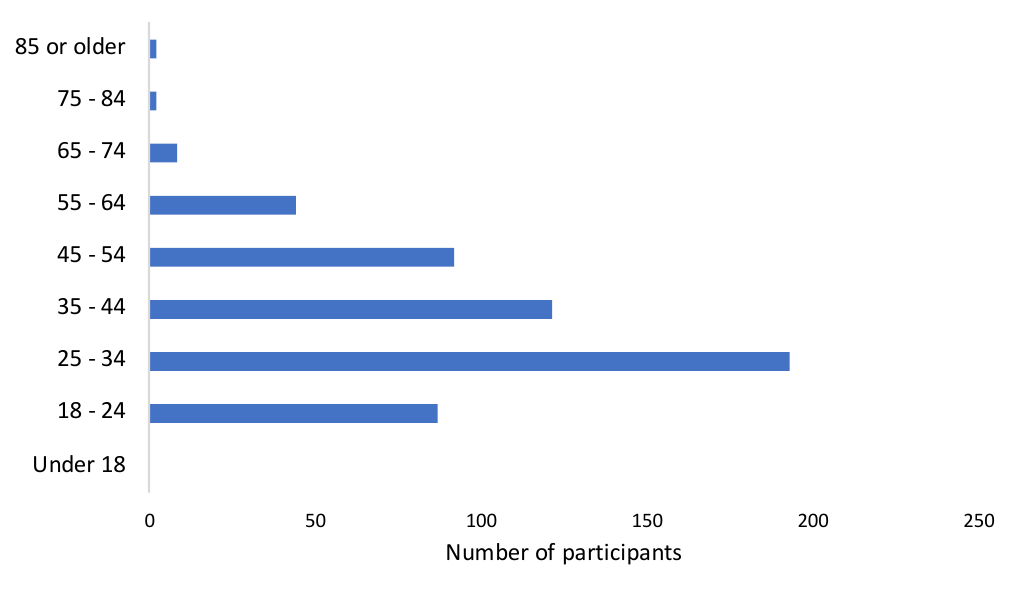}
    }
    \vspace{-2mm}
    \caption{\textbf{Age range.}
    \textmd{\footnotesize \scmath{51\%} of the participants are below 35, and overall \scmath{90\%} of the participants are between 18 and 54. Only \scmath{2\%} are older than 65.}} 
    \label{fig:age}
    \vspace{-3mm}
\end{figure}

\begin{figure}[!htbp]
\vspace{-3mm}
    \centering
    \centerline{
    \includegraphics[width=0.4\textwidth]{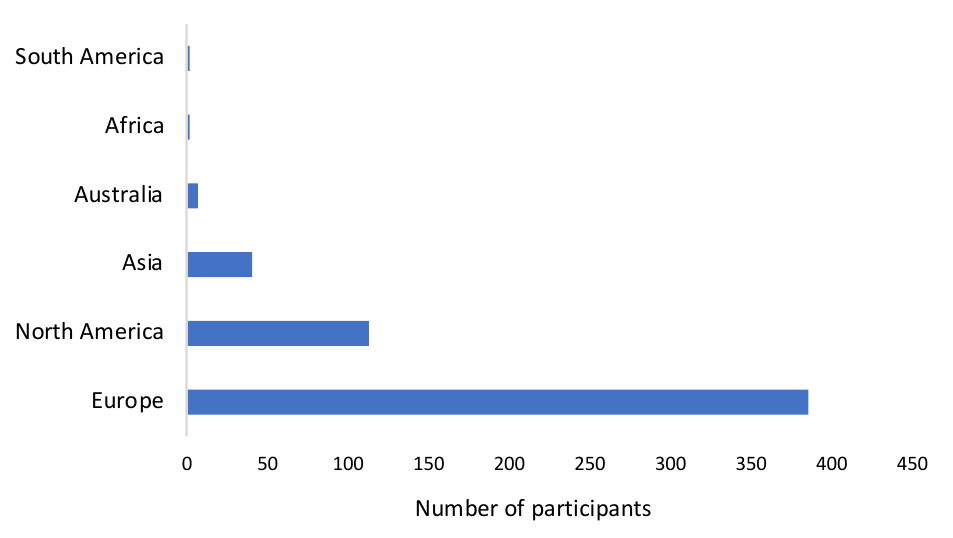}
    }
    \vspace{-2mm}
    \caption{\textbf{Residence based on continent.}
    \textmd{\footnotesize Most participants are from Europe (\scmath{70\%}) and North America (\scmath{21\%}). \scmath{7\%} are from Asia, while overall only \scmath{2\%} of all participants are from Australia, South America, or Africa.}} 
    \label{fig:continent}
    \vspace{-3mm}
\end{figure}

\begin{figure}[!htbp]
\vspace{-3mm}
    \centering
    \centerline{
    \includegraphics[width=0.4\textwidth]{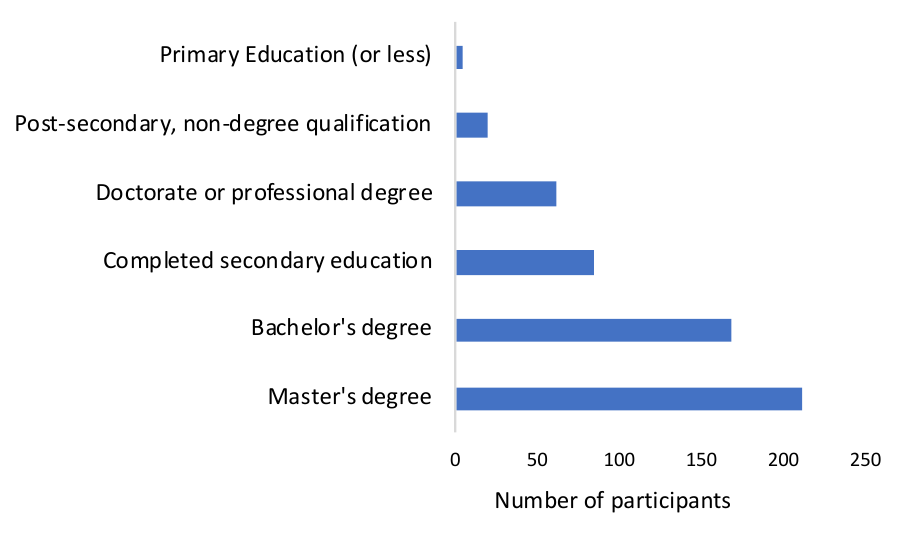}
    }
    \vspace{-2mm}
    \caption{\textbf{Education.}
    \textmd{\footnotesize \scmath{20\%} of the participants do not hold a university degree, while \scmath{31\%} hold a Bachelor's degree, and \scmath{50\%} a Master's degree or higher. }} 
    \label{fig:degree}
    \vspace{-3mm}
\end{figure}

\begin{figure}[!htbp]
\vspace{-3mm}
    \centering
    \centerline{
    \includegraphics[width=0.4\textwidth]{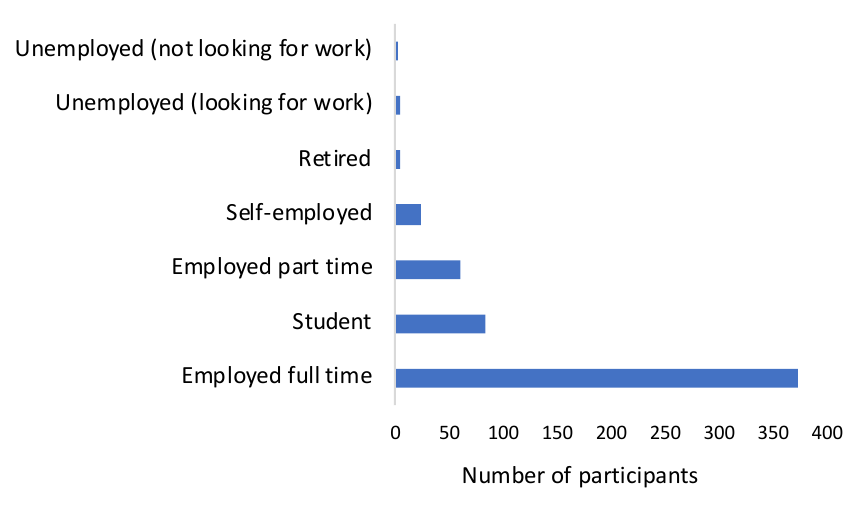}
    }
    \vspace{-2mm}
    \caption{\textbf{Employment.}
    \textmd{\footnotesize The majority of the participants are employed (\scmath{68\%}). \scmath{15\%} are students and \scmath{10\%} are self-employed. \scmath{2\%} are retired or not employed.}} 
    \label{fig:employment}
    \vspace{-1mm}
\end{figure}

\textbf{Demographics.} The detailed demographics of the survey participants are illustrated in Figs.~\ref{fig:age}, \ref{fig:continent},~\ref{fig:degree}, and~\ref{fig:employment}. We compared the demographics from our survey with the OECD demographics since \smamath{88\%} of the participants are from OECD countries. The goal is to identify any bias in our sample and include potential limitations when interpreting the results. We focus on the OECD since this is the largest group of countries covering our participants. The remaining \smamath{12\%} are from 17 distinct countries -- with one participant each from 14 different countries, 21 from India, 20 from Liechtenstein, and 4 from Qatar.\footnote{We rely on the data from the \href{https://data.oecd.org/pop/population.htm}{OECD demographics database} as of 2022 -- the latest available data. We refrain from analyzing the demographics from India, Liechtenstein, and Qatar due to the limited sample size.} The comparison of the participants' demographics from OECD countries to the overall population from the OECD is reported in Table~\ref{tab:oecd-comparison}.

\begin{table}[!htbp]

\centering
\caption{\textbf{Comparison of our sample to OECD population.}
    \textmd{\footnotesize The data is reported in \%. Compared to the OECD reference values, our sample includes fewer women and fewer individuals over 55. Also, our sample includes a comparatively high number of highly educated participants.}} 
\label{tab:oecd-comparison}
\vspace{-0.75mm}
\resizebox{0.6\columnwidth}{!}{
        \begin{tabular}{lcc}
        \toprule
        & \textbf{Survey Part. } & \textbf{OECD} \\
         & \textbf{from OECD} & \textbf{Reference} \\
        \midrule
        \textbf{Gender} & & \\
        \hspace{0.5cm} Female & 36.02 & 50.77 \\
        \hspace{0.5cm} Male & 63.15 & 49.23 \\
        \hspace{0.5cm} Other & 0.83 & N/A \\
        \midrule
        \textbf{Age} & & \\
        \hspace{0.5cm} 18 - 24 & 15.32 & 11.13 \\
        \hspace{0.5cm} 25 - 34 & 34.58 & 16.83 \\
        \hspace{0.5cm} 35 - 44 & 22.77 & 16.95 \\
        \hspace{0.5cm} 45 - 54 & 16.56 & 16.69 \\
        \hspace{0.5cm} 55 - 64 & 8.70 & 15.69 \\
        \hspace{0.5cm} 65 - 74 & 1.66 & 12.38 \\
        \hspace{0.5cm} 75 - 84 & 0.41 & 7.32 \\
        \hspace{0.5cm} 85 and over & N/A & 3.02 \\
        \midrule
        \textbf{Education} & & \\
        \hspace{0.5cm} Primary & 0.75 & 20.00 \\
        \hspace{0.5cm} Secondary & 11.03 & 40.00 \\
        \hspace{0.5cm} Tertiary & 88.22 & 40.00 \\ 
        \bottomrule
\end{tabular}
}

\vspace{-3mm}
\end{table}

\begin{figure}[!htbp]
    \centering
    \centerline{
    \includegraphics[width=0.42\textwidth]{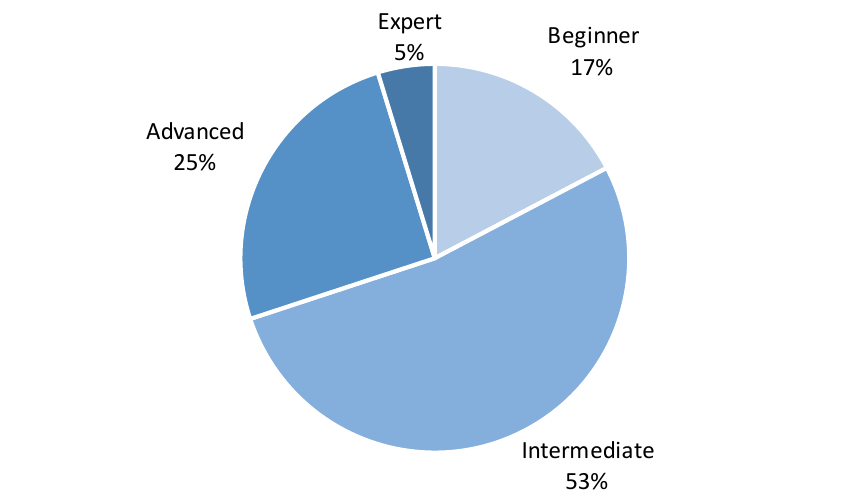}
    }
    \caption{\textbf{AI Expertise.}
    \textmd{\footnotesize \scmath{17\%} of the participants have little or no knowledge of AI, while around half of the participants have a basic understanding of AI, and only very few are experts.}} 
    \label{fig:AI-knowledge}
\end{figure}

\begin{figure}[!htbp]
    \centering
    \centerline{
    \includegraphics[width=0.42\textwidth]{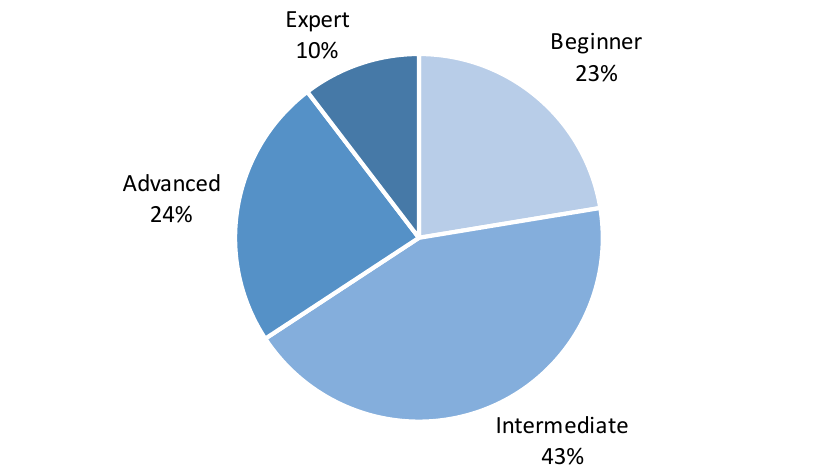}
    }
    \caption{\textbf{Cybersecurity Expertise.}
    \textmd{\footnotesize Only \scmath{23\%} of the participants have little or no knowledge of cybersecurity, while the majority of participants have at least an intermediate understanding of cybersecurity.}} 
    \label{fig:cybersec-knowledge}
\end{figure}

\textbf{Expertise.} Figs.~\ref{fig:AI-knowledge} and~\ref{fig:cybersec-knowledge} provide details on the participants' expertise in AI and cybersecurity. We asked the users to rate their knowledge of each field as either of the following:
\begin{enumerate}[label=\alph*)]
    \item Beginner (``I have little or no knowledge of AI/cybersecurity.'')
    \item Intermediate (``I have a basic understanding of AI/cybersecurity concepts.'')
    \item Advanced (``I have a solid understanding of AI/cybersecurity and its applications.'')
    \item Expert (``I have extensive knowledge and experience in AI/cybersecurity.'')
\end{enumerate}
We asked participants for a self-assessment of their knowledge to put their responses into further context based on their background. We chose not to assess participants' knowledge as it would add complexity to the survey and is not essential for our primary objective, i.e., collecting public opinions on the malicious use of AI. Indeed, prolonging the survey completion time rather contradicts our objective.

\textbf{Completion Statistics.}
Qualtrics recorded 596 survey initiations, with \smamath{82\%} completions. Dropouts mainly occurred either directly after giving consent or just before the requirement to provide qualitative input. On average users took four minutes to complete the survey (excluding outliers\footnote{Users taking more than 20 minutes are outliers: they likely did not complete the survey in a single session but instead paused and resumed.}). The survey was open for four months; however, most participants provided their answers in September 2023 (\smamath{68\%}), some in October (\smamath{7\%}), November (\smamath{14\%}), and December (\smamath{11\%}).

\textbf{Correlation Analysis.}
For the correlation analysis, we consider the user demographics as independent variables and the provided responses to the main body of the survey as dependent variables. First, we transform the qualitative responses into quantitative ones (i.e., we map ``yes'' or ``no'' to ``1'' or ''0"; whereas the knowledge level of AI/cybersecurity is represented by ordinal numbers, e.g., ``beginner'' to ``1'' and ``intermediate'' to ``2''). Second, we perform the Pearson correlation method, one of the most common ones~\cite{nettleton2014commercial}, and derive the corresponding statistic (in §\ref{sssec:user_quantitative}).

\textbf{How is it possible that some respondents who are not concerned about OAI think that AI will harm them?} In Fig.~\ref{fig:survey-responses-1}, it is intriguing to observe that there are 83 people who are ``not concerned about OAI'' and, among these, 21 believe that ``AI will harm me.'' There is a plausible explanation for this apparently nonsensical outcome. Indeed, these people do not worry about AI being used offensively---in the sense that they are not afraid (or are unaware) of the fact that cybercriminals may use AI to cause harm to them; however, these people may still be worried that AI may harm them because of, e.g., unintended AI failures or job displacements due to AI. As a matter of fact, our qualitative analysis found that many respondents provided similar concerns---which have little to do with cybercriminals. This is evidence that the term ``offensive AI'' may not be fully understood by some individuals, or they simply have more pressing concerns about AI which may cloud their vision about OAI-related problems.

\section{User and Expert Survey: Qualitative Analysis}
\label{app:user-survey-qualitative}

\noindent
We provide supplementary information on the qualitative analysis of the user survey (in §\ref{sec:5-userSurvey}) and the expert survey (§\ref{ssec:expert-survey}) based on the constructivist grounded theory~\cite{charmaz2014constructing} enhanced through two rounds of axial coding—termed horizontal and vertical coding. This methodology incorporates principles of constructivism and pluralism, leading to a more inclusive and comprehensive data analysis.

\textbf{Codebook.} We developed an ad-hoc codebook for our qualitative analyses (§\ref{ssec:2.3-userstudy}), which we derived after inspecting the responses to the (mutually exclusive) questions ``what are you most concerned about OAI?'' and ``why are you not concerned about OAI?'' of our survey. We present our codebooks in Table~\ref{tab:axial-codes-concerned} (Table~\ref{tab:axial-codes-NOT-concerned}), showing the horizontal axial codes (perspectives), along with their corresponding vertical axial codes and the associated related focused codes for each axial code for people who are (or not) concerned about AI. 

\begin{table}[!htbp]
    \centering
    \caption{\textbf{Axial Coding---Concerned.}
    \textmd{\footnotesize{Horizontal axial codes, vertical axial codes and related focused codes for people concerned about AI.}}}
    \label{tab:axial-codes-concerned}
    \vspace{-0.75mm}
    \resizebox{\columnwidth}{!}{
        \begin{tabular}{p{2cm}|p{3cm}|p{3cm}|p{3.5cm}}
            \toprule
            \textbf{Horizontal Axial Codes} & \textbf{Vertical Axial Code} & \textbf{Description} & \textbf{Related Focused Codes} \\
            \midrule
            \multirow{6}{*}{\parbox{2cm}{AI Impact and Societal Concerns}} & Ethical Implications of AI & Concerns about the ethical challenges posed by AI. & \textmd{\footnotesize AI in Healthcare, AI Surpassing Human Performance, Ethical Concerns, Military Applications and Warfare, Unintended Errors} \\
            \cline{2-4}
            & AI’s Societal Impact & Apprehensions about AI's broader societal effects. & \textmd{\footnotesize Negligent AI Development Practices, Loss of Control in Autonomous AI, Cybersecurity Attacks, Reduction in Human Learning, General Concerns, Job Displacement, Regulatory Deficiencies, Military Applications and Warfare, Spread of Misinformation, Adverse Social Impacts, Speed, Automation, and Ease of Use, Unintended Errors} \\
            \cline{2-4}
            & Privacy and Identity Concerns & Fears related to privacy, identity theft, and digital manipulation. & \textmd{\footnotesize Privacy, Profiling, and Manipulation, Data Misuse, Identity Theft and Deepfakes} \\
            \hline
            \multirow{6}{*}{\parbox{2cm}{Technological Risks and Threats}} & Risks of Advanced AI Capabilities & Concerns about AI surpassing human capabilities and control. & \textmd{\footnotesize AI Surpassing Human Performance, Loss of Control in Autonomous AI, Reduction in Human Learning, Job Displacement, Speed, Automation, and Ease of Use} \\
            \cline{2-4}
            & Cybersecurity and Data Risks & Fears of AI being exploited for cyberattacks or data abuse. & \textmd{\footnotesize Privacy, Profiling, and Manipulation, Data Misuse, Cybersecurity Attacks, Identity Theft and Deepfakes} \\
            \cline{2-4}
            & AI Development and Error Risks & Concerns regarding errors and issues in AI development practices. & \textmd{\footnotesize AI in Healthcare, Negligent AI Development Practices, Unintended Errors} \\
            \hline
            \multirow{6}{*}{\parbox{2cm}{AI in Domain-Specific Contexts}} & AI in Healthcare & Concerns about AI’s application in healthcare settings. & \textmd{\footnotesize AI in Healthcare} \\
            \cline{2-4}
            & AI in Military and Security & Apprehensions regarding AI's use in military and warfare contexts. & \textmd{\footnotesize Military Applications and Warfare} \\
            \cline{2-4}
            & AI's Impact on Work and Skills & How AI affects job security and human capabilities. & \textmd{\footnotesize AI Surpassing Human Performance, Reduction in Human Learning, Job Displacement, Speed, Automation, and Ease of Use} \\
            \hline
            \multirow{6}{*}{\parbox{2cm}{Human-AI Interrelations}} & AI Autonomy and Supremacy & Concerns about AI surpassing human capabilities and going out of control. & \textmd{\footnotesize AI Surpassing Human Performance, Loss of Control in Autonomous AI, Speed, Automation, and Ease of Use} \\
            \cline{2-4}
            & Job Displacement by AI & Concerns about AI replacing human jobs. & \textmd{\footnotesize AI Surpassing Human Performance, Job Displacement, Speed, Automation, and Ease of Use} \\
            \cline{2-4}
            & AI's Influence on Humans & The impact of AI on human learning and cognitive abilities. & \textmd{\footnotesize Reduction in Human Learning, General Concerns} \\
            \hline
            \multirow{6}{*}{\parbox{2cm}{AI Development and Regulatory Issues}} & AI Development Practices & Issues and concerns in the processes of developing AI. & \textmd{\footnotesize Negligent AI Development Practices} \\
            \cline{2-4}
            & Regulatory Frameworks for AI & The need for and absence of adequate AI regulation. & \textmd{\footnotesize Regulatory Deficiencies, General Concerns} \\
            \cline{2-4}
            & Operational Challenges of AI & Practical challenges in the deployment and operation of AI. & \textmd{\footnotesize Data Misuse, Speed, Automation, and Ease of Use, Unintended Errors, Loss of Control in Autonomous AI} \\
            \bottomrule
        \end{tabular}
    }
\end{table}

\begin{table}[!htbp]
    \centering
    \caption{\textbf{Axial Coding---Not Concerned.}
    \textmd{\footnotesize{Horizontal axial codes, vertical axial codes and related focused codes for people not concerned about AI.}}}
    \label{tab:axial-codes-NOT-concerned}
    \vspace{-0.75mm}
    \resizebox{1\columnwidth}{!}{
        \begin{tabular}{p{2cm}|p{3cm}|p{3cm}|p{3cm}}
            \toprule
            \textbf{Horizontal Axial Codes} & \textbf{Vertical Axial Code} & \textbf{Description} & \textbf{Related Focused Codes} \\
            \midrule
            \multirow{6}{*}{\parbox{2cm}{Technological Progression and Safety Perspective}} & Evaluation of AI Maturity & Assesses the current development stage and future potential of AI. & \textmd{\footnotesize AI Not Yet Mature} \\
            \cline{2-4}
            & Safety and Risk Management in AI & Focuses on the safety concerns and risk management strategies in AI. & \textmd{\footnotesize Adequate Protection and Prevention} \\
            \cline{2-4}
            & Balancing Progress and Safety & Examines the balance between technological progression and safety. & \textmd{\footnotesize Balanced Perspectives, Predominance of Benefits} \\
            \hline
            \multirow{6}{*}{\parbox{2cm}{Societal and Ethical Implications Perspective}} & AI's Societal Impact & Explores how AI impacts society, including its benefits and harms. & \textmd{\footnotesize Balanced Perspectives, General Lack of Concern, No Novel Concerns} \\
            \cline{2-4}
            & Ethical Considerations in AI & Discusses the ethical concerns related to AI development and use. & \textmd{\footnotesize Concerns Centered on Humans, Not AI, Adequate Protection and Prevention} \\
            \cline{2-4}
            & Public Attitudes towards AI & Analyzes public sentiment and attitudes towards AI. & \textmd{\footnotesize AI Not Yet Mature, Predominance of Benefits} \\
            \hline
            \multirow{6}{*}{\parbox{2cm}{Perceptual and Comparative Analysis Perspective}} & Comparative Analysis of AI & Weighs AI's benefits against its potential harms. & \textmd{\footnotesize Balanced Perspectives, Predominance of Benefits} \\
            \cline{2-4}
            & Public Nonchalance about AI & Captures the general lack of concern about AI's impact. & \textmd{\footnotesize General Lack of Concern, No Novel Concerns} \\
            \cline{2-4}
            & AI Readiness and Perception & Assesses perceptions of AI's readiness and potential benefits. & \textmd{\footnotesize AI Not Yet Mature, Predominance of Benefits} \\
            \hline
            \multirow{6}{*}{\parbox{2cm}{Human-Centric Approach Perspective}} & Human-Centric AI Impact & Focuses on AI's impact from a human-centered perspective. & \textmd{\footnotesize Concerns Centered on Humans, Not AI, Predominance of Benefits} \\
            \cline{2-4}
            & Perception of Protection Measures & Examines beliefs regarding the adequacy of AI protection measures. & \textmd{\footnotesize Adequate Protection and Prevention} \\
            \cline{2-4}
            & Balancing Human Concerns and AI Advancement & Weighs human concerns against the perceived advancements in AI. & \textmd{\footnotesize Balanced Perspectives, Predominance of Benefits} \\
            \bottomrule
        \end{tabular}
    }
\end{table}

\textbf{Application.} After deriving our codebook, we then apply it to analyze the corresponding responses---for both the non-expert survey (§\ref{sssec:user_qualitative}), as well as for the expert survey (§\ref{ssec:expert-survey}). We report the results in Table~\ref{tab:focused_codes_concerned} (Table~\ref{tab:focused_codes_not_concerned}), presenting the focused codes derived from the qualitative analysis, representing people's reasons and dimensions for being concerned about AI (not being concerned about AI). These focused codes were generated on the basis of the initial codes, with the frequency indicating how often each focused code emerged in relation to the initial codes.

\begin{table}[!htbp]
    \centering
    \caption{\textbf{Focused Codes---Concerned.}
    \textmd{\footnotesize{We present the focused codes and their frequencies w.r.t. the initial codes for people concerned about OAI.}}}
    \label{tab:focused_codes_concerned}
    \vspace{-0.75mm}
    \resizebox{0.9\columnwidth}{!}{
        \begin{tabular}{c|cc}
            \toprule
            \multirow{2}{*}{\textbf{Focused Code}} & \multicolumn{2}{c}{\textbf{Frequency}} \\
            & \textbf{User Survey} & \textbf{Expert Survey} \\
            \midrule
            AI in Healthcare & 4 & 0 \\
            AI Surpassing Human Performance & 14 & 0 \\
            Privacy, Profiling, and Manipulation & 19 & 3 \\
            Negligent AI Development Practices & 8 & 0 \\
            Data Misuse & 12 & 0\\
            Loss of Control in Autonomous AI & 20 & 0\\
            Cybersecurity Attacks & 141 & 4\\
            Reduction in Human Learning & 15 & 0\\
            Ethical Concerns & 19 & 0\\
            General Concerns & 24 & 0\\
            Identity Theft and Deepfakes & 61 & 1 \\
            Job Displacement & 16 & 0\\
            Regulatory Deficiencies & 9 & 0\\
            Military Applications and Warfare & 21 & 0\\
            Spread of Misinformation & 98 & 0\\
            Adverse Social Impacts & 11 & 0\\
            Speed, Automation, and Ease of Use & 11 & 1\\
            Unintended Errors & 10 & 0\\
            \bottomrule
        \end{tabular}
    }
\end{table}

\begin{table}[!htbp]
    \centering
    \caption{\textbf{Focused Codes---Not Concerned.}
    \textmd{\footnotesize{We present the focused codes and their frequencies w.r.t. the initial codes for people not concerned about OAI. (No participant of the expert survey was not concerned about OAI.)}}}
    \label{tab:focused_codes_not_concerned}
    \vspace{-2mm}
    \resizebox{0.7\columnwidth}{!}{
        \begin{tabular}{c|c}
            \toprule
            \multirow{2}{*}{\textbf{Focused Code}} & \textbf{Frequency} \\
            & \textbf{User Survey} \\
            \midrule
            AI Not Yet Mature & 5 \\
            Adequate Protection and Prevention & 21 \\
            Balanced Perspectives & 11 \\
            Concerns Centered on Humans, Not AI & 9 \\
            General Lack of Concern & 15 \\
            Predominance of Benefits & 8 \\
            No Novel Concerns & 11 \\
            \bottomrule
        \end{tabular}
    }
\end{table}
\section{Expert Statements: Source and Analysis}
\label{app:expert-statements}

\noindent
In Appendix~\ref{sapp:verbatim}, we list the statements provided by the experts, as indicated in our methodology (discussed in~§\ref{ssec:2.4-expert}). Then, in Appendix~\ref{sapp:expert-overview} we provide low-level details explaining how we objectively analyzed them (refer to §\ref{sssec:expert-overview}).

\subsection{Statements (Verbatim)}
\label{sapp:verbatim}

\noindent
We report the statements as written by each expert. The order is random, and we will not provide any information that can be used to identify the author of each statement.

\textbf{Statement from Expert 1:}
\emph{Is it AI-generated content or reality?}
To me, the scariest aspect of offensive AI is related to deepfakes and disinformation. Nowadays, we \textit{cannot trust} anything we see, hear, or read online anymore. As soon as I see a picture or video, I have to wonder: ``is it real or was it AI-generated?'' It is becoming increasingly harder to distinguish AI from reality. This forces individual users to take time to, e.g., look at a picture or video, and decide if it is real or not. Unfortunately, Web users do not always challenge what they encounter online: if they have seen a picture showing a specific fact, then they may assume that this fact definitely took place.

Attackers now use AI to create disinformation campaigns, fake reviews, and to manipulate crowds into believing that some AI-generated content is real. This has direct political and economical implications, e.g., election campaigns. One of the most important open problems in the field of OAI is therefore to be able to detect such deepfakes and disinformation campaigns.

\emph{Adverse effects of AI.}
Second, AI was initially meant to \textit{make our life easier}. Unfortunately, it also makes an attacker’s job easier, for example by using AI to write sophisticated phishing emails or advanced malware. A few years ago, phishing emails were quite easy to detect, e.g., because they were poorly written, full of grammar mistakes, and impersonal. Now, with ChatGPT and other LLMs, they look more authentic. Given that Web users also use such services to write (benign!) emails faster, it is becoming more challenging to detect such phishing emails and other AI-generated threats. While AI was meant to simplify some day-to-day tasks, it is also introducing new threats and challenges. A second open problem revolves around weighting the pros and cons: \textit{can an AI-based solution be misused?} It is critical to take into account potential adverse effects of AI before deployment, otherwise we end up in a world that is harder and more dangerous to live in than it was before AI.

\emph{Privacy-related attacks.}
Third, AI can also be misused to collect or infer sensitive information about Web users. For example, AI facilitates profile-matching across social networks, enables attribute-inference attacks, and deanonymization online. It can also have direct military implications when AI is leveraged for (unauthorized) surveillance. A third open problem consists in developing guidelines and defensive measures to better protect users and their (sensitive) data, but also to work on awareness campaigns to let users know about AI-based privacy-related threats.

\textbf{Statement from Expert 2:}
Given the rapidly increasing prevalence of AI, especially generative AI, in all walks of life, offensive AI is becoming an increasingly important topic. Within existing research, a rather positive image of AI often prevails. Discussions tend to focus on the potential positive aspects and impacts of AI. Although the dangers of AI are more and more being discussed, such discussions are often limited to side effects arising from predominantly benign use of AI, such as biases in benign AI systems or privacy violations during AI model training for benign AI systems. The active use of AI for malicious purposes has so far played a minor role in academic discourse but deserves more of our attention. In particular, I see the following three open problems as central challenges of offensive AI research that the academic community should address more closely.

First, research should specifically focus on offensive AI in the field of high-risk AI applications. One can certainly argue that offensive AI should generally be classified as a high-risk AI application. However, I believe that there are particularly high-risk areas of application for AI where offensive AI can be especially harmful. This applies in particular to applications of offensive AI that jeopardize democratic institutions and institutions, which uphold the rule of law (e.g. spreading false information to manipulate elections with the help of AI), or critical infrastructures (e.g. AI-based ransomware in medical facilities).

Second, from a socio-technical perspective, research should focus more on the interaction between humans and technology in terms of offensive AI. For example, research should delve deeper into how individuals and organizations can detect and counteract offensive AI attacks. This includes investigating what makes individuals and organizations particularly vulnerable to offensive AI and developing tools to help them recognize and fend off attacks with offensive AI (e.g., email clients that detect texts generated by generative AI and alert users).

Third, research should explore how fundamentally benign AI can turn into offensive AI and be used for malicious purposes. Typical examples include generating high-quality scam emails using generative AI. It is not surprising that fundamentally benign AI can also be misused for malicious purposes. Therefore, we should engage more in the systematic study of the potential dangers of benign AI in terms of its use as offensive AI and develop appropriate tools for assessing the risk potential for AI developers and researchers.

\textbf{Statement from Expert 3:}
In my opinion the primary open problems in the field of Offensive AI at this point in time are:

\emph{1) Lack of understanding regarding the potential and limitations of Offensive AI tools.} One of the greatest problems in this field is the lack of understanding regarding the real potential and limitations of AI on the offensive side. The vast majority of the published research presents success in achieving a specific goal (e.g., to jack a webserver using an AI tool) which stimulates our thinking regarding the potential of Offensive AI. On the other hand, research that shows failures and sheds light on the limitations of Offensive AI tools has yet to be published. A greater understanding of the boundaries of Offensive AI tools is required to shed light on the real potential and limitations of Offensive AI tools concerning integrated AI technology.
 
\emph{2) Lack of Effective Countermeasures.}	In some usecases, AI has reached the maturity required to orchestrate cyber attacks that could lead to financial losses (e.g., audio deep fakes). However, despite the rapid advancement made by the community on the offensive side, one of the greatest problems in this field is the fact that there needs to be more advancement on the opposite side of countermeasures. There are usecases in which effective countermeasures that could be used to detect, mitigate, and prevent AI-orchestrated attacks were not developed (e.g., detecting fake news). In other cases, the countermeasures that were developed were found ineffective in reality when simulated in real environments (e.g., audio deepfakes). There is a real need to develop countermeasures that are effective against the current threats posed by Offensive AI in real environments (e.g., audio deepfakes, text deepfakes, etc).
 
\emph{3) Lack of effective TARA standards for systems against Offensive AI.}	One of the greatest problems in this field is the inability to determine whether the risk posed to systems by the advancements published by new research is real. This happens because the practicality and the real outcome of the attacks are not clear from the research published in this field due to the facts that: (i) the vast majority of the research hasn’t been demonstrated against real systems and it is known that attacks against specific components may not affect the system itself, (ii) the level of expertise required to use the tools is not clear (layman? expert?).  
This requires the development of dedicated threat analysis and risk assessment standards (TARA) that will allow CISOs to objectively calculate the risk posed by offensive AI tools to their systems, taking into account the \href{https://en.wikipedia.org/wiki/Technology_readiness_level}{Technological Readiness Level} of a tool, the practicality of the attack (considering expertise, needed equipment, the window of opportunity, etc.), and the outcome of the attack in practice (considering validation against real systems). 
Such standards may help to quantify and shed light on the real risk posed by the developed offensive tools to organizations, putting things into the right perspective.   

\textbf{Statement from Expert 4:}
After reading the draft version of the SoK paper, I strengthened my conviction that humans are the most relevant target of offensive AI and, therefore, future research on offensive AI should prioritize the offensive use of AI to hack human beings by exploiting the large amount of data that can be collected on human behaviors and habits.

Therefore, in the following, I briefly describe the open problems in the field of offensive AI that I would like to give priority:

\emph{1) Cognitive biases and offensive AI.}
We have already clear evidence that AI can be used to hack human by performing contextualization and personalization in phishing attacks. It is therefore easy to conjecture that offensive AI could exploit cognitive biases of individuals to accomplish tasks that violate security and privacy. As an example, confirmation bias is a well-known and powerful cognitive bias that affects decision-making. Offensive AI could exploit this bias by creating echo chambers where users are only exposed to information that confirms their existing beliefs. This can lead to increased polarization, reduced critical thinking, and the spread of misinformation.

\emph{2) Offensive AI and behavioral economics.}
The field of behavioral economics pointed out well how cognitive biases strongly affect the financial decisions of individuals and how these decisions deviate from those predicted by classical economic theory. Given the above conjecture that offensive AI can exploit cognitive biases of individuals, it is a priority to investigate how offensive AI could be used to manipulate financial decisions of individuals at a large scale with serious financial consequences for the society. As an example, loss aversion is a well-known cognitive bias where the pain of losing is felt more acutely than the pleasure of gaining. Offensive AI could exploit this bias by spreading information targeted to each individual that emphasize potential losses, so prompting individuals to make irrational financial decisions to avoid these perceived losses. 

\emph{3) Human-AI Teaming against Offensive AI exploiting cognitive biases.}
I do believe that as offensive AI will become more sophisticated in exploiting cognitive biases such as confirmation bias and loss aversion, there will be a pressing need for defence measures that leverage both AI and human strengths. In fact, human-AI teaming can combine the rapid processing and pattern recognition capabilities of AI with contextual understanding and decision-making of humans. This line of research could intersect with the previous one on offensive AI and behavioral economics, given the seminal work of the Nobel Laureate Daniel Kahneman in behavioral economics on the analysis of comparative strengths and weaknesses of humans and machines in decision making; this previous work could be leveraged to combine humans and AI capabilities against offensive AI exploiting cognitive biases.

\textbf{Statement from Expert 5:}
Offensive AI can be used to cause harm to individuals, institutions and society at large. Among the open problems in the field of Offensive AI, the top three which need to be researched and solved are: detection of deepfakes, mitigating the use of AI to spread disinformation, and developing defensive AI mechanisms to counter AI-orchestrated cyberattacks.

Deepfake technology enables attackers to launch phishing and other forms of social engineering attacks by impersonating a victim through cloning their face or their voice. The variety of ways by which this can cause threats and harm makes it one of the most serious threats of Offensive AI. Deepfakes can be used to overcome biometric systems. They can be used to gain access to information and data. Attackers can use the technology to carry out fraud, to launch disinformation campaigns that seem credible, or for defamation. Mechanisms need to be developed to detect deepfakes to counteract their threat.

AI combined with social media can be used to accelerate the spread of disinformation for a variety of ulterior motives, such swaying public opinion and shaping behavior. Renee DiResta, of the Stanford Internet Observatory, says in relation to disinformation: “social media took the cost of distribution to zero, and generative AI takes the cost of generation to zero”~\cite{Economist-Disinformation}. New technology needs to be developed and incorporated into social media platforms in order to detect and prevent the spread of disinformation. This will not happen organically. It will only be propelled by the force of government legislation.

 Offensive AI can be used to enable new levels of cyberattack automation from launch to the various stages of attack propagation. It allows attackers to scale up their attack coverage and increase their success rate. This renders human controlled detection systems incapable of keeping up with the scale and speed of the attacks. Defensive AI mechanisms need to be developed which are capable of automatically detecting and counteracting their malicious twins.

\textbf{Statement from Expert 6:}
Open problems with offensive AI can be categorized into areas of focus based on the motivations of different expert groups. These groups may include offensive teams aiming to improve attacks, defense teams aiming to mitigate attacks, and trust and safety teams aiming to minimize harm to society. For red team purposes, the challenge lies in measuring the success and value of offensive AI attacks, as it may not always offer substantial increases in efficacy over traditional techniques. For blue teams, the challenge may be that offensive AI simply enables relatively trivial attacks to be scaled, with offensive AI being used to turn large groups of lower skilled adversaries into more highly skilled ones. For civil society, the challenge will be keeping up with new approaches, such as information/influence operations, and addressing the societal challenges they pose.

For offensive teams themselves, an obvious challenge is how to measure the success and value of attacks powered by offensive AI. It is not simply a case that offensive AI will always offer substantive increases in efficacy over other more traditional offensive techniques and if this is not the case, then questions should be raised as to the value of the approach. In particular, just as with other quantitative and qualitative measures of success, the value of AI can often be manipulated to show success and/or failure as desired. An example of this is that there is already research that purports to show how generative algorithms can find "0-days". This research needs to be reviewed against results using traditional non-AI based techniques, to determine whether the approach delivers the value initially claimed. The challenge can be summarised as finding suitable cross-domain experience and mechanisms to allow effective comparison and scoring of traditional and AI-powered offensive techniques.

For those trying to defend against existing adversaries, the challenge may simply be that offensive AI will likely be used to enable scaling of attacks. When we look at DoS for example, there are both technical attacks but also things that simply overwhelm the human at the other end of the process. What happens if it turns out that offensive AI is most effective when used to turn large groups of lower skilled adversaries into more highly skilled adversaries? For example, what tools and techniques will defenders need to spot and deal with these attacks? Assuming that generative algorithms are successful at complying with the prompts provided by their human users then their output will likely pass the requirements enforced by traditional security controls. This challenge can be summarised as finding ways to classify how particular offensive AI techniques may be weaponised and then defining solutions that are appropriate to each.

For society at large, the challenge will be less about measuring and improving efficacy of offensive AI usage or stopping individual attacks but rather trying to keep up with new approaches. Offensive AI will likely result in new types of techniques that don't fit conventional cyber security definitions. For example, are information/influence operations considered as part of offensive AI and how can they best be addressed?

\textbf{Statement from Expert 7:}
Researchers are beginning to develop AI for various offensive use cases that will present challenges to defensive systems and processes in the potency and speed of cyber campaigns via offensive copilots, scaling social engineering attacks, and enhancing offensive operations. This statement outlines these challenges and the gaps that need to be addressed for these techniques to be considered effective.

\emph{Offensive Copilots: Reducing Time to Impact.} 
Offensive copilots are AI systems that can significantly reduce the time required for threat actors to execute an attack by automating labor-intensive tasks that typically demand specific expertise or extensive exploration. For example, AI can expedite the discovery, development, and delivery of exploits or polymorphic malware. The primary gap for attackers in achieving this capability is largely an engineering problem of agents systems to tools (e.g., reversing tools, CVE lookups, executable instrumentation) through agent systems, checks to validate successful outcomes, and the tedious optimization of system instructions in generative AI for the copilots actions to produce reasonable results.

\emph{Scaling Social Engineering Attacks with Minimal Direction.}
AI's potential to scale social engineering attacks is another developing attack vector. Threat actors can use generative AI to create deepfakes and other convincing forms of fake identities, which can be used in automated and interactive phishing or scamming operations. Unlike other automated social engineering tools, the AI-driven attacks are adaptive and interactive. The gap for attackers in achieving this lies in the engineering of agent-based systems with optimized playbooks for carrying out such operations. To be convincing, especially in interactive situations, further research is needed in realistic interaction techniques. This includes developing scoped questions and responses, achieving alignment without providing the idiosyncratic generic responses of AI like ChatGPT, and mimicking human-like selectivity in responses, including delays in responding. The primary research required is to enhance these systems' realism and effectiveness in social engineering contexts.

\emph{Scaling Offensive Operations.}
Offensive AI enhances the targeting and scaling of cyber operations beyond traditional automation. AI-driven attacks can achieve higher precision and impact by integrating multiple automated steps from frameworks like MITRE ATT\&CK, which maps out various cyber adversaries' tactics and techniques. By integrating generative AI agent frameworks with preexisting tools, attackers can orchestrate complex operations that tackle large portions of an attack lifecycle seamlessly. However, achieving the scale and sophistication required for these operations presents significant engineering challenges. Additionally, there may be insufficient data on complex attacks to generate novel end-to-end attacks effectively. Fine-tuning large language models (LLMs) to cyber-operations might be necessary so that the AI can selectively focus on key indicators in long contexts and predict appropriate next steps accurately.

\textbf{Statement from Expert 8:}
In the realm of threat intelligence research, state-sponsored threat actors have been reported to leverage offensive AI in their attack campaigns\footnote{\url{https://openai.com/index/disrupting-malicious-uses-of-ai-by-state-affiliated-threat-actors/}}. These actors use AI services for reconnaissance (identifying targets, researching tools, translating technical papers) and weaponization (scripting, creating social engineering content).

From this perspective, several open research questions remain. One question is whether offensive AI has been employed in other stages of the cyber attack kill chain~\cite{Cyberkillchain}. For example, while there has been research on using AI for code obfuscation, it is unclear if threat actors have adopted AI-obfuscated malware in the wild. Can threat actors enhance their fuzzing capabilities with neural networks to find software vulnerabilities? Defensive Endpoint Detection and Response (EDR) products use AI to analyse logs and construct attack paths and timelines—can AI also be leveraged to reverse this process? Can threat actors use AI to design and automate lateral movement within compromised networks? These are intriguing questions that warrant further exploration.

Another important question is whether we can detect and mitigate attacks launched with offensive AI tools, and how these approaches differ from existing detection techniques. For instance, can AI-model obfuscated malware still be detected through entropy measurements\footnote{\url{https://redcanary.com/blog/threat-detection/threat-hunting-entropy/}}? Can phishing emails generated by AI be identified through content analysis?

Lastly, there are critical questions regarding AI service providers. What techniques can detect and prevent platform abuse? To date, there have been few reports from AI service providers about their tools being used for offensive attacks. Despite potential conflicts of interest with company reputation, increased transparency and disclosure would benefit the industry and the public, leading to stronger collective defense. Beyond detection techniques, what forms of collaboration or communication channels should be established between service providers and industry partners? What information can be shared to enhance collaboration? These topics are essential for ongoing discussion.
In conclusion, while offensive AI might present more challenges for the defenders, it also opens up new avenues for research and collaboration. Addressing these open questions will require a concerted effort from researchers, industry professionals, and AI service providers.

\textbf{Statement from Expert 9:}
Technological advancements are usually considered in terms of how they can be used by humans to benefit their tasks or life in general. This common approach towards understanding the positive effects of technology has in particular dominated inquiries in the interdisciplinary field of Information Systems (IS), which traditionally investigates the use of digital systems from a socio-technical perspective. Only in recent years this attention has somewhat shifted towards investigating the “dark side” of technology to understand adverse usage outcomes, albeit still mainly considering the (non-)achievement of predetermined goals. There is an emerging stream of research in IS on responsible AI with a particular emphasis on principles needed to mitigate AI related risks. Notwithstanding, the current discourse in IS has largely overlooked the offensive potential of AI, as considered in this study, which demonstrates that AI clearly provides or improves capabilities of adversaries for many unknown or weakly understood use cases that have the potential to disrupt not only the lives of people but also organizations and entire societies. 

The following three research problems represent broad OAI capabilities that can be seen most threatening from a cost/benefit perspective, as they offer high rewards (or harm) and little cost to adversaries in terms of enabling or enhancing their attacks, and are difficult to mitigate from the defenders’ standpoint. These include (1) social engineering, (2) information gathering and misinformation, and (3) vulnerability exploitation. Regarding (1), humans have been generally seen to be the weakest link in private and organizational contexts. AI will most likely aggravate this situation by, e.g., allowing for quickly building fake personas/profiles that can be used in spear phishing or other attacks on targets that have been cost-effectively identified via AI. From the IS perspective, open research issues comprise, e.g., how humans process or act on these attacks (e.g., heuristically vs. systematically), safeguards or countermeasures such as awareness building and associated training methods, or the use of automated AI systems to counter AI-enabled attacks. Regarding (2), information gathering and dissemination also scale well via AI and can provide the basis for a number of attacks to, e.g., use AI for reconnaissance, which is particularly hard to prevent, and efficiently disseminating content crafted or fabricated by AI that can be used to, e.g., polarize or reduce trust in institutions or science, thereby adversely affecting public opinion. Regarding (3), from an adversarial perspective an attacker can abuse AI systems by exploiting AI vulnerabilities through white-box, black-box, or gray-box techniques, which at least partially can be achieved with little technical knowledge. From the IS perspective, important safeguards to investigate may for example include data governance practices to prevent unauthorized access to sensitive (training) data and data biasing. 

These research problems are of course neither exclusive nor exhaustive. From a more holistic view, these and other issues necessitate research on the effective government of AI, accounting for all stakeholders (as shown in this study), which should help mitigating many problems. For example, it will be necessary to investigate the role of global laws and regulations, and human oversight in dictating and testing the functions of AI, which in turn should reduce the capabilities of AI for offensive purposes in many use-cases. Also, respecting fairness, transparency and accountability (FAT) principles, currently debated in work on responsible AI, in the early design stage of AI systems development, e.g., through guiding frameworks or standards would constitute another important safeguard. In any case, a cautious and step-wise approach toward the implementation of AI-based solutions in organizations and society, providing for sufficient time for testing and corrections, is highly advised considering the potential of OAI.

\textbf{Statement from Expert 10:}
In my opinion, these are the three most important problems:
 
\emph{1) Measuring AI usage in real-world attacks.}
An important open question is how to reliably measure the use of offensive AI in real-world attacks. So far, the community has studied a range of theoretically possible AI attacks, but we don’t have sufficient understanding of whether and how often such attacks are carried out by attackers in practice, who are running these attacks, and what strategies attackers have to integrate/use AI. These questions cannot be answered without conducting empirical measurements by collecting and analyzing real-world attack data. Answering these questions is critical to the community to define accurate threat models and develop effective countermeasures. A key technical challenge is to distinguish AI-generated offensive content (text, images, voices, network traces, software code) within real-world data with limited “ground truth”. Whether it is malware samples, network traces, or data from social media, it may be possible to manually label “offensive content” (e.g., malware) but it’s still challenging to further label the AI-generated offensive content (e.g., AI-written malware) from offensive content generated by traditional methods (e.g., malware written by hackers). Solving this problem is key to running real-world measurements to plot out the threat landscape.
 
\emph{2) Defending against AI use in disinformation and online deception.}
Disinformation is a major threat to our society today, and it is difficult to address this threat with technical means only. Disinformation has been a problem even before the take-off of generative AI (e.g., with manually crafted fake news, and manually altered photographs). The key difference is generative AI significantly reduces the technical barriers/costs of generating false content, which, to some extent, democratizes cyber offense.  Today, lay users (without any programming experience) are able to craft high-quality “deepfakes” using basic text prompts. In addition to detecting such AI-generated content, a bigger challenge is to determine the “intent” of the deepfake, effectively flagging the content with a malicious intention from that of benign-intended (e.g., AI content for entertainment). This process may require collaborative efforts from human users/moderators and automated detection techniques.   
 
\emph{3) Using offensive AI to enhance existing defense.}
Offensive AI has the potential to be used positively to improve our defense. A concrete example is to use AI methods (e.g., Large Language Models or “LLM”) to scan software code bases to detect bugs/vulnerabilities and augment traditional tools such as fuzzing. However, as an offensive technique (for vulnerability discovery), it can also be used by malicious parties to find zero-day vulnerabilities to facilitate system compromise. The question is how to tip the scales to benefit the defenders even more.

\textbf{Statement from Expert 11:}
\emph{1) Privacy attacks beyond membership inference.}
One of the most well-researched uses of AI to attack privacy are attacks on training data, including membership and attribute inference~\cite{shokri2017membership}. However, AI can also enable other privacy attacks, such as linking of separate data items, e.g., for cross-device tracking, or fingerprinting of encrypted network traffic, which has been demonstrated for websites, apps, and voice commands. Understanding how these attacks work and how effective they are is a crucial precondition for the next steps: designing effective countermeasures and protections.

\emph{2) Detection of AI-based attacks.}
Detecting ongoing attacks and correctly attributing their source is particularly important for attacks that target humans, such as misinformation or phishing, which rely on AI-generated content. If we can attribute this content to AI in general, or even to the specific model that powers the attack, we can create awareness-based protections, such as labels and warnings in user interfaces, that may be easier to deploy than technical countermeasures that aim to prevent the attack. At the same time, awareness-based protections are desirable because they empower users to defend themselves. Existing approaches for detection rely on AI to label content as real or generated~\cite{sha2023fake}, however, it is not clear how futureproof and generalizable these approaches are.

\emph{3) Quantification.}
The effectiveness of attacks and defenses is commonly quantified with traditional machine learning metrics such as precision and recall. While these metrics are useful to compare the effectiveness of new attacks/defenses with existing attacks/defenses in controlled experimental settings, they are not sufficient to evaluate the effects in realistic settings. Additional metrics are needed, for example, to evaluate the feasibility of attacks, to estimate economic effects to inform regulators and lawmakers, or to quantify population-level privacy harms.

\textbf{Statement from Expert 12:}
\emph{1) Research incentives.}
Research on offensive AI traditionally focuses on exploring the potential of AI to automate and enhance a wide range of malicious processes. These include vulnerability discovery, exploit development, penetration testing, evasive malware generation, encrypted traffic and hardware side-channel analyses, device fingerprinting, inference of private data, and more.
While scientific articles exploring and prototyping these attacks are abundant, the impact of the research direction as a whole may be threatened by certain common practices. The field has largely stepped away from the underlying goal of offensive research -- understanding and mitigating potential threats --, and instead centers around advancing simulated attacks until they surpass state-of-the-art performance. While the technical novelties behind AI-based attacks can be impressive and educating, aiming exclusively for superior attack performance may have harmful consequences on science. There is a tangible risk that research that does not delve into the exact workings of the novel attack vector to sufficiently inform mitigations does little to progress the field beyond the existing knowledge base. If offensive capabilities outpace defensive measures in research, this may inadvertently encourage a competitive mindset that prioritizes breaking systems over securing them, leaving systems chronically more vulnerable overall.
To collectively address this open problem, it is imperative for the community to consolidate new practices, where an in-depth understanding of the novel AI-based attack mechanism and suggesting (and possibly evaluating) potential mitigations receive more attention and value (instead of being perceived as a limitation of the attack). Note that while this concern affects the entire offensive security field, it is especially challenging in the context of AI, where data-driven attacks are automated and inherently opaque, demanding targeted explainability measures.

\emph{2) Realistic attack simulation.}
In many offensive AI applications common evaluation practices have been established for the sake of simplicity, reproducibility, and compatibility with prior work. The consequences are as follows: (i) most of the studies do not or cannot aim for realistic estimations of threat severity, and (ii) the simplifying assumptions are often unacknowledged or inadvertently omitted, thus hindering the objective assessment of results.
To yield results that more accurately reflect real-world or that can be more reliably assessed, the simplifying assumptions behind the data, the chosen metrics and the evaluation settings, as well as assumptions behind baseline attacker capabilities, need to be gradually overcome and systematically acknowledged.

\emph{1) Threat of offensive generative AI (GenAI).}
Offensive GenAI is a novel research direction that presents an unprecedented level of urgency due to the wide-spread adoption. The quality of AI-generated content has recently surpassed any expectations, demonstrating blasting performance against the weakest link in secured systems: humans. Social engineering, phishing attacks, spread of misinformation and fake content, and other kinds of AI-assisted manipulation demand focused attention from the scientific community.
The core goals can be: (i) developing research practices that allow reliable simulations of the novel threat to reflect real-world infrastructures; (ii) encouraging interdisciplinary collaborations with non-technical fields (policymakers, educators, sociologists, psychologists, ethicists, etc.) for correct study designs and accurate interpretation of human behaviour; (iii) directly utilizing findings from offensive GenAI research to inform not just automated defenses, but, crucially, effective awareness campaigns.

In summary, conscious and effective design of defenses is a shared ultimate goal of offensive and defensive research teams and relies on raising our standards when harvesting solid insights from AI-based attacks.

\subsection{Systematic Analysis via NLP: methods and results}
\label{sapp:expert-overview}

\begin{figure*}[!htbp]
    \vspace{-3mm}
    \centering
    \centerline{
    \includegraphics[width=\textwidth]{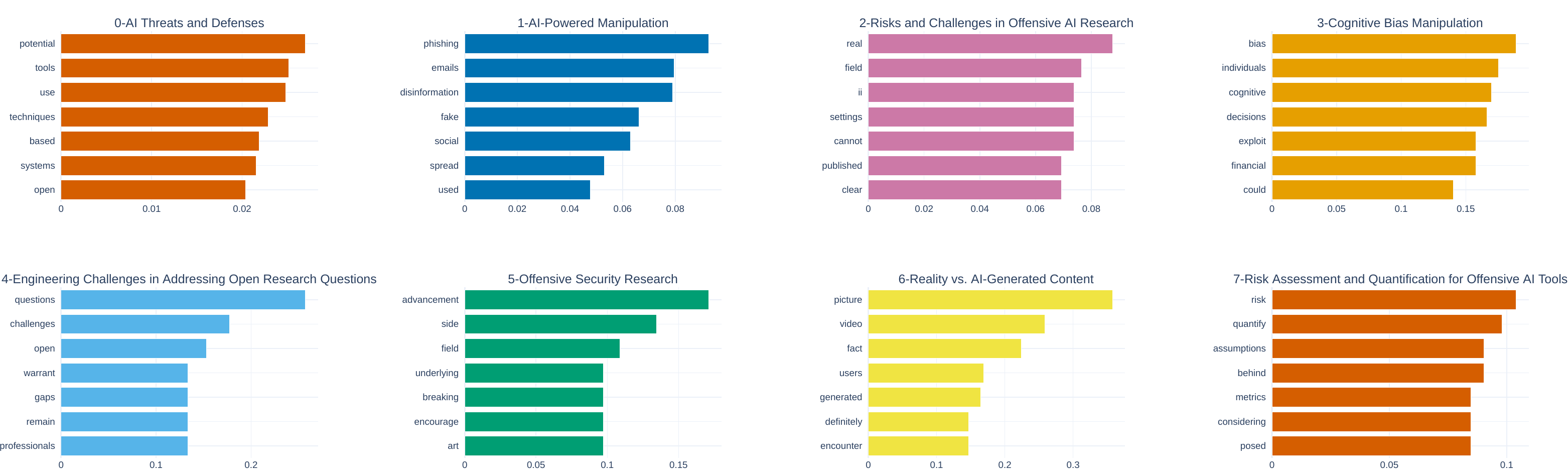}
    }
    \vspace{0mm}
    \caption{\textbf{Topics identified by BERTopic.}
    \textmd{\footnotesize We use BERTopic to analyze our expert statements and output the 8 most relevant topics. Each plot in the figure refers to a topic (title), wherein the y-axis shows the six most relevant words in the topic, and the x-axis denotes the weight of each word.}} 
    \label{fig:topics}
    \vspace{-2mm}
\end{figure*}

\noindent
We provide low-level technical detail (as well as the complete results) of the natural language processing (NLP) methods we used to analyze the expert statements.

\subsubsection{\textbf{N-grams analysis}}
\label{ssapp:ngram}
As a preliminary check, we process the entire statements and extract the 20 most common bigrams/trigrams. The results are as follows: as we expected, the most common n-gram is ``offensive ai'' (61 occurrences); the second most common is ``use ai'' (11 occurrences), the third is ``generative ai'' (10 occurrences), followed by ``social engineering,'', ``ai generated'' and ``ai based'' (9 occurrences each). Then, we have ``real world'' and ``cognitive biases'' (7 occurrences each). Next, with 6 occurrences, there are ``threat actors'', ``problems field'', ``offensive ai tools'', ``ai used'', ``ai tools'', ``ai systems''. With 5 occurrences, there are ``open problems'', ``generated content'', ``benign ai'', ``ai generated content''. Finally, with 4 occurrences, there are ``social media'' and ``social engineering attacks''.

\subsubsection{\textbf{Keyword Extraction.}}
\label{ssapp:keyword}
We then analyze each statement individually by extracting the most relevant keywords, using KeyBERT~\cite{grootendorst2020keybert}, a popular text-mining technique (used also, e.g., in~\cite{elsharef2024facilitating}). Specifically, KeyBERT takes some text as input, and returns a list of keywords, each provided with a number (normalized between 0 and 1) representing its relevance in the text provided as input. Before running KeyBERT, we remove the common stopwords: {\small [``offensive'', `'ai'', `'artificial'', ``intelligence'', ``malicious'', ``attack'', ``attacks'', ``security'', ``harmful'', ``research'', ``problems'']}. The following are the top-5 keywords returned by KeyBERT for each expert statement (ES);
\begin{itemize}[leftmargin=1cm]
    \item[ES1:] {\small [(`biases', 0.2294), (`dangers', 0.2205), (`discourse', 0.2051), (`increasingly', 0.1979), (`academic', 0.1916)]} 
    \item[ES2:] {\small[(`malware', 0.304), (`threat', 0.296), (`defense', 0.2564), (`detection', 0.2564), (`vulnerabilities', 0.243)]}, 
    \item[ES3:] {\small[(`countermeasures', 0.3488), (`cyber', 0.332), (`tools', 0.3243), (`threats', 0.3089), (`tool', 0.2919)]}, 
    \item[ES4:] {\small[(`technological', 0.4223), (`technology', 0.4061), 
    (`exploitation', 0.3831), (`safeguards', 0.3655), (`exploiting', 0.365)]}, 
    \item[ES5:] {\small[(`phishing', 0.3615), (`exploiting', 0.3552), (`biases', 0.3445), (`bias', 0.3247), (`behavioral', 0.2992)]}, 
    \item[ES6:] {\small[(`attackers', 0.3865), (`malware', 0.3828), (`hackers', 0.3294), (`vulnerabilities', 0.3008), (`defending', 0.2907)]}, 
    \item[ES7:] {\small[(`phishing', 0.3917), (`threats', 0.3853), (`deepfake', 0.3744), (`deepfakes', 0.3721), (`cyberattacks', 0.3679)]}, 
    \item[ES8:] {\small[(`defense', 0.4248), (`defend', 0.3772), (`adversaries', 0.3539), (`defenders', 0.3315), (`mitigate', 0.2803)]}, 
    \item[ES9:] {\small[(`attackers', 0.4356), (`threat', 0.3449), (`exploits', 0.336), (`malware', 0.3313), (`tactics', 0.3311)]}, 
    \item[ES10:] {\small[(`disinformation', 0.3484), (`reality', 0.3289), (`surveillance', 0.3198), (`threats', 0.3067), (`attackers', 0.2937)]}, 
    \item[ES11:] {\small[(`privacy', 0.4319), (`protections', 0.3869), (`phishing', 0.373), (`countermeasures', 0.358), (`defenses', 0.3413)]}, 
    \item[ES12:] {\small[(`defenses', 0.4028), (`threats', 0.3804), (`exploit', 0.3801), (`malware', 0.3642), (`attacker', 0.3562)]]}
\end{itemize}
Our repository includes the code to generate this output~\cite{repository}.

\begin{figure*}[!htbp]
    \vspace{-1mm}
    \centering

    \begin{subfigure}[!htbp]{0.45\columnwidth}
        \centering
        \frame{\includegraphics[width=1\columnwidth]{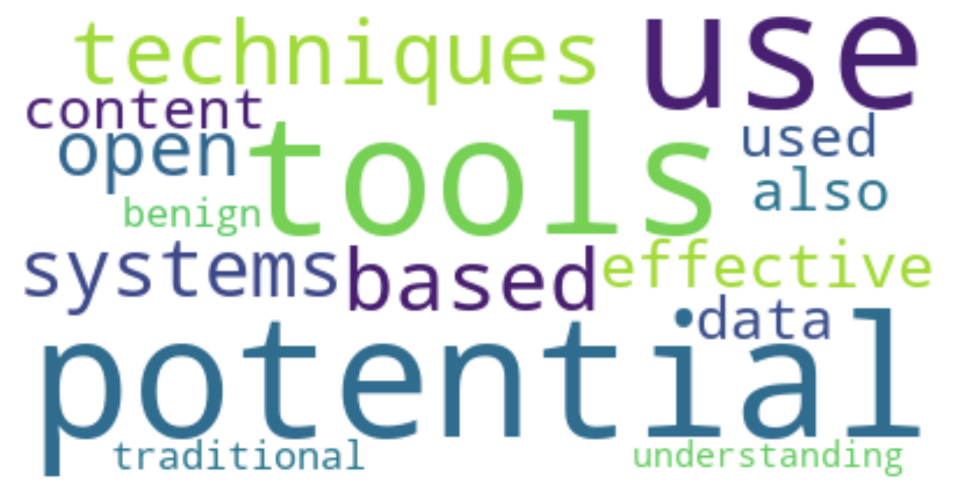}}
        \caption{0: AI Threats and Defenses}
        \vspace{3mm}
        \label{sfig:wordcloud_0}
    \end{subfigure} 
    \begin{subfigure}[!htbp]{0.45\columnwidth}
        \centering
        \frame{\includegraphics[width=1\columnwidth]{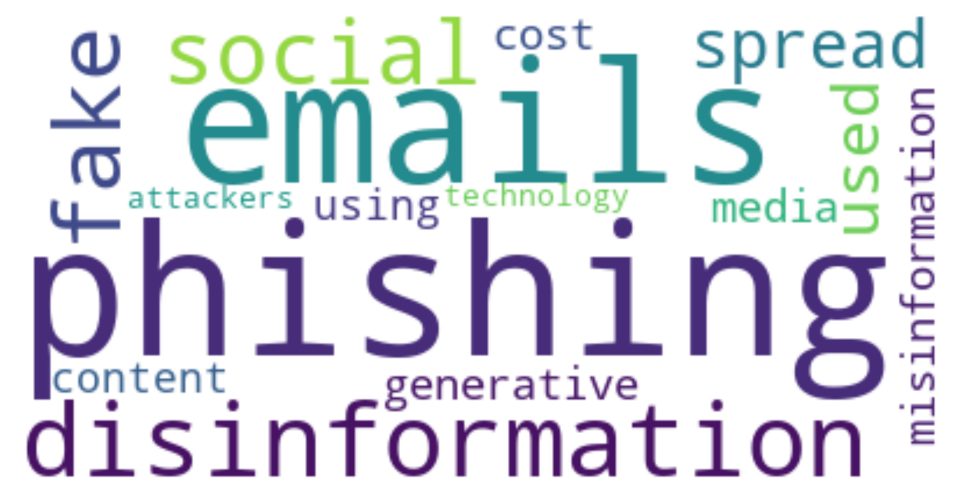}}
        \caption{1: AI-Powered Manipulation}
        \vspace{3mm}
        \label{sfig:wordcloud_1}
    \end{subfigure} 
    \begin{subfigure}[!htbp]{0.45\columnwidth}
        \centering
        \frame{\includegraphics[width=1\columnwidth]{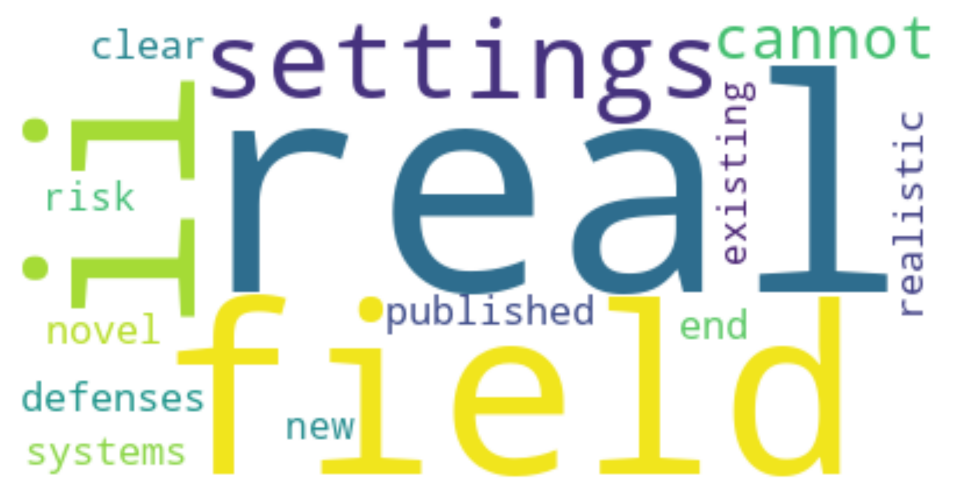}}
        
        \caption{2: Risks and Challenges in Offensive AI Research}
        \vspace{0mm}
        \label{sfig:wordcloud_2}
    \end{subfigure} 
    \begin{subfigure}[!htbp]{0.45\columnwidth}
        \centering
        \frame{\includegraphics[width=1\columnwidth]{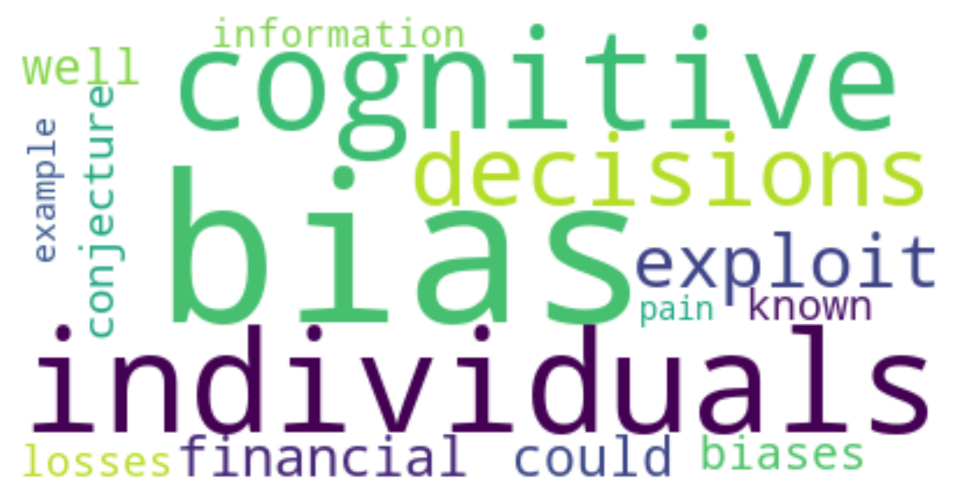}}
        \caption{3: Cognitive Bias Manipulation}
        \vspace{3mm}
        \label{sfig:wordcloud_3}
    \end{subfigure} 
    
    \begin{subfigure}[!htbp]{0.45\columnwidth}
        \vspace{2.5mm}
        \centering
        \frame{\includegraphics[width=1\columnwidth]{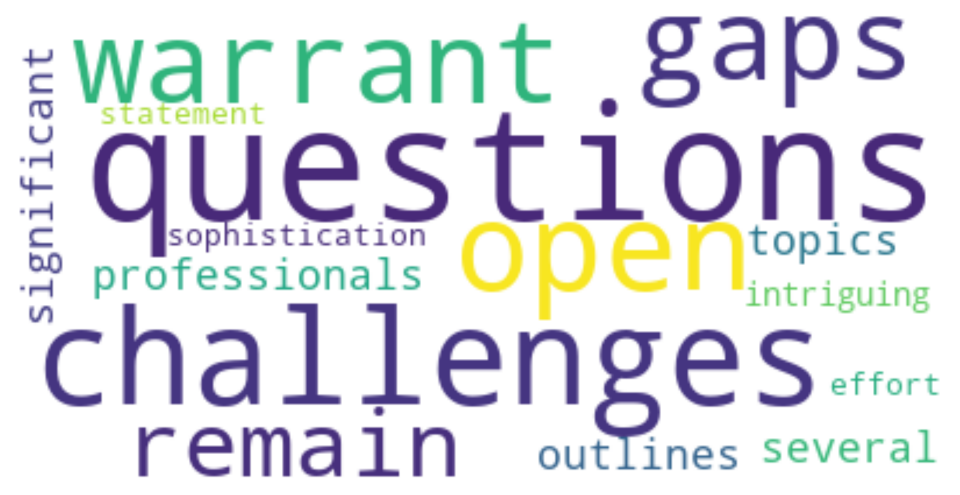}}
        \caption{4: Engineering Challenges in Addressing Open Research Questions}
        \label{sfig:wordcloud_4}
    \end{subfigure} 
    \begin{subfigure}[!htbp]{0.45\columnwidth}
        \centering
        \frame{\includegraphics[width=1\columnwidth]{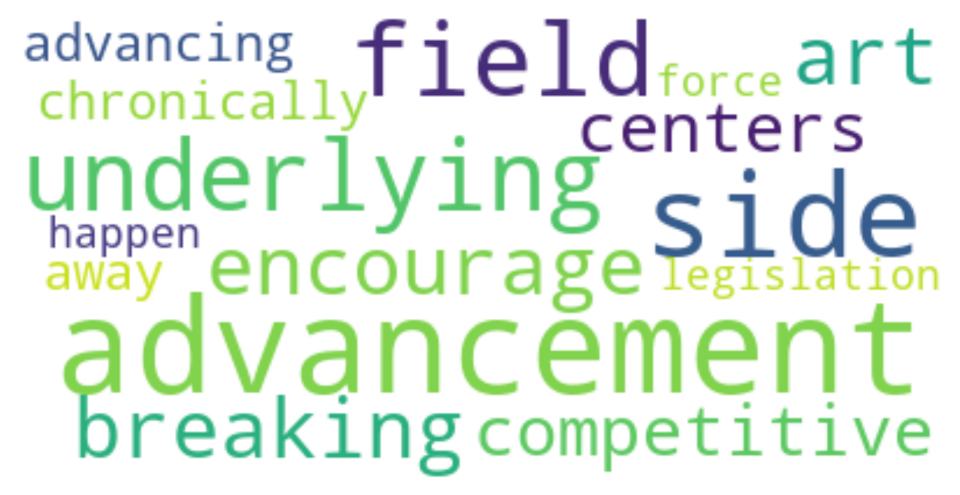}}
        \caption{5: Offensive Security Research}
        \label{sfig:wordcloud_5}
    \end{subfigure} 
    \begin{subfigure}[!htbp]{0.45\columnwidth}
        \centering
        \frame{\includegraphics[width=1\columnwidth]{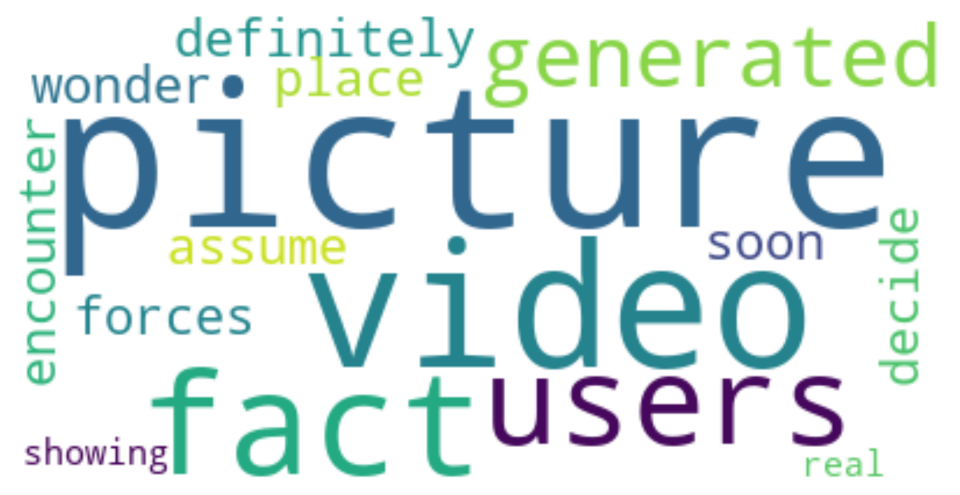}}
        \caption{6: Reality vs AI-Generated Content}
        \label{sfig:wordcloud_6}
    \end{subfigure} 
    \begin{subfigure}[!htbp]{0.45\columnwidth}
        \centering
        \vspace{2.5mm}
        \frame{\includegraphics[width=1\columnwidth]{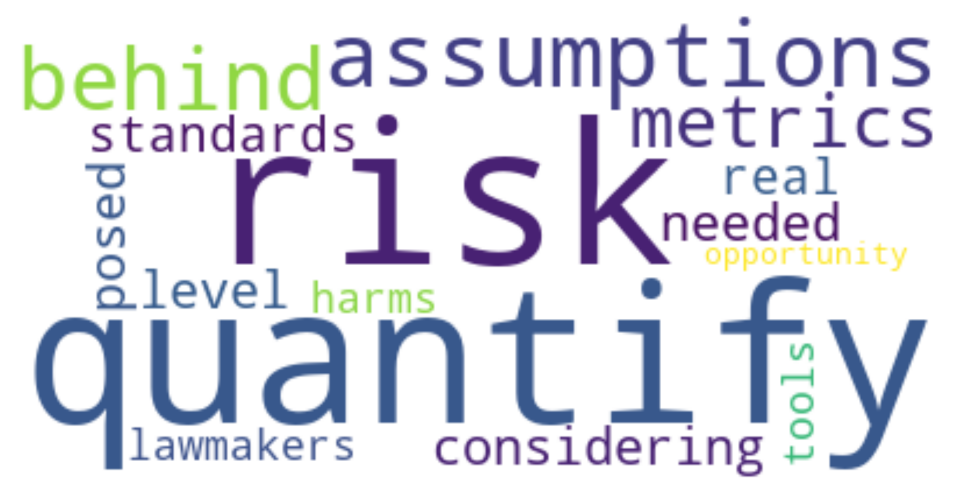}}
        \caption{7: Risk Assessment and Quantification for Offensive AI Tools}
        \label{sfig:wordcloud_7}
    \end{subfigure} 
    
    \caption{\textbf{Word Clouds of the topics identified by BERTopic.}
    \textmd{\footnotesize Each subfigure reports the word cloud of each of the 8 topics identified by BERTopic.}} 
    \label{fig:wordclouds}
    \vspace{-2mm}
\end{figure*}

\begin{figure*}[!htbp]
    \vspace{-3mm}
    \centering
    \centerline{
    \includegraphics[width=1\textwidth]{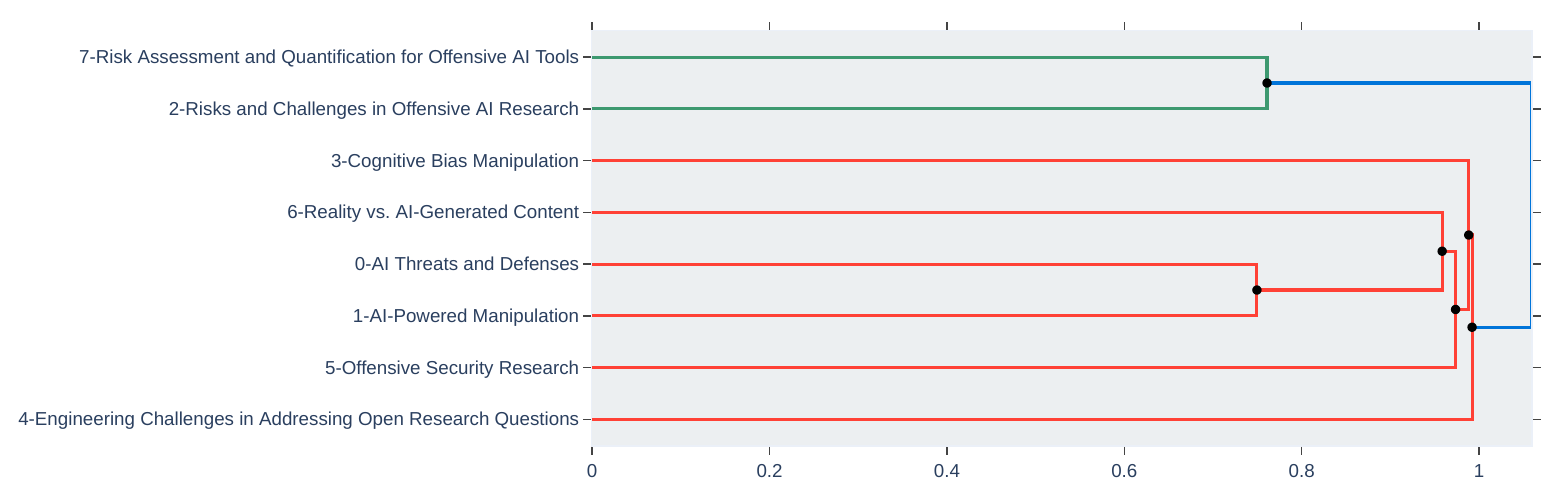}
    }
    \vspace{-2mm}
    \caption{\textbf{Hierarchical clustering of the topics identified by BERTopic.}
    \textmd{\footnotesize We visualize the 8 topics identified by BERTopic to discern similarities.}} 
    \label{fig:clustering}
    \vspace{-2mm}
\end{figure*}

\subsubsection{\textbf{Topic Modeling}}
\label{ssapp:topic}
Finally, we use topic modeling to extract the most relevant ``topics'' envisaged by our experts. To this end, we rely on BERTopic~\cite{grootendorst2022bertopic} (used in, e.g.,~\cite{govindarajan2023forecasting}). Specifically, BERTopic takes as input a collection of documents, and returns as output a finite number of ``topics:'' each topic is a list of words which collectively represent a certain concept; each word of a topic has also a certain weigh in the overall definition of the overarching concept. To apply BERTopic, we proceed as follows:
\begin{itemize}[leftmargin=*]
    \item We take all the expert statements and split them into individual sentences, totaling in 218 sentences (5,082 words).
    \item After preprocessing these sentences (using~\cite{bge_embedding}), we send them to BERTopic~\cite{grootendorst2022bertopic}, specifying to output the 8 most relevant topics (we follow the default configuration of~\cite{grootendorst2022bertopic}).
    \item We take the 8 topics provided by BERTopic, remove noisy stopwords (the same we considered for the keyword extraction), and use LLAMA2~\cite{touvron2023llama} to ``label'' each topic.
\end{itemize}
The results of these operations are summarized in Fig.~\ref{fig:topics}, showing the 8 topics (labeled by LLAMA2) provided by BERTopic, alongside the weights of the six most-relevant words for each topic. We also report in Figs.~\ref{fig:wordclouds} the word clouds defining each topic. Finally, we show in Fig.~\ref{fig:clustering} the visualization of the 8 topics after having been analyzed via an hierarchical clustering algorithm (integrated in BERTopic~\cite{grootendorst2022bertopic}), allowing to discern how similar these topics are to each other. We provide the low-level source code of the abovementioned operations in our repository~\cite{repository}.

\subsection{Authorship and Credits}
\label{sapp:authorship}

\noindent
This SoK is the result of a collective effort stemming from many individuals---including the 12 experts that provided their statements. However, the contributions of these 12 experts go beyond the mere 300--500 words paragraph outlining open problems on OAI---which, we remark, represents the basis of one of this SoK's major contributions (\smabb{C}3).

First, the experts provided a first-round of feedback after reading our draft (which ended after the current Section~\ref{sec:5-userSurvey}, and for which Section~\ref{sec:2-preliminaries} was still incomplete) alongside providing their statements. No critical issues had been identified with our overarching goal and research methodology, but their remarks were invaluable in identifying and addressing some problems in the presentation and scope of our work. Then, the experts also provided a second-round of feedback on a revised version of our draft, which included the systematization of the experts' statements into the set of open problems and concerns on OAI (i.e., the current Section~\ref{sec:experts}) as well as the discussion and conclusions (Sections~\ref{sec:discussion} and~\ref{sec:conclusions}). The experts' remarks also encompassed the appendices. All of these contributions improved the quality of our SoK tremendously.
In addition, the experts also contributed by: {\small \textit{(i)}}~assisting in the revision of our SoK after receiving the reviews for another version of this work submitted to another top-tier security venue; {\small \textit{(ii)}}~approving the ``submitted' version of this paper for SaTML25; and {\small \textit{(iii)}}~assisting in the rebuttal phase of SaTML25, including the preparation of this ``revised'' version of our SoK. 

Due to the above mentioned reasons, the 12 experts have been listed among the co-authors of this paper. Every co-author of this work fulfills the authorship criteria embraced by IEEE~\cite{ieeeauthorship}. For transparency, we provide below our CrediT statement~\cite{elseviercredits}, outlining how each co-author contributed to this SoK; we also report the full details of each author (some of which were omitted from this paper's first page).

\begin{itemize}[leftmargin=*]
    \item \textbf{Saskia Laura Schröer} (\textit{saskia.schroeer@uni.li}): Conceptualization, Methodology, Software, Formal analysis, Validation, Investigation, Data Curation, Writing (Original Draft, Review, Editing), Visualization.
    \item \textbf{Giovanni Apruzzese} (\textit{giovanni.apruzzese@uni.li}): Conceptualization, Methodology, Validation, Visualization, Writing (Original Draft, Review, Editing), Supervision
    \item \textbf{Soheil Human} (\textit{soheil.human@wu.ac.at}): Conceptualization, Methodology, Software, Validation, Formal Analysis, Investigation, Writing (Original Draft, Review, Editing), Visualization
    \item \textbf{Pavel Laskov} (\textit{pavel.laskov@uni.li}): Conceptualization, Formal Analysis, Resources, Writing (Original Draft, Review, Editing), Validation, Supervision, Project Administration, Funding acquisition
    \item \textbf{Hyrum S. Anderson} (\textit{hyrum@robustintelligence.com}): Writing (Review \& Editing), Validation
    \item \textbf{Edward W. N. Bernroider} (\textit{edward.bernroider@wu.ac.at}): Writing (Review \& Editing), Validation
    \item \textbf{Aurore Fass} (\textit{fass@cispa.de}) [affiliated with CISPA Helmholtz Center for Information Security]: Writing (Review \& Editing), Validation
    \item \textbf{Ben Nassi} (\textit{nassiben@technion.ac.il}) [affiliated with Technion - Israel Institute of Technology]: Writing (Review \& Editing), Validation
    \item \textbf{Vera Rimmer} (\textit{vera.rimmer@kuleuven.be}) [affiliated with DistriNet @ KU Leuven]: Writing (Review \& Editing), Validation
    \item \textbf{Fabio Roli} (\textit{fabio.roli@unige.it}): Writing (Review \& Editing), Validation
    \item \textbf{Samer Salam} (\textit{ssalam@cisco.com}): Writing (Review \& Editing), Validation
    \item \textbf{Ashley Shen} (\textit{ashlshen@cisco.com}): Writing (Review \& Editing), Validation
    \item \textbf{Ali Sunyaev} (\textit{sunyaev@kit.edu}): Writing (Review \& Editing), Validation
    \item \textbf{Tim Wadhwa-Brown} (\textit{twadhwab@cisco.com}): Writing (Review \& Editing), Validation
    \item \textbf{Isabel Wagner} (\textit{isabel.wagner@unibas.ch}): Writing (Review \& Editing), Validation
    \item \textbf{Gang Wang} (\textit{gangw@illinois.edu}) [affiliated with the University of Illinois Urbana-Champaign]: Writing (Review \& Editing), Validation
\end{itemize}
In the authors' list, the 12 experts have been alphabetically ordered.

\section{Extra Information on our Research Methods}
\label{app:extra_figures}

\subsection{Papers and Briefings: comparison and discipline}
\label{app:sok}
\begin{figure}[!htbp]
\vspace{-3mm}
    \centering
    \centerline{
    \includegraphics[width=0.4\textwidth]{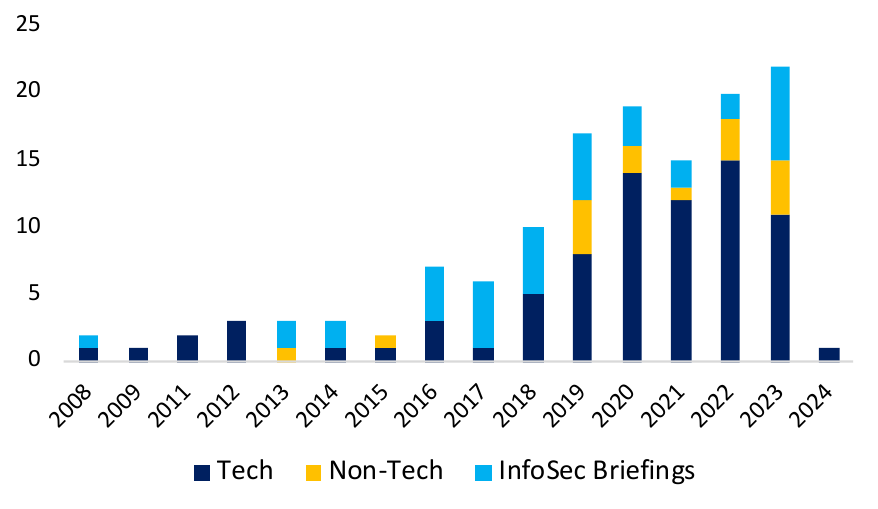}
    }
    \vspace{-2mm}
    \caption{\textbf{All works on OAI per year.}
    \textmd{\footnotesize We present the yearly distribution of the works on OAI considered in this SoK, distinguishing technical and non-technical academic publications (§\ref{sec:3-academic}) from InfoSec briefings (§\ref{sec:4-infosec}).}} 
    \label{fig:works}
    \vspace{-3mm}
\end{figure}

\noindent
We provide in Fig.~\ref{fig:works} the yearly distribution of the academic papers and InfoSec briefings over time. 

Moreover, we find it instructive to analyze the \textit{discipline} of the venues of each work included in our literature systematization. Indeed, recall that our literature search encompassed repositories (§\ref{ssec:2.1-literature}) of a wide range of scientific disciplines. It is hence insightful to highlight such a diversity---especially given that OAI is a theme that can be tackled from diverse perspectives. To infer the discipline of each venue in an objective way, we relied on the Scimago database~\cite{scimago}: by querying this database with the name of a given venue, the database returns metadata of such venue---including its ``subject area and category,'' such as Computer Science (CS) or Engineering. We hence query this database for each venue of the 95 papers included in our literature systematization, and report the ``primary'' subject area and category according to Scimago. The results are in the ``Discipline'' column in Tables~\ref{tab:technical} and~\ref{tab:related-non-tech}. We make some remarks.
\begin{itemize}[leftmargin=*]

    \item \textit{Multi-disciplinary works.} Some venues in the Scimago database are associated with more than one subject area. For instance, ``Applied  Sciences'' (the venue in which, e.g., the paper by Yu et al.~\cite{yu2019low} was published) is associated to Engineering, Physics, and CS (among others). In these cases, we infer the most appropriate discipline depending on the corresponding paper. E.g., for the work by Yu et al.~\cite{yu2019low}, we assigned CS since it was the closest match.
    
    \item \textit{Indexing.} The Scimago database is large and extensively curated, but some venues are not indexed. For instance, we could not find any entry for the ``European Interdisciplinary Cybersecurity Conference'' (which is the venue of~\cite{ozturk2023new}). In these cases, we assigned a discipline after qualitatively scrutinizing the works published in such venue. Nonetheless, lack of indexing should not be taken as poor value of any work in our systematization: first, because some venues are still supported by reputable scientific organizations (e.g., the ``European Interdisciplinary Cybersecurity Conference'' is affiliated to the ACM~\cite{acmconf}); second, because 94 out of 95 papers have passed peer-review (the only ones for which no published version exists are the paper by Tran et al.~\cite{tran2021deep}, which currently has over 45 citations on Google Scholar; and the one by Toemmel~\cite{toemmel2021catch}).

    \item \textit{Ranking.} For most venues, Scimago also provides a ranking, expressed in terms of ``quartiles'' (e.g., Q1--Q4, with Q1 being the highest rank). We preferred not to report this information in our Tables. First, because venue ranking is not necessarily proof of a paper's quality (there are fundamental flaws even in papers accepted to IEEE S\&P~\cite{apruzzese2023real}). Second, because it would be misleading since some disciplines may have different rankings: for instance, the work by Iqbal et al.~\cite{iqbal2023chatgpt} was published in ``Frontiers in Communication and Networks'', which is a journal whose primary field is CS, but for which the ranking is provided for two categories of CS (Q1 for ``Computer Networks and Communications'' and Q2 for ``Signal Processing''). Third, because some venues (especially conferences) do not have any quartile: for instance, the work by Anand et al.~\cite{anand2018keyboard} was published in CODASPY'18 which is not ranked (an alternative would be to use other rankings, e.g., CORE, but this would create issues for multidisciplinary venues). Therefore, we preferred not to provide any ranking-related information in our Tables.
\end{itemize}
By observing the Disciplines in Table~\ref{tab:technical}, we see that most works are from CS: only one is from Economics~\cite{khan2021offensive}, and two from Engineering~\cite{neal2021reinforcement, confido2022reinforcing}. This is expected given that the papers in Table~\ref{tab:technical} are highly technical. In contrast, for non-technical papers in Table~\ref{tab:related-non-tech}, there are eleven works from CS, but four works are published in Social Sciences venues, and one even for Medicine~\cite{de2023chatgpt}. Therefore, we endorse future research to build on our systematization by considering papers from diverse fields.

\subsection{History of the term ``offensive AI''}
\label{sapp:background-OAI}

\begin{figure}[!htbp]
\vspace{-3mm}
    \centering
    \centerline{
    \includegraphics[width=0.4\textwidth]{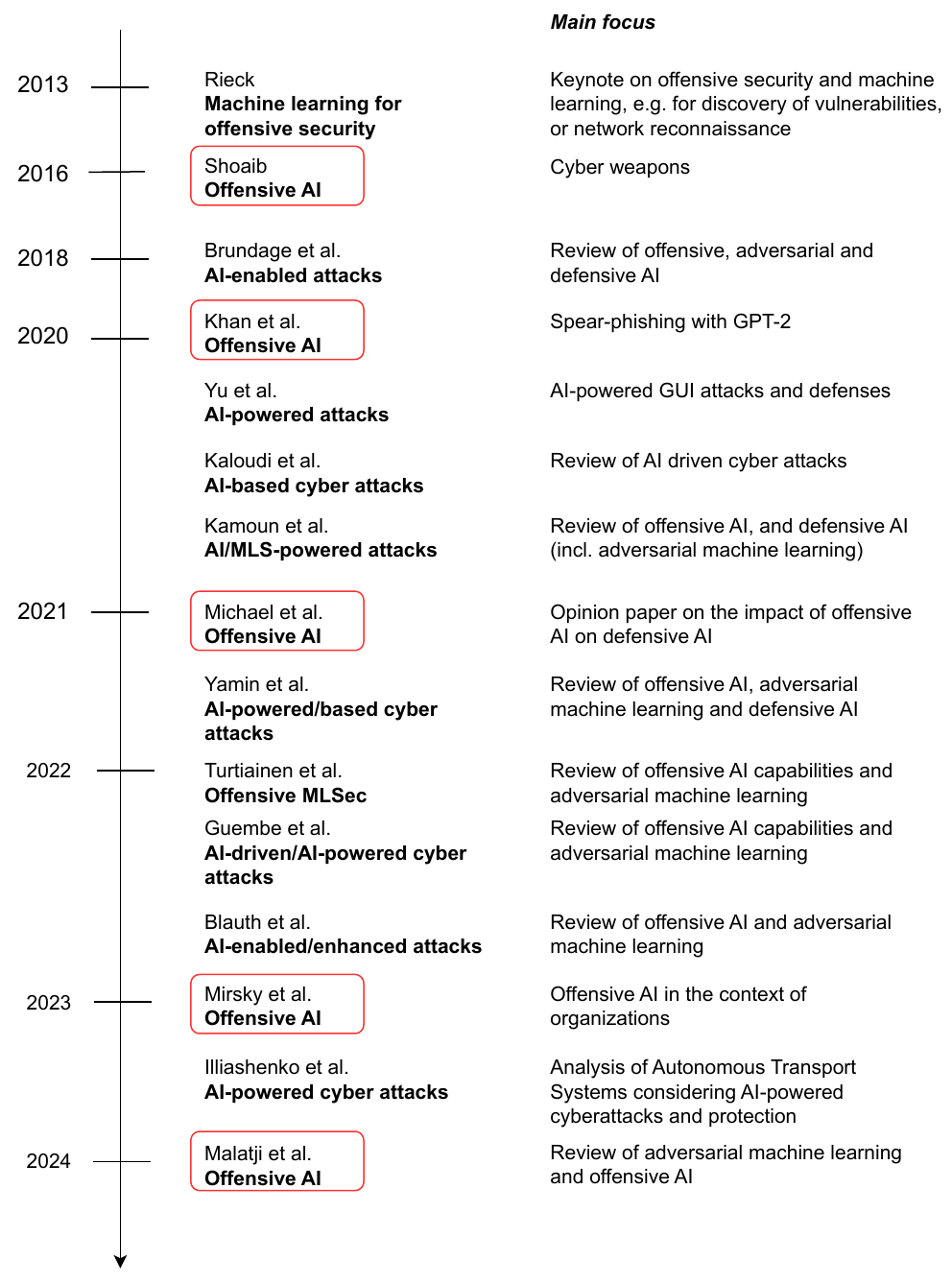}
    }
    \vspace{-2mm}
    \caption{\textbf{The evolution of the term ``offensive AI.''}
    \textmd{\footnotesize The term offensive AI (and similar related terms) has substantially evolved over time.}}
    \label{fig:oai-term}
    \vspace{-3mm}
\end{figure}

\noindent
In Figure~\ref{fig:oai-term} we provide a supplementary illustration of the different terms that have been used to refer to offensive AI.

\begin{figure*}[!htbp]
    \vspace{-3mm}
    \centering
    \centerline{
    \includegraphics[width=0.99\textwidth]{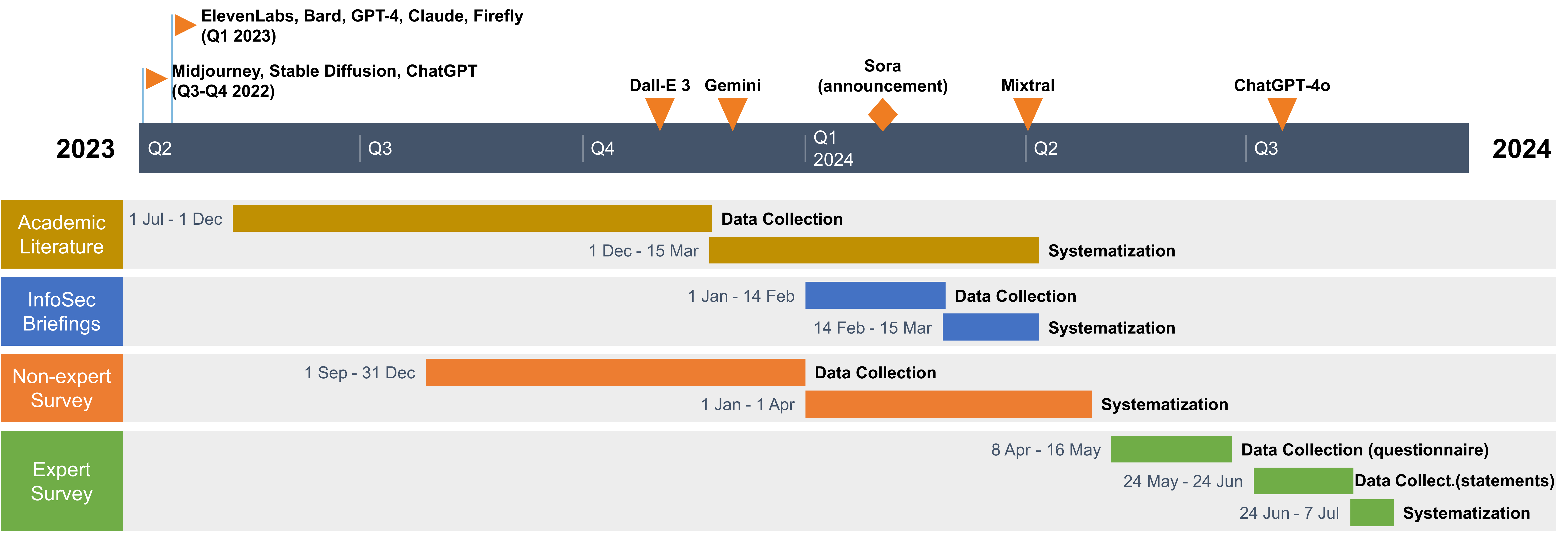}
    }
    \vspace{-2mm}
    \caption{\textbf{Timeline of our Research.}
    \textmd{\footnotesize We began working on this SoK at the beginning of 2023. Throughout the entire activities shown in the figure, we have also had frequent meetings to steer the direction of our research and also to revise the paper. For instance, gaps in the ``expert survey'' swimlane are due to the necessity of analysing the experts' initial input before giving them the draft of our SoK. The timeline also shows important milestones in the context of OAI, represented by the (public) release of popular large language models that can process text, audio, images, or video content. Sources:~\cite{midjourney, stablediffusion, chatgpt, elevenlabs, googlebard, claude, firefly, dalle3, googlegemini, sora, mixtral, chatgpt4, gpt4}}} 
    \label{fig:timeline}
    \vspace{-3mm}
\end{figure*}

\subsection{Timeline of our research (and challenges)}
\label{sapp:timeline}

\noindent
To realize our SoK, we carried out diverse research activities. We provide a timeline in the Gantt chart shown in Fig.~\ref{fig:timeline}. 
In what follows, we will describe the temporal evolution of our activities, describing also some challenges we encountered.

We began with a literature review---as is the case when approaching novel research directions. We started to do this in Summer 2023. During this investigation, we realized that the term ``offensive AI'' was not widespread in the literature, and that many papers proposed AI applications that could be used offensively---but did not use the term ``offensive AI'' (or a derivative) in the paper. Such preliminary findings prompted us to adopt a more systematic approach which entailed a qualitative assessment: after identifying a set of candidate papers, we would scrutinize such papers to determine if they fell in our own definition of the term ``offensive AI.'' As an additional sanity check, we have also tried to replicate the procedure followed by some prior work~\cite{arp2022and}, i.e., by looking only at papers published in top-tier venues---intent in finding the coverage of OAI in this select number of venues. To this end, we considered all publications in the top-4 security conferences (NDSS, CCS, S\&P, USENIX SEC) between 2021--2023, and we carried out a simple keyword search. Specifically, we looked for any full paper that had the term ``offensive AI'' / ``offensive artificial intelligence'' / ``offensive machine learning'' / ``offensive ML'' in either the abstract or title. We found only one paper that matched this criteria~\cite{si2022so}. We believe this number underestimates the real number of publications discussing Offensive AI (as evidenced, e.g., by Mirsky et al.~\cite{mirsky2021threat}). Therefore, such a finding confirms that our choice of {\small \textit{(i)}}~carrying out a larger search (encompassing four popular scientific repositories) that considers ``any'' type of peer-reviewed work (irrespective of the ``ranking'' of any given venue); and then {\small \textit{(ii)}}~qualitatively analyzing each returned paper to determine if it fell in our definition of OAI, to be appropriate for our goals. Notwithstanding, the data collection phase of our literature analysis (for which we considered papers matching our queries and taken from IEEE Xplore, the ACM DL, arXiv, and Google Scholar) spanned between July and December 2023: during this timeframe we also progressively screened the papers returned by our search queries. We ultimately stopped our literature search, in December 2023, at which point we obtained a set of 95 papers that fell in our definition of OAI. We began systematizing these papers with the goal of identifying relevant aspects related to OAI---captured by our checklist; this analysis required several months of work by multiple authors, and terminated in March 2024. 

The second major activity we have carried out was the user study with non-experts. The idea of carrying out this study was inspired by the initial findings of our literature analysis. Specifically, we conjectured that the potential applications of OAI were so diverse that it could be insightful to ask ``non-experts'' about their opinion on OAI. We designed our questionnaire and began disseminating it in September 2023. We stopped collecting responses at the end of December 2023. We then analyzed the collected responses and derived our codebook (which we knew would be used also later for analyzing the experts' input).

The third activity we have carried out is the analysis of InfoSec briefings---which spanned between Jan. and March 2024. We posited that these venues could provide a complementary perspective on the ``practical'' use cases of OAI in the real world---especially given that not many research papers showcased real-world demonstrations of OAI. This procedure was not trivial---despite the existence of a much lower number of InfoSec briefings than research papers. Indeed, while \textit{finding} the briefings was simple, \textit{ascertaining} whether a briefing is about OAI required us to watch the entire video (\smamath{\sim}30--40m long) of the presentation, in some cases. Moreover, to systematically analyze the OAI-related briefings, we also had to rely on the video. This may explain why considering InfoSec briefings is uncommon in the SoK literature.

The last activity we have carried out is the user study with experts. First, we stress that most of the experts we contacted accepted our request to participate in our research. After finding an agreement with 12 experts, we distributed our questionnaire; we did this in April 2024. Then, in the following weeks, we analyzed the responses we collected and also assembled what was the initial draft of our SoK. We shared this draft with the experts at the end of May 2024, and gave each expert 3 weeks of time to provide their statement of 300--500 words (one expert was late and submitted their contribution 1 week later). At the end of June 2024, we received all the statements and systematically analyzed them. We then shared an improved version of our SoK with the experts so that they could provide feedback---which was invaluable to clarify misunderstandings and improve the clarity of our work.

\vspace{1mm}
{\setstretch{0.7}
\textbox{{\small \textbf{Real-world developments.} In Fig.~\ref{fig:timeline}, we have also reported some major real-world events in the context of OAI. Specifically, we consider that \textit{the public release of LLM}, which undoubtedly enable anyone (including evildoers) to use AI, represent important milestones. We observe that \textit{all our research activities have been carried out well after} the public release of powerful LLMs (such as Bard, ChatGPT, Claude, Stable Diffusion). Therefore, we find it unlikely that, e.g., the rollout of Dall-E3 or Gemini may have impacted the responses of our non-expert survey; at the same time, the release of ChatGPT-4o is also unlikely to have substantially affected the ideas of our experts. Ultimately, the AI field is very fast paced, and educated people are likely aware of this fact.}}}
\vspace{-1mm}

\end{document}